%% file: paperfluxesv15.tex
\documentclass[11pt,a4paper]{article}
 \usepackage[latin1]{inputenc}
 \usepackage[UKenglish]{babel}
 \usepackage{myway}
 \usepackage{multirow}
 \usepackage{graphicx}
 \usepackage{epstopdf}
 \usepackage{subfigure}
	
\hypersetup{pdftitle={Gauge Fluxes in F-Theory and Type IIB Orientifolds}, 
pdfauthor={Sven Krause, Christoph Mayrhofer, Timo Weigand}, 
pdfsubject={String theory model building, F-theory},
bookmarksopen, bookmarksnumbered, bookmarksopenlevel=2}

\newcommand{\beq}{\begin{equation}} 
 \newcommand{\eeq}{\end{equation}}
\newcommand{\bal}{\begin{aligned}}  
 \newcommand{\eal}{\end{aligned}}
\newcommand{\bea}{\begin{eqnarray}} 
 \newcommand{\eea}{\end{eqnarray}}
 \def\ov{\overline}

\newcommand{\ccP}{\mathcal{P}}

\newcommand{\bbP}{\mathbb{P}}
\newcommand{\bbZ}{\mathbb{Z}}

\newcommand{\tw}{\text{w}}

\newcommand{\cW}{{\cal W}}
\newcommand{\cB}{{\cal B}}
\newcommand{\cF}{{\cal F}}
\newcommand{\aK}{\bar{\cal K}}
\newcommand{\tY}{\tilde{Y}_4}

\newcommand{\tG}{\tilde{G}_4}
\newcommand{\tGP}{\tG^{\,X}(\ccP)}

\newcommand{\eps}{\epsilon}
\newcommand{\tD}{\tilde{D}}
\newcommand{\tF}{\tilde{F}}
\newcommand{\cP}{{\cal P}}

\newcommand{\tsp}{\rule{0pt}{2.6ex}}

\newcommand{\executeiffilenewer}[3]{%
 \ifnum\pdfstrcmp{\pdffilemoddate{#1}}%
 {\pdffilemoddate{#2}}>0%
 {\immediate\write18{#3}}\fi%
}

\title{Gauge Fluxes in F-theory and Type IIB Orientifolds}

\preprint{}

\author%
{Sven Krause,%\note[1]{\href{mailto:S.Krause@ThPhys.Uni-Heidelberg.de}{S.Krause@ThPhys.Uni-Heidelberg.de}}%
}
\author%
{Christoph Mayrhofer%\note[2]{\href{mailto:C.Mayrhofer@ThPhys.Uni-Heidelberg.de}{C.Mayrhofer@ThPhys.Uni-Heidelberg.de}}%
}
\author%
{and Timo Weigand%\note[3]{\href{mailto:T.Weigand@ThPhys.Uni-Heidelberg.de}{T.Weigand@ThPhys.Uni-Heidelberg.de}}%
}
\affiliation{Institut f\"ur Theoretische Physik, Universit\"at Heidelberg, Philosophenweg 19, D-69120 Heidelberg\vspace{0.1cm}}

\emailAdd{S.Krause@ThPhys.Uni-Heidelberg.de} 
\emailAdd{C.Mayrhofer@ThPhys.Uni-Heidelberg.de}
\emailAdd{T.Weigand@ThPhys.Uni-Heidelberg.de}

\abstract{ 

We provide a detailed correspondence between $G_4$ gauge fluxes in F-theory compactifications with $SU(n)$ and $SU(n) \times U(1)$ gauge symmetry and their Type IIB orientifold limit. Based on the resolution of the relevant F-theory Tate models, we classify the factorisable $G_4$-fluxes and match them with the set of universal D5-tadpole free $U(1)$-fluxes in Type IIB. 
Where available, the global version of the universal spectral cover flux corresponds to Type IIB gauge flux associated with a massive diagonal $U(1)$. In $U(1)$-restricted Tate models extra massless abelian fluxes exist which are associated with specific linear combinations of Type IIB fluxes.
Key to a quantitative match between F-theory and Type IIB is a proper treatment of the conifold singularity encountered in the Sen limit of generic F-theory models. 
We also shed further light on the brane recombination process relating generic and $U(1)$-restricted Tate models.
}
\begin{document}

\maketitle
\newpage

%%%%%%%%%%%%%%%%%%%%%%%%%%%%%%%%%%%%%%%%%%%%%%%%%%%%%%%%%%%%%%%%%%%%%%%%%%%%%%%%%%%%%%%%%%%%%%%%%%%%%%%%%%%%%%%%%%%%%%%%

\section{Introduction}

Our understading of four-dimensional F-theory \cite{Vafa:1996xn,Morrison:1996na,Morrison:1996pp} vacua as compactifications on  singular elliptic fibrations supplemented by consistent $G_4$-fluxes has improved considerably over the past few years. 
Apart from representing interesting examples of non-perturbative string dynamics in their own right, F-theory compactifications on singular Calabi-Yau four-folds have attracted a lot of recent attention in the context of 4-dimensional GUT model building \cite{Donagi:2008ca,Beasley:2008dc,Beasley:2008kw,Donagi:2008kj,Hayashi:2008ba}.
The singularities of the internal elliptic Calabi-Yau four-fold $Y_4$ are in one-to-one correspondence with the gauge groups along 7-branes \cite{Bershadsky:1996nh}, the matter fields at the intersection of 7-branes \cite{Katz:1996xe} and their Yukawa interactions \cite{Beasley:2008dc, Donagi:2008ca,Hayashi:2009ge}, as reviewed e.g.~in \cite{Denef:2008wq,Weigand:2010wm}. 
Of particular importance for a well-defined dimensional reduction is the resolution of these singularities in terms of a smooth Calabi-Yau four-fold $\hat Y_4$. Inspired by  $SU(5)$ GUT model building,  fully-fledged Calabi-Yau  four-folds and their resolutions have been constructed via toric methods  in \cite{Blumenhagen:2009yv,Grimm:2009yu} and \cite{Chen:2010ts,Knapp:2011wk,mayrhofer:diss} (see \cite{Marsano:2009ym} for the construction of a base space of such fibrations). Subsequently, a more detailed analysis of higher codimension singularities in $SU(5)$ models has been provided in the resolution schemes of \cite{Esole:2011sm,Marsano:2011hv,Krause:2011xj,Grimm:2011fx}.\footnote{Recent progress concerning six-dimensional F-theory vacua has been achieved e.g.~in \cite{Kumar:2009ac,Grassi:2011hq,Braun:2011ux,Park:2011ji,Bonetti:2011mw,Morrison:2012td}.}

\subsection{The Quest for Gauge Fluxes}

In this article we focus on a better understanding of the flux sector of F-theory compactifications. Gauge fluxes are key to model building because they are responsible, among other things, for a chiral matter spectrum. Various types of chirality inducing gauge fluxes have been constructed recently in terms of $G_4$-fluxes defined on a resolved four-fold $\hat Y_4$ \cite{Braun:2011zm,Marsano:2011hv,Krause:2011xj,Grimm:2011fx}. 
While this marks an important step forward in F-theory model building, a number of open questions concerning the flux sector remain. This has to do with the fact that $G_4$-fluxes are given by harmonic four-forms in $H^{2,2}(\hat Y_4)$, while the gauge flux along a 7-brane $W$ is thought of as a two-form $F \in H^{1,1}(W)$. 
In fact, most of the information on the gauge flux $F$ is encoded only very implicitly in $G_4$, and the precise relation between fluxes in F-theory and Type IIB is not obvious. Typical open questions concern the localization of fluxes on individual branes, the appearance of D-terms or the role of the D5-brane tadpole \cite{Grimm:2011tb}.
Our aim is to systematize the construction of gauge fluxes in $SU(n)$ and $SU(n) \times U(1)$ F-theory models and to gain a better intuition for these fluxes by relating them to gauge fluxes in the  Type IIB limit. Along the way we will also make contact with the semi-local description of gauge fluxes via the spectral cover~\cite{Donagi:2008ca,Hayashi:2008ba,Donagi:2009ra,Marsano:2009gv,Blumenhagen:2009yv,Marsano:2011nn,Wijnholt:2012fx}. Somewhat surprisingly, we will identify the universal spectral cover fluxes with specific fluxes in Type IIB. 

Before summarizing our main findings in section \ref{intro-sec:results}, we now set out to provide some more background on the construction of fluxes in Type IIB, via spectral covers and as $G_4$-fluxes.

The probably most intuitive and familiar way to think about gauge theories on 7-branes is in terms of a Type IIB orientifold on a Calabi-Yau $X_3$ with stacks of 7-branes along holomorphic divisors.
Chirality inducing gauge fluxes arise as the $U(1)$-fluxes associated with the diagonal $U(1)$ of the $U(N)$ gauge groups realised on the 7-branes at generic position.
To the extent that each such brane stack comes with its own $U(1)$-gauge potential and thus its own set of fluxes, gauge fluxes are localised on the branes. A more mathematical way to phrase this is that they are simply the first Chern class of the $U(1)$-bundles over the various 7-branes. The flux degrees of freedom assemble, at first sight, into a vector $F \in H^{1,1}(D_A)$ on each brane $D_A$.
Being related to a $U(1)$-symmetry, each such flux induces a field dependent Fayet-Iliopoulos term (in the usual abuse of nomenclature) entering the D-term condition for the respective 7-brane. While chiral indices at the brane intersection depend only on the flux difference between the intersecting branes, topological invariants such as the 3-brane tadpole are sensitive to the local flux data along each individual 7-brane.

If we uplift a Type IIB model to F-theory we can also consider the semi-local neighbourhood of one of the non-abelian brane stacks, thereby making contact with the spectral cover approach  to F-theory model building. A generic, say, $SU(5)$ model is described semi-locally by a Higgs bundle over the GUT brane $S$, whose structure group is the commutant $SU(5)_\perp$ within an underling $E_8$ gauge group, $E_8 \rightarrow SU(5) \times SU(5)_\perp$  \cite{Donagi:2008ca,Hayashi:2008ba,Donagi:2009ra,Wijnholt:2012fx}. 
Naively the associated fluxes seem of a completely different nature than the ones in Type IIB models: First, the role of the complementary $SU(5)_\perp$ is \emph{sans pareil}  in Type IIB. Also the number of flux degrees of freedom do not seem to match:
The spectral cover fluxes are described  in terms of a two-form in $\eta \in H^2(S)$, but this two-form is completely fixed once the geometry is specified. The degrees of freedom of these fluxes are merely one overall discrete parameter to be chosen in such a way that the fluxes are well-quantised. To confuse us even more,
since there is no extra $U(1)$ contained within $SU(5)$, the fluxes do not induce any D-term, again in contrast to Type IIB expectations. 
Does this mean that such spectral cover fluxes are truly non-perturbative and do not exist in Type IIB at all?
For non-generic, so-called split spectral covers, another set of fluxes arises \cite{Donagi:2009ra,Marsano:2009gv,Blumenhagen:2009yv,Marsano:2011nn,Wijnholt:2012fx}. These do induce a D-term and are thus closer to the IIB picture, but it is not clear at all where they are localized --- if that notion is appropriate in the first place.

In global F-theory compactifications gauge fluxes are encoded, via duality with M-theory, by $G_4$-fluxes.\footnote{In addition, $G_4$-fluxes describe the analogue of bulk fluxes in Type IIB, which are relevant for moduli stabilization. See e.g.~\cite{Lust:2005bd,Alim:2009bx,Grimm:2009ef} for an incomplete list of recent investigations of various aspects of such fluxes and their superpotentials in F-theory.}
As anticipated above, these are specified by certain elements of  $H^{2,2}(\hat Y_4)$ subject to a number of constraints which will be reviewed at the beginning of section \ref{subsec_G41}. In this picture the localization of gauge fluxes familiar from Type IIB models has become completely obscure.
A notable exception are fluxes associated with the Cartan $U(1)$s of a non-abelian gauge group $G$ along a divisor ${\cal W}$ on the base: The corresponding four-forms are of the type $F \wedge \tw_i$, where the two-forms $\tw_i, i=1, \ldots, {\rm rk}(G)$ are associated with the $\mathbb P^1$s in the fiber over ${\cal W}$ needed to resolve the singularity (more precisely, they are Poincar\'e dual to the divisor obtained by fibering these over ${\cal W}$), and $F\in H^{1,1}(B)$. This agrees with Type IIB intuition if we identify $F|_{\cal W}$ with the flux along the 7-brane. A recent discussion in particular of such Cartan fluxes has been given in \cite{Grimm:2012rg}. For non-Cartan fluxes, however, which are the fluxes that we are interested in here, the geometric picture sketched above does not apply, and an identification of $G_4$ with a flux $F$ along some 7-brane is difficult.

\subsection{Summary of Results} \label{intro-sec:results}

To match $G_4$ gauge fluxes with the analogous objects in Type IIB orientifolds and in the language of spectral covers, we will work in a  specific type of F-theory models.
Concretely, we construct the Tate model and its resolution corresponding to an $SU(n)$ gauge group along a divisor $W$ as well as the $U(1)$-restricted Tate model \cite{Grimm:2010ez}  leading to gauge group $SU(n) \times U(1)$.
Our first task in section \ref{sec:topological_invariants} is to compute the resolution divisors, their detailed intersection structure and the topological invariants of the resolution space, thereby generalising our previous analysis in \cite{Krause:2011xj}, which was valid for $n=5$. This serves two purposes: First, it provides the necessary topological data to quantitatively compare the F-theory geometry to its Type IIB weak coupling limit; second, the construction of $G_4$-fluxes hinges upon control over the four-forms of the resolved space ${\hat Y_4}$.  We then analyze the Type IIB limit of this class of F-theory models  in section \ref{sec:sen_limit}.  The pure $SU(n)$ models correspond to a Type IIB orientifold with one $U(n)$ brane stack and its image as well as one invariant 7-brane of Whitney-brane type \cite{Collinucci:2008pf}.
In the $SU(n) \times U(1)$ model, the latter splits into a brane/image-brane pair. 
When considering the Sen limit, an interesting complication arises that had first been observed in \cite{Donagi:2009ra} for $SU(5)$ models and in fact holds more generally: The Type IIB three-fold $X_3$ associated with a generic F-theory model with $SU(n)$ gauge group exhibits a conifold singularity. For $SU(5)$ models this singularity is related to the existence of the $E_6$-point at which the ${\bf 10 \, 10 \, 5}$ Yukawa couplings are realised. In the presence of such a conifold singularity, F-theory and Type IIB are not smoothly connected.  Our strategy is to impose certain constraints on the topology of the models such that this singularity is absent. Indeed for such models typical topological data such as the Euler characteristic on the one hand and the curvature induced 3-brane charge on the other match, as is verified in section \ref{sec:TopInv}. 

Having established a solid geometric foundation we can analyze the F-theory $G_4$ gauge fluxes in our class of models.
 On a Calabi-Yau four-fold $\hat Y_4$ there are two types of harmonic (2,2)-forms: four-forms which factorise into the wedge product of two-forms and those which do not. The subspace of $H^{2,2}(\hat Y_4)$ spanned by  linear combinations of factorisable four-forms is called primary vertical subspace $H^{2,2}_{\rm vert.}(\hat Y_4)$ \cite{Greene:1993vm,Mayr:1996sh}.
  In section \ref{subsec_G41}, we classify the primary vertical $G_4$-fluxes on the resolution four-folds associated with our $SU(n) [\times U(1)]$ models. For generic $SU(n)$ models, i.e.~for models with no extra $U(1)$-factor, no such fluxes exist if $n <5$ apart from the $SU(n)$ Cartan fluxes. By contrast, for $SU(n) \times U(1)$ models one can always construct gauge flux associated with the extra $U(1)$ as in \cite{Grimm:2010ez,Krause:2011xj,Braun:2011zm,Grimm:2011fx}. These are related to a class of non-factorisable fluxes in generic $SU(n)$-models via brane recombination. Beginning with $n=5$, as another class of fluxes we recover the so-called universal fluxes first observed in \cite{Marsano:2011hv}, which in fact correspond to universal spectral cover fluxes under heterotic duality. We work out the detailed topological signatures such as induced chiralities and 3-brane charges in section \ref{sec:FSU5U1} and comment on the quantisation condition for the fluxes. For definiteness we focus on restricted $SU(5) \times U(1)$ and generic $SU(5)$ models. In particular, we find agreement with the (split) spectral cover approach in section \ref{sec:FSCC}.  

In section \ref{sec:GFIIB} we compare the $G_4$-fluxes to gauge fluxes in Type IIB. 
Finding match between the two pictures relies crucially on the D5-tadpole cancellation condition. In F-theory, this condition is already built in, while in Type IIB models it must be imposed by hand and significantly reduces the number of consistent fluxes.  
For the two types of brane set-ups corresponding to the $SU(n)$ and the $SU(n) \times U(1)$ models, we identify,  in section \ref{sec:generic_flux_config}, a generating  set for all universally present gauge fluxes satisfying the D5-tadpole cancellation condition and find only two such inequivalent types of fluxes. 
We further analyze these fluxes in section \ref{sec:IIBSU5U1} by specifying to $(S)U(5) \times U(1)$.
One of the two sets of fluxes
corresponds to the gauge flux associated with the linear combination of $U(1)$s that remains massless with respect to the geometric St\"uckelberg mechanism. All its topological characteristics such as D3-charge, chiral spectrum and the D-term match with the $U(1)$-flux in $SU(n) \times U(1)$ F-theory models. 
In particular, this explains the apparent ``delocalisation'' of this type of $G_4$ flux because the corresponding Type IIB fluxes are a linear combination of fluxes on both brane stacks.  
In addition, we identify a special gauge flux associated with the geometrically massive diagonal $U(1) \subset U(n)$. For $n=5$ this flux matches exactly the universal spectral cover flux, which is rather surprising given the very different origin of this flux in the two pictures. This means that the fluxes used in heterotic spectral cover models and the diagonal Type II fluxes are really the same objects. We stress, though, that our conclusions hold in this direct form for the special type of models with no conifold singularity and thus no $E_6$-point. This $E_6$-point would violate the perturbative Type IIB selection rules associated with the diagonal $U(1)$. 
The identification of the diagonal $U(1)$-flux with the spectral cover flux is also surprising in view of the analysis of \cite{Grimm:2011tb}, according to which the diagonal $U(1)$ is decoupled from the low-energy spectrum and described by certain non-harmonic two-forms. In section \ref{sec:Massive} we put our findings in perspective with the analysis of \cite{Grimm:2011tb}.

Finally, in section \ref{sec:BraneRecomb} we shed some more light on the brane recombination process that relates the restricted $SU(n) \times U(1)$ and the generic $SU(n)$ Tate models.  Gauge fluxes can obstruct this recombination if the spectrum of recombination modes is purely chiral. 
We analyze necessary conditions for recombination to be possible in the presence of fluxes and match these with restrictions on the gauge flux on the recombined side  as found previously in \cite{Braun:2011zm}.

Some open questions are summarized in section \ref{sec:Concl}. Most of the technicalities that occurred in the course of our analysis have been relegated to the appendices.

\section{F-theory Four-folds Versus Type IIB  Brane Configurations}
\label{sec:geometry_topology}
In this section we construct the class of F-theory four-folds and their Type IIB limits which serves as our laboratory to compare the respective gauge fluxes. Starting with the singular F-theory fibrations corresponding to $SU(n)$ or $SU(n) \times U(1)$ gauge theories, we calculate characteristic topological invariants that allow us to quantitatively match these geometries with their Type IIB orientifold limit. We then derive the analogous brane configurations in Type IIB theory by taking the Sen limit and find agreement between the geometric D3-tadpole contributions in both pictures.  As we will see, %this requires a proper treatment of certain
this will requires to exclude models
which would encounter a conifold singularity in the Sen limit.
%with conifold singularities encountered in the Sen limit.

\subsection{The Geometry of F-theory \texorpdfstring{$SU(n) (\times U(1))$}{SU(n)(xU(1))}  Models}
\label{sec:topological_invariants}

The geometry of a four-dimensional F-theory  compactification is given by an elliptically fibered four-fold $Y_4$. We define the four-fold as a divisor in an ambient five-fold $X_5$ by describing it via a Weierstrass model in Tate form
\begin{equation}\label{tate_poly}
 P_T = \{y^2 + a_1 x y z + a_3 y z^3 = x^3 + a_2 x^2 z^2 + a_4 x z^4 + a_6 z^6\}.
\end{equation}
The coordinates $(x,y,z)$ are homogenous coordinates of the fibre ambient space ${\mathbb P}_{2,3,1}$ fibered over a three-dimensional base $B$. The Tate polynomial coefficients $a_i(u_i)$ depend on local coordinates on $B$ such as to form sections of $\aK^i$, where $\aK$ denotes the anti-canonical bundles of $B$.\footnote{Cohomology classes on the base $B$ of the F-theory four-fold will be denoted by caligraphic letters. For simplicity we use the same symbols for the base classes and their pull-back to the four-fold as well as for classes in cohomology and their Poincar\'e dual homology classes.}  We stress that we explicitly assume the existence of suitable sections $a_i(u_i)$ such that the Tate form (\ref{tate_poly}) is well-defined globally. More general Weierstrass models which cannot be put in Tate form globally are possible \cite{Katz:2011qp} and our analysis does a priori not apply to these.

As is well-known the 7-brane locus in this set-up is given by the vanishing of the discriminant of the Weierstrass model. In models with an $SU(n)$-singularity in the fiber over a base divisor  $\cW: w=0$ the discriminant factorises as 
\bea \label{Delta1}
\Delta = w^n \Delta'.
\eea
This is achieved by restricting the sections $a_i$ in (\ref{tate_poly}) in a manner determined by application of the Tate algorithm \cite{Bershadsky:1996nh}. The vanishing orders of $a_i$ along $w=0$ are collected in eq.~\ref{app_vi} in the appendix. 
In such models the fibre over $w=0$ degenerates to an $\tilde{A}_{n-1}$-singularity.  In absence of further gauge groups the fibre above generic points on $\Delta'=0$ acquires merely an $I_1$-singularity. The singularity type of the fibre enhances over the intersection curves of the divisor $w=0$ and the $I_1$-locus, which is also where massless matter charged under $SU(n)$ is located.

In the sequel we assume that the divisor $\cW: w=0$ itself is smooth and connected. In particular, this excludes the possibility that the discriminant locus self-intersects in a curve contained in $\cW$. Such self-intersections host extra matter --- here in the symmetric representation of $SU(n)$  --- and have been analyzed recently in \cite{Morrison:2011mb}.

It may be phenomenologically preferable to have additional $U(1)$-symmetries in the F-theory model, e.g.~to engineer specific selection rules in $SU(5)$ based GUT models.  
A certain class of models with such extra abelian gauge symmetries is provided by fibrations with extra sections. As an example of such geometries we consider here, in addition to the above $SU(n)$ model,  the $U(1)$-restricted Tate model, obtained by the additional requirement that the Tate polynomial coefficient $a_{6}$ vanish everywhere on the base \cite{Grimm:2010ez}. 
This introduces an additional $SU(2)$-singularity along a self-intersection curve of the divisor $\Delta'$.

Since the enhanced singularity of the $SU(n)$ or $SU(n) \times U(1)$ model renders the entire four-fold $Y_4$ singular (rather than just the fibre), special care must be taken in calculating topological invariants on it. It is easiest to resolve the singularities first and then perform our calculations on the resulting smooth resolution manifold $\hat Y_4$.\footnote{Alternatively one can apply the technology of singular cohomology and continue to work on $Y_4$.} The resolution can be done by pasting in (possibly weighted) $\mathbb{P}^2$s into the ambient space, which reduce to $\mathbb{P}^1$s on the Calabi-Yau four-fold located exactly at the fibre's singular points. In particular, to resolve the $SU(n)$ singularity one introduces a set of so-called exceptional divisors $e_i$ along with a scaling relation for each $e_i$,
\beq \label{eq:scalings}
 (x, y, e_0, e_i) \sim (\lambda^{v_3} x, \lambda^{v_4} y, \lambda e_0, \lambda^{-1} e_i), 
\eeq
where $e_0=0$ denotes the proper transform of $w=0$. For $SU(n)$-singularities one can show that the powers $v_3$, $v_4$ are intrinsically related to the vanishing orders of the Tate polynomial coefficients $a_3$, $a_4$. We leave the details to appendix \ref{app:res_struc}. The intersection structure of the various $E_i$ (the divisor classes of the divisors $\{e_i=0\}$ on $\hat Y_4$) is directly related to the $SU(n)$-Cartan matrix via
\bea
\int_{\hat Y_4} E_i\,E_j\,{\cal B}_a\,{\cal B}_b = C_{ij} \int_\cW {\cal B}_a\,{\cal B}_b,
\eea
where ${\cal B}_k$ are base divisor classes.

In the $SU(n) \times U(1)$ model, one must resolve in addition the mentioned curve of $SU(2)$-singularity along the base curve
\beq
 C_{34} = \{a_{3,v_3} = 0\} \cap \{a_{4,v_4} = 0\}.
\eeq
To resolve it, a $\bbP^1$ is pasted in along the submanifold $\{x=0\}\cap \{y=0\}$ in the ambient space, $(x,y) \rightarrow (\tilde{x}s, \tilde{y}s)$. Reducing to the four-fold, only those $\bbP^1$s which are fibred over $C_{34}$ remain. The blow-up introduces another divisor $S: s=0$, which is actually a section, together with a new scaling relation $(x, y, s) \sim (\lambda x, \lambda y, \lambda^{-1} s)$.

We should stress an important point: The above construction of a resolved four-fold $\hat Y_4$ works irrespective of the details of the base space $B$ --- provided $B$ has enough sections such as to form a Tate model in the first place \emph{without creating any singularities apart from the ones accounted for in the Tate model itself}. This is a condition that must be checked in concrete examples. The advantage of his method is that it allows us to reduce all expressions in our analysis to general quantities defined directly on the base space $B$. This is key for a comparison with Type IIB orientifolds on the double-cover of $B$ via the Sen limit. Alternatively, one may directly construct the singular four-fold $Y_4$ fibered over a concrete base $B$ and resolve it via toric methods, as in \cite{Blumenhagen:2009yv,Grimm:2009yu,Chen:2010ts,Knapp:2011wk,Grimm:2011fx}. While this is  particularly powerful for explicit model building, base independent computations are less immediate to perform.

In the remainder of this section we sketch the logic behind the computation of the intersection forms on $\hat Y_4$ and of topological quantities such as $c_2(\hat Y_4)$ and $\chi(\hat Y_4)$. All details are provided in appendix \ref{app:DetGeom}.

Since the blow-up divisors as well as the fibre $\mathbb{P}_{231}$ form a toric subspace of the ambient manifold, it is possible to use toric methods to deduce certain relations concerning the intersection structure of these divisors. In particular, one can derive a base-independant subset of the generator set of non-intersecting brane configurations of the ambient space and, in combination with the proper transform of the Tate polynomial, compute a subset of the Stanley-Reisner ideal of the Calabi-Yau four-fold. The Stanley-Reisner ideal encodes the  sets of non-intersecting divisors. For example, in all elliptic fibrations of the form described above, $xyz$ is always an element of the Stanley-Reisner ideal, indicating that the three divisors $\{x=0\}$, $\{y=0\}$, $\{z=0\}$ do not intersect in the ambient space. Put differently, there is no patch for this manifold on which all three of these variables are allowed to vanish.

From the various elements of the Stanley-Reisner ideal, it is in turn possible to express double intersections of two exceptional divisors as linear combinations of double intersections involving base divisors. In particular, the double intersections of two $SU(n)$ resolution divisors are expressible as 
\beq \label{eiej_relations}
 E_i\,E_j = C_{ij}\,({ Z}+\aK)\,\cW + w_m\,E_m\,\cW + k_m\,E_m\,\aK + b\,E_2\,E_4,
\eeq
where $b=0$ for $SU(n)$ models with $n<5$. The details of the derivation, and in particular the coefficients $w_m$ and $k_m$ are provided in appendix \ref{app:intersec_struc}. In all cases the coefficients of the first term are the $ SU(n)$ Cartan matrix entries.

The relations just described reduce the number of independent products of two-forms. This property will be useful later on, when we consider potential flux configurations for F-theory models. They are also useful in order to express the second Chern class of the Calabi-Yau four-fold, which enters the flux quantisation condition, as well as to express the fourth Chern class, which is needed to evaluate the Euler characteristic. The latter two points are what we focus on here.

From the scaling relations described in (\ref{eq:scalings}) and the general adjunction formula, one can deduce the relationship between the Chern class of the resolution manifold and the Chern class of the original manifold. 
Here, we outline the generic $SU(n)$ model. More details as well as a derivation of the analogous quantities in the $SU(n) \times U(1)$ model can be found in appendix \ref{app:topological_invariants}.

The scaling relation (\ref{eq:scalings}), introduced by each blow-up, implies that in each case the class of the divisor $\{x = 0\}$ changes according to ${ X} \rightarrow \pi^{\ast} { X} - v_3^{\,i} E_i$, where $\pi^{\ast}$ denotes the pullback of the original class to the resolution manifold. Similarly one can read off the changes for the divisors $\{y=0\}$, $\{e_0=0\}$. All other divisors remain in the pullbacks of the original classes, while the class of the vanishing locus of the Tate divisor changes by $-(v_3^{\,i} + v_4^{\,i}) E_i$. The sum of the changes is zero, which, along with the adjunction formula, implies that the first Chern class of the resolution manifold will be the same as that of the original manifold. This demonstrates that the resolution manifold is indeed Calabi-Yau, if we start with a Calabi-Yau manifold. These properties then also imply that the change in the second Chern class can be expressed in terms of the above changes. We leave the explicit expressions for the $SU(n)$-cases with $n \leq 5$ to appendix \ref{app:topological_invariants}.

From the expressions for the change of the second Chern class one can in turn derive the change in the Euler characteristic of the four-fold. Since the arithmetic genus $\chi_0 = \tfrac{1}{720} \int_{Y_4} \left(c_4 - 3\,(c_2)^2\right) = 2$   is the same for all Calabi-Yau four-folds (see e.g.~\cite{Collinucci:2010gz}, where this was exploited in a similar context), the change in $c_4$ is related to the change in $c_2$ by
\beq\label{eq:delta_c4_CY-4-fold}
 \Delta c_4 = 3\, \left(2\,c_2^{ns}\,\Delta c_2 + (\Delta c_2)^2 \right).
\eeq
Here we define
\beq
 \Delta c := c\left(Y_4^{\rm resolved}\right) - c\left(Y_4^{\rm non-singular}\right)
\eeq
and $c_2^{ns}$ is the pullback of the second Chern class of the original, non-singular four-fold to the resolution manifold. In fact, this is given by $c_2^{ns} = 12\,({ Z}+\aK)^2 + [c_2(B_3)] - \aK^2$. Noting that, in the $SU(n)$-cases, the change in the second Chern class is perpendicular to $Z$ as well as to the intersection of two base divisors, it is clear that the first term in~\eqref{eq:delta_c4_CY-4-fold} vanishes. Then in each of the $SU(n)$-cases, the change in the Euler characteristic reduces to an integral over the GUT-brane. The precise formulae are again collected in appendix \ref{app:topological_invariants}. Here we note that we reproduce the spectral cover formulae of \cite{Blumenhagen:2009yv} for $SU(n), n=2,3,5$, and provide a similar formula for $SU(4)$, which is not derivable via spectral covers. The $SU(5)$-case has already been computed in \cite{Marsano:2011hv}. The total Euler characteristic of the resolution manifold is then simply the sum of $\Delta \chi$ and the well-known Euler characteristic of the non-singular, elliptically fibred Calabi-Yau four-fold
\beq \label{chins}
 \chi_{ns} = 360 \, \aK^3 + 12 \, \aK \, \left[ c_2(B) \right].\\
\eeq

The analogous expressions for restricted $SU(n) \times U(1)$ models are derived by the same logic and are listed in appendix \ref{app:topological_invariants}, see in particular table \ref{delta_c2_sunu1} for the change in $c_2(\hat Y_4)$ in that case. Note furthermore that the Euler characteristic drops by an additional 
\bea \label{tadchange}
 \Delta^{su_n \times u_1}_{su_n} \chi(Y_4) = 3\,\chi(C_{34})
\eea
as a consequence of the $U(1)$-restriction.

\subsection{Sen Limit and Conifold Points}
\label{sec:sen_limit}

The connection between the F-theory $SU(n)$-models of the previous sections and their perturbative formulation as Type IIB orientifolds is made  via the well-known Sen limit. As we will discuss, in order to carry out this limit without encountering singularities we must impose certain restrictions on the models under consideration. These will turn out crucial also for a correct identification of the F-theory and Type IIB gauge fluxes. 

\subsubsection*{Generalities of the Sen limit}

Recall from \cite{Sen:1997gv} that for an elliptic fibration in Weierstrass form
\beq
y^2 = x^3 + f x z^4 + g z^6
\eeq
with $f$ and $g$ sections of $\aK^4$ and $\aK^6$, respectively, the orientifold limit corresponds to letting  $\epsilon \rightarrow 0$ in the parametrisation
\beq
f = -3 h^2 + \epsilon \,  \eta, \qquad \quad g = -2 h^3 + \epsilon h \eta - \frac{\epsilon^2}{12} \chi.
\eeq
In this limit the discriminant locus  $\Delta = 4 f^3 + 27 g^2$ factorises as
\beq
\Delta  = -9  \epsilon^2 h^2 (\eta^2 - h \chi) + {\cal O}(\epsilon^3).
\eeq

To apply this to a Weierstrass model in Tate form as in (\ref{tate_poly}),  we note that the sections $f$ and $g$ are related to the Tate polynomial $a_i$ via
\beq
  f=-\frac{1}{48}( b_2^2 -24\, b_4), \qquad
  g=-\frac{1}{864}( -b_2^3 + 36 b_2 b_4 -216 \, b_6) 
\eeq
with 
\beq
  b_2=a_1^2 + 4\, a_2, \qquad b_4=a_1\, a_3 +2 a_4, \qquad b_6=a_3^2+4\, a_6.
\eeq
To perform the Sen limit one therefore identifies \cite{Donagi:2009ra}
\beq
  b_2= - 12 \, h, \qquad b_4 = 2 \, \epsilon \, \eta , \qquad b_6= - \frac{\epsilon^2}{4} \chi,
\eeq
corresponding to a rescaling 
\bea
a_3\to \epsilon \, a_3, \qquad a_4\to \epsilon \, a_4, \qquad a_6\to \epsilon^2 \, a_6.
\eea
The discriminant takes the form
\begin{equation}
 \Delta = -{\epsilon^2} \, b_2^2 \, b_8 + {\cal O}(\eps^3), \qquad b_8 =  \frac{1}{4} ( b_2 b_6 - b_4^2).
\end{equation}

The Calabi-Yau three-fold $X_3$ on which the Type IIB orientifold is defined takes the form of the double cover of the F-theory base manifold $B$. It is given by the hypersurface equation
\bea
X_3: \xi^2 = b_2,
\eea 
where $b_2(u_i)$ depends on the local coordinates $u_i$ of $B$.
The orientifold involution acts as 
\bea
\sigma: \xi \longrightarrow - \xi.
\eea
The O7-plane therefore corresponds to the fix point locus $b_2=0$, while the vanishing locus of $b_8$ represents the 7-brane configuration.

\begin{figure} 
 \centering
 \def\svgwidth{0.8 \linewidth}
 \executeiffilenewer{double-cover.svg}{double-cover.pdf}%
 {inkscape -z -D --file=double-cover.svg %
  --export-pdf=double-cover.pdf --export-latex}%
   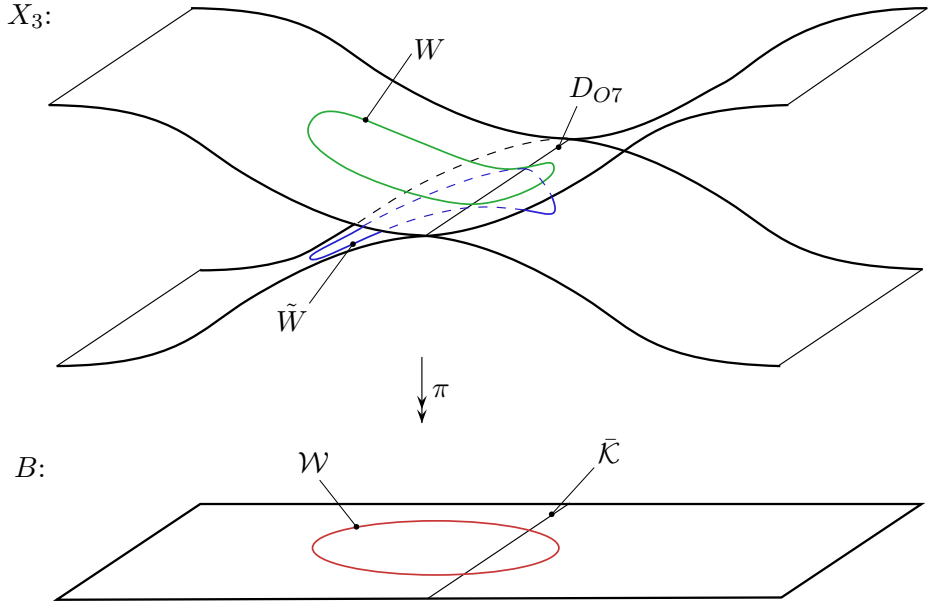%

 \caption{Schematic drawing of the Calabi-Yau $X_3$ and its orientifold projection $\pi$ to $B$. The Calabi-Yau is embedded as a hypersurface in the $\mathcal O(\aK)$-bundle over $B$.}
\label{fig:double-cover}
\end{figure}

The Calabi-Yau three-fold is related to the F-theory base $B$ through a two-to-one map
\bea
\pi: X_3 \rightarrow B\,,
\eea
see also  Figure~\ref{fig:double-cover}.
This induces a natural action on (co)homology. In particular, to each element $\omega_B \in H^2(B)$ we can assign a two-form $\omega_{X_3} \in H^2(X_3)$ by pullback,
\bea
\omega_{X_3} = \pi^* (\omega_B).
\eea
Finding the exact relation between the forms on $X_3$ and $B$ is part of the task of constructing an F-theory uplift of a given Type IIB model. There are two types of behaviour for a divisor class on $X_3$ under the orientifold involution $\sigma$: It can be mapped to itself or it can be exchanged with another class. The F-theory uplift of such exchange involutions has previously been considered in \cite{Collinucci:2008zs,Collinucci:2009uh,Blumenhagen:2009up}. 
If the orientifold involution $\sigma$ on $X_3$ exchanges two classes $D$ and $\tilde D$ it is useful to introduce the combinations
\bea
D_\pm = D \pm \tilde D.
\eea
The class $D_+$ survives the projection from $X_3$ to $B$.
The preimage of $D_+$ under $\pi^*$ in $H^2(B)$ will then be denoted by $\cal D$, such that
\bea \label{piD}
\pi^*({\cal D}) = D_+.
\eea
In addition, one must determine the precise relation between invariant classes on $X_3$ and their analogue on $B$.
For example the Sen limit implies that 
\bea \label{pibarK}
\pi^*(\bar{\cal{K}}) = [D_{O7}],
\eea
where $D_{O7}$ is the class of the orientifold plane. This allows one to deduce further relations between invariant classes on $X_3$ and their descendents on $B$.
The intersection numbers of elements in $H^2(B)$ and their pullbacks on $X_3$ are related as
\bea \label{integral-match}
 \int_{X_3}  \pi^*({\cal D}_a) \wedge \pi^*({\cal D}_b) \wedge \pi^*({\cal D}_c)=\int_{\pi(X_3)} {\cal D}_a \wedge {\cal D}_b \wedge {\cal D}_c  =2\int_{B} {\cal D}_a \wedge {\cal D}_b \wedge {\cal D}_c ,
\eea
where the factor of $2$ arises because `$\pi(X_3)=2\,B$'. Note that this is consistent with the explicit uplift models considered in \cite{Blumenhagen:2009up} and the general analysis in \cite{Grimm:2011tb}.

\subsubsection*{Sen limit for $SU(n)$ models and conifold points}

We are now ready to apply this reasoning to the  $SU(n)$ Tate model of the previous section by specifying the vanishing behavior $a_i = a_{i,v_i}  w^{v_i}$ encoded in eq.~\eqref{app_vi} in the appendix. On general grounds,  in Type IIB orientifolds an $(S)U(n)$ gauge symmetry is realized by a brane/image-brane pair exchanged by the orientifold involution. Therefore the 7-brane stack on the divisor $w=0$ on the F-theory base $B$ must correspond to a stack/image-stack pair on $X_3$.  Similarly, for the restricted Tate model with $a_{6,n} = 0$ one expects another brane/image-brane pair corresponding to the extra $U(1)$ factor appearing the theory.

To see how this comes about we first observe that one can re-express $b_8$ as follows, where we distinguish whether $n$ is even or odd,
\begin{equation} \label{b8_1}
 b_8 = \begin{cases}
     &w^{n} \, \left[a_{6,n} b_2 - \left(a_{4,k} + \frac{a_1 + \xi}{2} a_{3,k} \right) \left(a_{4,k} + \frac{a_1 - \xi}{2} a_{3,k} \right) \right], \qquad \qquad n = 2k,\\
     &w^{n} \, \left[a_{6,n} b_2 - w \left(a_{4,k+1} + \frac{a_1 + \xi}{2 w} a_{3,k} \right) \left(a_{4,k+1} + \frac{a_1 - \xi}{2w} a_{3,k} \right) \right], \quad n = 2k+1.
       \end{cases}
\end{equation}
For $n=2k +1$ the second term in square brackets is no longer symmetric if one pulls the $w$-factor into one of the two round brackets. On the other hand, as pointed out, one would expect that in the restricted case ($a_{6,n} = 0$) $b_8$ factors into $w^n$ times two factors which are exchanged under the involution so as to represent a brane/image-brane pair. To remedy this apparent problem, we note that the divisor $w=0$ does indeed split into two on $X_3$: From the Calabi-Yau equation
\beq\label{eq:CY-threefold}
 P_{X_3} = (\xi - a_1) (\xi + a_1) - a_{2,1} w = 0
\eeq
it follows that $P_{X_3}|_{w=0}$ factorises. Let us thus define
\bea
 \omega &= \{w=0\} \cap \{\xi - a_1 = 0\}, \\
 \tilde{\omega} &= \{w=0\} \cap \{\xi + a_1 = 0\}\,.
\eea
These two divisors are exchanged under the involution $-\xi \leftrightarrow \xi$. Then $\omega$ and $\tilde{\omega}$ define the brane/image-brane pair as expected to account for $SU(n)$ gauge symmetry. Note that while in the ambient space they both lie in the same divisor class, on the Calabi-Yau three-fold $X_3$ they lie in different classes.

Since, on the three-fold, the section $w$ factors into two components, it is possible to write the second term in square brackets of the odd-$n$-case of eq.~(\ref{b8_1}) as 
\begin{equation} \label{b8_2}
 \left(a_{4,k+1} \omega  + \tfrac{a_1 + \xi}{2 \tilde{\omega}} a_{3,k} \right) \left(a_{4,k+1} \tilde{\omega} + \tfrac{a_1 - \xi}{2 \omega} a_{3,k} \right).
\end{equation}
Now the two terms are exchanged under the involution, leading to a brane/image-brane pair in the $U(1)$-restricted case. 
Note that, similarly to the GUT brane/image-brane pair, the brane and its image do not necessarily lie in the same class.

One might object that $\omega$ and $\tilde{\omega}$ appear in the denominator of the above expressions, and that these are therefore not everywhere well-defined. 
However, from the definitions of $\omega$, $\tilde{\omega}$ and the Calabi-Yau equation one can see that the expression $\tfrac{\xi - a_1}{\omega}$ corresponds to the algebraic cycle given by the intersection of $\{\xi-a_1\}$ and $\{a_{2,1}\}$ in the ambient space, and similarly $\tfrac{\xi+a_1}{\tilde{\omega}}$ corresponds to $\{\xi+a_1\} \cap \{a_{2,1}\}$.
Hence, we obtain two well-defined expressions for the additional branes in the case of a splitting. \\

We now come to an important subtlety. For general base manifolds $B$ the Calabi-Yau equation \eqref{eq:CY-threefold} exhibits  a conifold singularity at~\cite{Donagi:2009ra}
\begin{equation}
a_{1}=a_{2,1}=w=0.
\end{equation}

In the presence of such a conifold point it is not possible to smoothly interpolate between the F-theory picture and the Type IIB orientifold regime. In particular, it is not clear that topological invariants should agree on both sides.
Therefore, we only consider the Sen limit in cases where this intersection is not realised on the base. 
Note that for $n=5$ this conifold point is precisely the point of $E_6$ singularity enhancement ~\cite{Donagi:2009ra} at which the top Yukawa couplings ${\bf 10 \, 10 \, 5}$ of an $SU(5)$ model are located, which are perturbatively absent in Type IIB models.  
We thus demand that we do not have an
``$E_{6}$''-point (or its generalizations in general $SU(n)$ models) on the F-theory side. 
This enforces the special relation
\bea\label{KWW}
2 \int_B {\cal \aK}^2 \cW = \int_B {\cal \aK} \cW^2.
\eea
The corresponding relation  on the IIB side is that
\begin{align}\label{eq:no-E6-point-relation}
\int_{X_3} D_{O7}\,(2\,D_{O7}-W_+)\,W_+=\int_{X_3} D_{O7}\,(2\,D_{O7}-W-\tilde{W})(W+\tilde{W})= & 0\,.
\end{align}
In fact, from equation \eqref{eq:CY-threefold} we observe that the loci $\{\xi=0\}\cap\{\omega=0\}$ and $\{\tilde{\omega}=0\}\cap\{\omega=0\}$ are identical. Written in terms of classes, this becomes
\begin{equation}\label{eq:brane-image-inter-relation}
 D_{O7}\, W=W\,\tilde W=D_{O7}\, \tilde W \qquad \Rightarrow \qquad W_+^{\,2} - W_-^{\,2} = 2 D_{O7} W_+.
\end{equation}
Note that \eqref{eq:brane-image-inter-relation} implies  \eqref{eq:no-E6-point-relation} because $D_{O7} W_- =0$, i.e.~the pullback of an orientifold odd class to the fix point locus vanishes. 
Thus in a Type IIB orientifold arising as the Sen limit of an F-theory Tate model (\ref{tate_poly}) with a smooth $SU(n)$ divisor without encountering a conifold point, the $U(n)$ brane and image-brane stack intersect only over the O7-plane. Furthermore, the F-theory uplift of a $U(n)$ orientifold with this latter property automatically satisfies relation (\ref{KWW}) as was observed in the examples of \cite{Blumenhagen:2009up,Collinucci:2009uh}. We will only consider models of this type.
%Therefore, F-theory models which arise as the uplift of smooth Type IIB orientifolds  do automatically  not admit the above type of conifold points, as was observed in the examples of \cite{Blumenhagen:2009up,Collinucci:2009uh}.
As will become evident in the next section, the conditions  (\ref{eq:brane-image-inter-relation}) and (\ref{KWW}) are crucial for a quantitative match between the topological invariants of the F-theory and the Type IIB models. \\

\begin{figure}
 \centering
\subfigure[Non-restricted Models]
{
 \def\svgwidth{0.45 \textwidth}
 \executeiffilenewer{non_restricted.svg}{non_restricted.pdf}%
 {inkscape -z -D --file=non_restricted.svg %
  --export-pdf=non_restricted.pdf --export-latex}%
   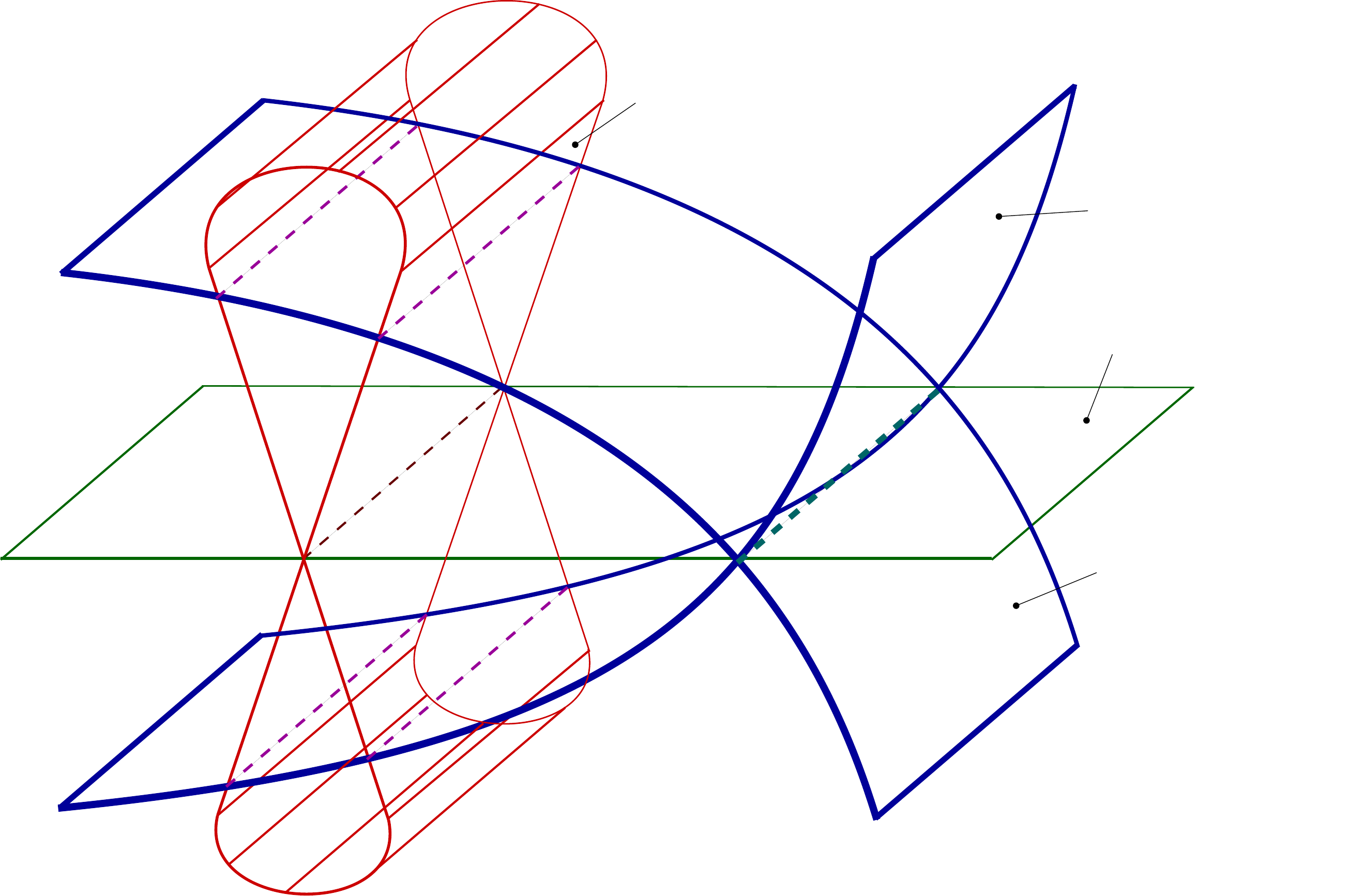%

 \label{fig:non-restricted}
}
\subfigure[$U(1)$-Restricted Models]
{
 \def\svgwidth{0.45 \textwidth}
 \executeiffilenewer{restricted.svg}{restricted.pdf}%
 {inkscape -z -D --file=restricted.svg %
  --export-pdf=restricted.pdf --export-latex}%
   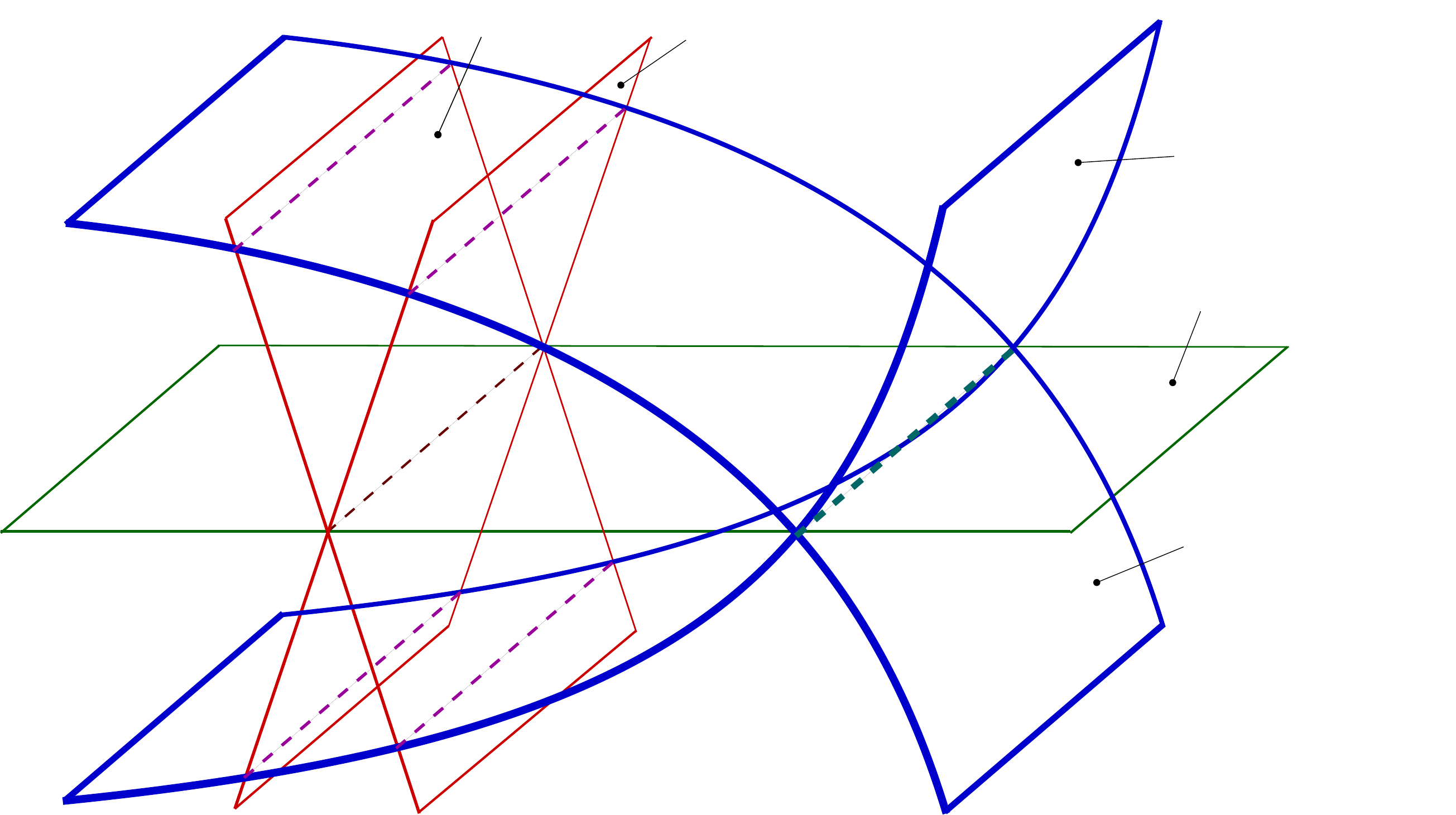%

 \label{fig:restricted}
}
\caption[Different Tybe IIB Set-ups]{Brane configurations for \subref{fig:non-restricted} $SU(n)$- and \subref{fig:restricted} $SU(n)\times U(1)$- Type IIB set-ups. Note that $V$ and $\tilde{V}$ lie in the same class in the latter case if $n$ is even.}
\label{fig:IIb-set-ups}
\end{figure}

Let us summarize the brane set-ups as the Type IIB limits of the $SU(n)$- and $SU(n) \times U(1)$- F-theory models:
The brane contents corresponding to the $SU(n)$ F-theory models, with generic non-vanishing sections $a_{6,n}$, is that of a $U(n)$ brane/image-brane stack along $W$ and $\tilde W$ together with a single connected 7-brane along the divisor in square-brackets in (\ref{b8_2}). This latter divisor is singular and of so-called Whintey type. Whitney branes of this form have been analyzed in detail in \cite{Collinucci:2008pf} and further in  \cite{Collinucci:2010gz, Braun:2011zm}. In appendix \ref{app:Whitney_brane_charges} we provide a generalisation of the description of  \cite{Collinucci:2010gz, Braun:2011zm}, which is in the context of models with extra non-abelian gauge groups up to $Sp(1)/SU(2)$, to general $SU(n)$ models. We will find some novelties which are important for a quantitative match with F-theory. 
Since the Whitney brane carries trivial gauge group, the gauge group in Type IIB is
$SU(n) \times U(1)_a$, where the latter denotes the diagonal abelian factor in $U(n)$. This $U(1)_a$ is massive by means of the St\"uckelberg mechanism since the stack and image-stack are in different homology classes on $X_3$. The models are therefore precisely of the type considered recently in \cite{Grimm:2011tb}. 

On the other hand, for the $U(1)$-restricted Tate models, with $a_6 \equiv 0$, the extra brane in  square-brackets in (\ref{b8_2}) splits into a brane/image-brane pair. If $n = 2k$, brane and image are in the same homology class, whereas for $n=2k+1$ they are not. The special situation of $SU(2)$ has already been analyzed in \cite{Braun:2011zm}. The gauge group for odd $n$ is $SU(n) \times U(1)_a \times U(1)_b$, where both $U(1)$-factors are individually massive, but a massless linear combination of them exists. For $n$ even, the second $U(1)_b$-factor is by itself massless.

A schematic drawing of the various brane set-ups is given in Figure~\ref{fig:IIb-set-ups}. In both cases, the GUT stack is situated on the divisor $\{\omega = 0\}$ and its image on $\{\tilde{\omega} = 0\}$. The remaining brane of the non-restricted case lies in the class $D_{V + \tilde{V}} = 8 D_{O7} - n \left(W + \tilde{W}\right)$ as is required to cancel the $D7$-charge induced by the O-plane. In the restricted case, the remaining brane splits into a brane/image-brane pair with associated divisor classes $V$ and $\tilde{V}$. For $n$ even, these two classes are the same, $V = \tilde{V} = \tfrac12 D_{V + \tilde{V}}$, however the divisors themselves do not necessarily coincide. On the other hand, in the odd case $V$ and its image are different classes: $V = 4 D_{O7} - k W - (k+1) \tilde{W}$ where $k$ is defined via $n = 2k+1$.

\subsection{Topological Invariants of Type IIB Brane Configurations} \label{sec:TopInv}
As a first and important check of the Sen limit of our F-theory $SU(n)$ model, we compute the induced D3-brane charges of the Type IIB 7-branes and orientifold plane and compare them with the Euler characteristic of the resolved F-theory four-fold $\hat Y_4$.

Recall that the D3-brane tadpole cancellation condition for a Type IIB orientifold takes the form
\bea \label{D3tad-gen}
Q_{\rm gauge, \, IIB}  + N_{D3} = \frac{1}{4} N_{O3}  + Q_{D3}^g, 
\eea
where $Q_{\rm gauge, \, IIB}$
represents the gauge flux induced 3-brane charge, $N_{D3}$ and $N_{O3}$ counts the number of D3-branes and O3-planes and $Q_{D3}^g$ is the curvature induced 3-brane charge from the O7-plane and the D7-branes. In this section we focus on the latter.
The contribution of a brane/image-brane pair to $Q_{D3}^g$ depends on the Euler characteristic $\chi(D_i)$ of the smooth divisor $D_i$,
\bea \label{QigIIBa}
Q_i^g = \frac{n_i}{48} \Big( \chi(D_i) +  \chi(\tilde D_i) \Big), \qquad \quad \chi(D) = \int_{X_3}  \Big( D^3 \,+\, c_2\,D  \Big).
\eea
For an invariant brane along a divisor of Whitney type, on the other hand, the singular geometry must be resolved and the 3-brane charge is proportional to the Euler characteristic of the resolved space \cite{Collinucci:2008pf},
\bea \label{QigIIBb}
Q_i^g = \frac{1}{48}  \chi_0(D_i) =     \frac{1}{48}   \int_{X_3} \Big( D_i^3 \,+\, c_2\,D_i - 3\,D_i\,D_{O7}\,(D_i-D_{O7})   \Big).
\eea
Note that the expressions (\ref{QigIIBa}) and (\ref{QigIIBb}) are often referred to as the ``downstairs'' picture because we compute all quantities for each brane and its image (or, in the case of the Whitney brane, for the entire invariant brane on $X_3$)  and divide by 2.

The geometric tadpole contribution for non-restricted $SU(n)$-models is then given by
\bea
 24\,Q_{D3}^g = \frac{n}{2}\,\left( \chi(W) + \chi(\tilde{W})\right) + \frac{1}{2} \chi_0(8D_{O7} - n\,W_+) + 2\,\chi(D_{O7}). 
\eea
The first term is due to the $SU(n)$ stack and its image, the second due to the Whitney brane and the last term represents the contribution from the O7-plane.
This is readily evaluated and to be compared with $\chi(\hat Y_4)$ of the resolved F-theory four-fold as given by the sum of (\ref{chins}) and the results collected in  table \ref{chi_FT}. 
Note that to obtain agreement, we crucially need  the relations (\ref{eq:no-E6-point-relation}) and (\ref{eq:brane-image-inter-relation}). 
With the help of the identification (\ref{pibarK}) and \eqref{piD} as well as the factor $\frac{1}{2}$ in (\ref{integral-match}) we find agreement for the D3-tadpole for $SU(n), \, n = 3,4,5$.

In a similar manner one finds perfect match between the Euler characteristic of the $U(1)$-restricted $SU(n)$ Tate models and their Sen limit. More details are provided in appendix \ref{app_tad}.

\subsubsection*{$SU(2)$ versus $Sp(1)$}
As an aside we note that the above expressions do not quite match in the $SU(2)$-case; they would  if $W_+\left(W_+^{\,2} - W_-^{\,2}\right)$ were zero in this case.
To understand the mismatch, we calculate the D3-charge of a brane set-up which corresponds to an $Sp(1)$-gauge theory. This consists of a stack of two branes in some invariant divisor class $W$ and a Whitney brane in the class $8 D_{O7} - 2 W$. The corresponding tadpole formula,
\bea
 24\,Q_{D3}^g = \frac{n}{2} \chi(W) + \frac{1}{2} \chi_0(8D_{O7} - n\,W) + 2\,\chi(D_{O7}),
\eea
(with $n=2$) leads to an expression which matches the F-theory D3-tadpole upon the usual identifications.

This suggests an identification of the $SU(2)/Sp(1)$ F-theory model with the $Sp(1)$- rather than the $SU(2)$-brane set-up in Type IIB.
This roots in the resolution algorithm as presented in section \ref{sec:topological_invariants}. Everything worked out there relies on divisor classes and is therefore 'blind' to cases where the hypersurface is restricted to a special form. However, this happens for $SU(2)$ where $a_2$ is restricted to $a_{2,1}w$ although a general $a_2$ would be allowed by the multi-degree of $P_T$.

For Calabi-Yau manifolds embedded as complete intersections in a toric variety we can make this more explicit. There the patch structure of the ambient space can be encoded in terms of a (N-lattice) polytope and its triangulation. Each lattice point inside this polytope corresponds to a toric divisor of the ambient space. From this one can construct a dual (M-lattice) polytope  encoding the monomials of the hypersurface.
Restricting the Tate polynomial coefficients, $a_i = a_{i,k} w^k$, translates into removing points from this dual polytope. The convex hull of the remaining set of points then forms a smaller dual polytope. If this is reflexive, the dual serves as an ambient space in which the singularities of the hypersurface are resolved. By comparing this to the original manifold, one can read off the additional divisors of the ambient space.

Now, moving from an $Sp(1)$- to an $SU(2)$-singularity corresponds to restricting the second Tate polynomial coefficient to $a_2 = a_{2,1} w$. Since both singularities are of rank one, one does not expect an additional divisor from this restriction. Furthermore, since the restrictions necessary to induce an $Sp(1)$-singularity are also required to induce an $SU(2)$-singularity, the $SU(2)$-dual polytope has to be contained within the $Sp(1)$-dual polytope. In other words, those points of the $Sp(1)$-dual polytope which are removed in moving to an $SU(2)$-singularity are either interior points with no effect on the actual ambient space or they are such that the resulting $SU(2)$-dual polytope is not reflexive any more. In both cases the deformation cannot be resolved by the canonical procedure.
We thus conclude that the above derivation for the D3-tadpole contribution of the F-theory model does actually correspond to an $Sp(1)$-model rather than an $SU(2)$-model.

\section{Gauge Fluxes in F-theory \texorpdfstring{$SU(n)$}{SU(n)} and \texorpdfstring{$SU(n) \times U(1)$}{SU(n) x U(1)} Models}

In this section we classify the set of factorisable gauge fluxes in F-theory models with $SU(n)$ and $SU(n) \times U(1)$ gauge symmetry. 
Specialising for concreteness to $n=5$ we provide a number of characteristic topological invariants which will allow us to compare these fluxes both to their spectral cover description and, in section \ref{sec:GFIIB}, to make contact with fluxes in the Type IIB limit. For a discussion of $G_4$-gauge fluxes in the recent literature on globally defined four-folds see \cite{Braun:2011zm,Marsano:2011hv,Krause:2011xj,Grimm:2011fx}.

\subsection{`Vertical' \texorpdfstring{$G_4$}{G4}-Fluxes in \texorpdfstring{$SU(n) (\times U(1))$}{SU(n) x U(1)} F-theory Models}\label{subsec_G41}

Gauge flux in F-theory is described in terms of the flux associated with the field strength $G_4= dC_3$ of the dual M-theory dimensionally reduced on $\hat Y_4$.
A first constraint on $G_4$ is that it must be vertical, i.e.~orthogonal both to the section ${Z: z=0}$ of $\hat Y_4$ and to the pullback of any two base divisors ${\cal B}_a$,
\bea \label{verticality}
\int_{\hat Y_4} G_4 \wedge { Z} \wedge {\cal B}_a = 0, \qquad \quad \int_{\hat Y_4} G_4 \wedge {\cal B}_a \wedge {\cal B}_b = 0.
\eea

Some further consistency conditions must be met.
First, 
 according to \cite{Witten:1996md} $G_4$ must be (half-)integer quantised
\bea \label{Qcond}
G_4 + \frac{1}{2} c_2({\hat Y_4}) \in H^4(\hat Y_4, \mathbb Z)\,.
\eea 
The F-term supersymmetry condition requires that in addition $G_4 \in H^{2,2}(\hat Y_4)$. In the presence of abelian gauge symmetries extra D-term conditions arise, as will be discussed further below.

As anticipated in the introduction, a Calabi-Yau four-fold admits two different types of such elements of  $H^{2,2}(\hat Y_4)$. Forms that can be expressed as the sum of the wedge product of two elements of $H^{1,1}(\hat Y_4)$ are denoted as elements of the primary vertical subspace $H^{2,2}_{\rm vert}(\hat Y_4)$  \cite{Greene:1993vm,Mayr:1996sh}. Such fluxes automatically satisfy the F-term supersymmetry condition for all choices of complex structure moduli and therefore exist everywhere in complex moduli space. Typically, the vast majority of four-forms cannot be written as the sum of such factorisable four-forms. 
In this section we wish to classify the available gauge fluxes in $H^{2,2}_{\rm vert}(\hat Y_4)$ for the $SU(n)$ and $SU(n) \times U(1)$ resolution four-forms.
In a first step this requires finding a basis of factorisable four-forms satisfying (\ref{verticality}). In a second step we discuss which of these leave the non-abelian part of the gauge group invariant so that they can be switched on in phenomenological models e.g.~of $SU(5)$ GUT type to induce a chiral spectrum without affecting the gauge symmetry. 

\subsubsection*{$SU(n)$-Models}

An obvious set of fluxes in $H^{2,2}_{\rm vert}(\hat Y_4)$ that satisfy (\ref{verticality}) is given by the Cartan fluxes, i.e.~four-forms of the type $[E_i] \wedge [{\cal F}_i]$ for some base divisor classes ${\cal F}_i \in H^{1,1}(B)$. Another possibility are combinations of $[E_i]\wedge [E_j]$-terms which are orthogonal to the intersection of two base divisors. Orthogonality to $Z\,{\cal B}_a$ follows immediately from the Stanley-Reisner ideal elements $z e_i$, derived in appendix \ref{app:intersec_struc}. On the other hand, the intersection of $[E_i]\wedge [E_j]$ with two base divisors always involves the Cartan matrix entries $C_{ij}$, cf.~\eqref{eiej_relations}. It is thus easy to see that potential flux candidates are given by linear combinations of $[E_i]\wedge [E_j]$ whose corresponding Cartan-matrix entries add up to zero. As an example consider $SU(5)$. In view of the specific form of (minus) the $SU(n)$-Cartan matrix,
\beq
 C_{ij} = \left( \begin{matrix} -2 & 1 & 0 & \cdots & 0 \\ 1 & -2 & 1 & \cdots & 0 \\ 0 & 1 & -2 & \cdots & 0 \\ \vdots & \vdots & \vdots & \ddots & \vdots \\ 0 & 0 & 0 & \cdots & -2 \end{matrix} \right),
\eeq
 viable combinations include for instance $E_2 E_4$ or $E_2 E_3 - E_3 E_4$, while for example $E_2 E_3$ is not a viable flux candidate.\footnote{If the divisor $\cW$ has vanishing intersection with every other divisor class, more vertical fluxes exist. Clearly this situation is not of interest to applications and will thus be discarded.}

It is simple enough to derive a basis of linearly independent viable combinations, and we list one such basis for $SU(n), n \leq 5$ in table \ref{tab:eiej_basis} in the appendix.
From the derivations in subsection \ref{sec:topological_invariants}, in particular from (\ref{eiej_relations}), one notes that the forms listed therein can be re-expressed entirely in terms of Cartan fluxes $E_i\,\aK$, $E_i\,\cW$ and --- in the case of $SU(5)$ --- in terms of $E_2\,E_4$. In particular, none of the combinations depends on $(Z+\aK)\cW$, because the coefficients of this term in (\ref{eiej_relations}) are just the Cartan matrix elements, and we have chosen combinations whose Cartan matrix elements add up to zero. Since there is no linear combination of Cartan fluxes that is orthogonal to all possible Cartan fluxes, linear combinations of $E_i \, E_j$-terms do not add new elements to the space of potential fluxes for $SU(n)$ with $n<5$. This, in turn, implies that there are no factorisable gauge fluxes leaving the $SU(n)$ gauge symmetry intact for these $SU(n)$, $n=2,3,4$.

For the $SU(5)$-case, on the other hand, there appears one additional element in (\ref{eiej_relations}) and we had previously chosen to use $E_2 E_4$. With the help of the intersection numbers
\beq
 \int_{\hat Y_4} E_2\,E_4\,E_i\,{\cal B}_a = \int_{\cW} (1,-1,1,-1)_i \,  \aK\,{\cal B}_a, \qquad \int_{\hat Y_4} E_2\,E_4\,E_2\,E_4 = \int_{\cW} -{\cal \aK\,\cW}
\eeq
we can then use this element in combination with Cartan fluxes to arrive at a combination of four-forms which is orthogonal to all Cartan fluxes,
\beq \label{universal_flux_1}
G_4^{\,\lambda} = \lambda \Big( E_2\,E_4 + \tfrac15 (2, -1, 1, -2)_i \, E_i \, \aK \Big).
\eeq
This flux has been noted previously in \cite{Marsano:2011hv}, albeit with a different derivation. Our overall normalisation has been chosen for later convenience.
We use the $\lambda$-label in anticipation of the fact that this flux is identified with a so-called universal spectral cover flux, whose parameter space simply consists of one scaling parameter. The discrete parameter $\lambda$ will later on be constrained such that the flux quantisation condition is satisfied.

\subsubsection*{$SU(n) \times U(1)$-Models}
For Weierstrass models with the additional restriction $a_{6,n} = 0$, the resolution of the induced singularity along $C_{34}$ results in an additional divisor class $S$. Its intersection properties were analyzed in \cite{Krause:2011xj} and include two particularly useful properties,
\beq
\int_{\hat Y_4} (S - { Z} - \aK)\,Z\,{\cal B}_a\,{\cal B}_b = 0, \qquad \int_{\hat Y_4}  (S - { Z} - \aK)\,{\cal B}_a\,{\cal B}_b\,{\cal B}_c = 0.
\eeq
Thus, it is clear that expressions of the type $[S-{ Z}-\aK] \wedge [{\cal F}]$ (where $[{\cal F}]$ is a two-form of the base) form a set of additional vertical flux candidates. In order to find additional fluxes that preserve the gauge group, this can be combined with the usual Cartan fluxes to construct linear combinations which are orthogonal to all Cartan fluxes. For the various gauge groups and corresponding resolution manifolds considered here, we list the results in table \ref{tab:Cartan_orthogonal} in the appendix.

For future reference, it is useful to define $\tw_X$ to be precisely this linear combination:
\begin{equation}
 \tw_X =  (S - { Z} - \aK) + t^i \, E_i
\end{equation}
with $t^i$ given in table \ref{tab:Cartan_orthogonal}. 
Again, the rationale behind our overall normalisation, which differs from the one we used in \cite{Krause:2011xj} by a factor of $\frac{1}{n}$, will become clear later.
We note that the condition, for this kind of flux, to be orthogonal to all Cartan fluxes translates into
\beq
 \,\delta_{jN} + t^i\,C_{ij} = 0\,.
\eeq
Here, $N$ is the index of the exceptional divisor which intersects the additional divisor $\{s=0\}$, see also the definition below table \ref{index_conversion} in the appendix. This property is useful in the analysis of the D3-tadpole contribution, chirality and D-term induced by this type of flux.\\

\subsection{Specialisation to \texorpdfstring{$G_4$}{G4} Fluxes in \texorpdfstring{$SU(5) (\times U(1))$}{SU(5) x U(1)} Models}\label{sec:FSU5U1}
For concreteness we restrict ourselves in the following to the analysis of $SU(5) \times U(1)$ models and, towards the end of  this section, to generic $SU(5)$ models.

As established above, a basis for factorisable gauge fluxes compatible with the $SU(5) \times U(1)$ gauge symmetry is given by the two fluxes $G_4^{\,X}({\cal F}) = -\tw_X \wedge {\cal F}$ and $G_4^{\,\lambda}$,  
\begin{align}
&&&&&&&&&&& G_4^{\,X}({\cal F})    &&= \quad -\left( S - { Z} - \aK + \tfrac15 (2,4,6,3)_i E_i \right) \wedge {\cal F}, &&&&&&&&&&\\
&&&&&&&&&&& G_4^{\,\lambda} &&= \quad \lambda \left( E_2 \wedge E_4 + \tfrac15 (2,-1,1,-2)_i E_i \wedge \aK \right).&&&&&&&&&&
\end{align}
Note again that fluxes of the type $G_4^{\,\lambda}$ do not exist for $SU(n)$, $n=2,3,4$. The flux $G_4^{\,X}({\cal F})$, on the other hand, is associated with the extra abelian gauge factor, which we call $U(1)_X$ in the sequel.
To better understand the nature of these fluxes and to compare them to gauge fluxes in Type IIB orientifolds we now compute 
the induced chiralities, $D3$-tadpole contribution and D-terms of these fluxes.\\

\subsubsection*{Chiralities in $SU(5) \times U(1)$ models}
The chiral spectrum in models with $G_4^{\,X}$ has already been computed in \cite{Krause:2011xj,Grimm:2011fx}. 
The matter spectrum of the $U(1)$-restricted $SU(5)$ GUT model consists of chiral multiplets in representations $\mathbf{10}_1, \, \mathbf{{5}}_{3}, \, \mathbf{{5}}_{-2}, \, \mathbf{1}_{-5}$, where the subscripts denote the $U(1)_X$ charges. These states in representation $R_q$ arise from membranes  wrapping certain combinations of resolution $\mathbb P^1$s in the fibre over the matter curves $C_{R_q}$ in the base. To each component of the representation one can assign a matter surface $S_{R_q}^k$ by fibreing the corresponding combination of $\mathbb P^1$s over $C_{R_q}$. See also \cite{Esole:2011sm,Marsano:2011hv} for a description of these matter surfaces for non-restricted $SU(5)$ models and \cite{Grimm:2011fx} for a toric approach.

The chirality induced by  $G_4^{\,X}$ is then computed by integrating the flux over the corresponding matter surfaces \cite{Donagi:2008ca}. The result is identical for each component of the representation. For $G_4^{\,X}$ this can be shown to yield \cite{Krause:2011xj,Grimm:2011fx} 
\begin{align}
 \chi(R_q) = \int_{S_{R_q}^k} G_4^{\,X}({\cal F}) = \frac{q}{5} \int_{C_{R_q}} {\cal F}.
\end{align}
In order to compare these expressions to the Type IIB-picture later on, we also list the classes of divisors whose intersection defines the various matter curves inside the base:
\beq \label{Fmattercurves}
 \begin{aligned}
 &C_{\mathbf 5_{3}} = \cW (3 \aK - 2 \cW),  \qquad \qquad C_{\mathbf{10}_{1}} = \cW \aK, \\ &C_{\mathbf 5_{-2}} = \cW (5 \aK - 3 \cW) ,\qquad \quad \, \, \,C_{\mathbf 1_{-5}} = (4 \aK - 3 \cW) (3 \aK - 2 \cW).
 \end{aligned}
\eeq

By the same strategy the chiral index induced by the universal flux $G_4^\lambda$ can be computed. This requires the intersection numbers derived in the previous sections. We collect the chiralities of both types of fluxes in table \ref{Fchir}.

\begin{table}[htb]
 \begin{center}
  \begin{tabular}{c | c c c |c c c}
  \text{State}       &\quad& Chirality under \,  $G_4^\lambda$                                &\quad& Chirality under $  G_4^{\,X}({\cal F})$\\
  \hline \\[-8pt]
  $\mathbf{10}_{1}   $&&$  \tfrac15 \lambda \int_{C_{\mathbf{10}_1}} -6 \aK + 5 \cW$          && $ \tfrac15 \int_{C_{\mathbf{10}_{1}}} {\cal F}$\\
  $\mathbf{5}_{3}     $&&$  \tfrac15 \lambda  \int_{C_{\mathbf{5}_3}} 2 \aK$                  && $ \tfrac35 \int_{C_{\mathbf{5}_{3}}} {\cal F}$\\
  $\mathbf{5}_{-2}    $&&$  - \tfrac15 \lambda \int_{\cW} \cW \bar{{\cal K}}$                 && $ -\tfrac25 \int_{C_{\mathbf{5}_{-2}}} {\cal F}$\\
 $ \mathbf{1}_{-5}    $&&$ 0 $                                                                && $ -\int_{C_{\mathbf{1}_{-5}}} {\cal F}$
 \end{tabular}
 \caption{Chiral index for massless matter. In the case of $G_4^\lambda$ the chirality for $\mathbf{5}_{-2}$ was derived via anomaly cancellation.}
 \label{Fchir}
 \end{center}
\end{table}

\subsubsection*{D3-tadpole in $SU(5) \times U(1)$ models}

The D3-tadpole cancellation condition in F-theory takes the well-known form 
\bea
Q_{\rm gauge, F} + N_{D3} = \frac{\chi(\hat Y_4)}{24}, \qquad \quad Q_{\rm gauge, F} = \frac{1}{2} \, \int_{\hat Y_4} G_4 \wedge G_4.
\eea
The flux contribution  $Q_{\rm gauge, F}$ due to $G_4^X$ and $G_4^\lambda$ is easily evaluated with the intersection numbers at hand,
\beq \label{QgaugeF}
 Q_{\rm gauge, F} = - \int_B \left[(\aK - \tfrac35  \cW) \, \cF^2 - \tfrac15 \lambda \, \aK \, \cW \, \cF + \tfrac12 \lambda^2 \,  \aK \, \cW\, (\tfrac65  \aK - \cW) \right].\\
\eeq
Note in particular the cross-term linear in $\lambda$.

\subsubsection*{The D-Term in $SU(5) \times U(1)$ models}

Since the $U(1)$-restricted Tate model exhibits an abelian gauge symmetry $U(1)_X$ with gauge potential $A_X$ as in $C_3 = A_X \wedge \tw_X + \ldots$,
switching on gauge fluxes $G_4$  entails a field-dependent Fayet-Iliopoulos D-term of the form ~\cite{Donagi:2008kj,Grimm:2010ez,Grimm:2010ks,Grimm:2011tb}
\bea\label{xiXF}
\xi_X(G_4) \simeq \frac{1}{2 {\cal V}_B} \int_{\hat Y_4} \tw_X \wedge J \wedge G_4
\eea
with ${\cal V}_B$ the volume of the F-theory base $B$.
The D-term for $G_4^X$   and $G^\lambda_4$  is therefore proportional to
\bea \label{DXGX}
\xi_X(G^X_4) \simeq -\frac{2}{{\cal V}_B}  \int_{B_3} J \wedge {\cal F} \wedge   \left(3  \, \cW - 5\, \aK \right), \quad \xi_X(G^\lambda_4) \simeq \frac{\lambda}{{\cal V}_B}  \int_{B_3} J \wedge \cW \wedge \aK.
\eea

Clearly for the non-restricted $SU(5)$ model no such $U(1)_X$ D-term arises as the abelian symmetry is higgsed. We will have more to say about D-terms in section \ref{sec:Massive}.

\subsubsection*{Flux quantisation in $SU(5) \times U(1)$ models}

The general quantisation condition for fluxes in M/F-theory on a Calabi-Yau four-fold $\hat Y_4$ is given in eq.~(\ref{Qcond}). A possible shift from integrality of the flux would be due to curvature contributions encoded in $\frac{1}{2} c_2(\hat Y_4)$.
From our explicit computation of the topological invariants for $\hat Y_4$ it follows that  $\frac{1}{2} c_2(\hat Y_4)$ takes the form
\bea \label{c2mod2a}
\frac{1}{2} c_2(\hat Y_4) = \frac{5}{2} \cW \wedge \tw_X      \qquad\qquad {\rm mod} \quad  {\mathbb Z}.
\eea
Since $5 \tw_X$ is an integer form, a \emph{sufficient} condition for the quantisation of $G_4^X({\cal F}) = {\cal F} \wedge \tw_X$ is
\bea
\frac{1}{5} {\cal F} + \frac{1}{2} {\cal W} \in H^{2}(B, \mathbb Z).
\eea
However, this condition may well be too strong. 
To do better we need an explicit basis $\{\tilde \omega_\alpha\}$ of the integer cohomology $H^{4}(\hat Y_4, \mathbb Z)$. Given such a basis one requires that 
\bea \label{integer1}
\int_{\hat Y_4}\Big( G_4 + \frac{1}{2} c_2(\hat Y_4)\Big)  \wedge \tilde \omega_\alpha \in \mathbb Z.
\eea

At a general level, i.e.~for an $SU(n) (\times U(1))$ Tate model over a generic base $B$, we do not have a  basis of $H^4(\hat Y_4, \mathbb Z)$.
What we can do at this stage is deduce \emph{necessary} conditions on the fluxes by integrating $G_4 + \frac{1}{2} c_2(\hat Y_4)$ against those integral four-forms which we \emph{have} constructed. These are first of all the elements in $H^{2,2}_{\rm vert.}(\hat Y_4, \mathbb Z)$ given by the product of two two-forms and second the four-forms Poincar\'e dual to the matter surfaces $S^k_R$, which are examples of non-factorisable four-forms. 

In particular, we can test for a possible shift in the quantisation condition due to $\frac{1}{2} c_2(\hat Y_4)$
by integrating (\ref{c2mod2a}) against the above-mentioned set of four-forms.  In appendix \ref{app:Quantisation} we find that the only potential half-integer contribution can be expressed as the following integral defined entirely on the base,  
\bea \label{c2-contr}
\frac{1}{2} \int_B \cW^2 {\aK}. 
\eea
If  we now specialise to models which are  smoothly connected to a Type IIB orientifold by  imposing the constraint (\ref{KWW}), the integral (\ref{c2-contr}) is manifestly integer and there is no curvature induced shift in the quantisation condition. 
In particular, this guarantees that all chiral indices in table \ref{Fchir} are integer, as they must be, because the matter surfaces $S^k_R$ are among the integer four-forms appearing in the constraint (\ref{integer1}) and  $\frac{1}{2} c_2(\hat Y_4)$ has no effect. 

Note that for smooth Weierstrass models, the contribution $\frac{1}{2} c_2(\hat Y_4)$ \emph{is} always integer, as shown in \cite{Collinucci:2010gz}.
We stress, however, that our analysis above merely gives rise to \emph{necessary} conditions because it has not been settled that the set of integral four-forms we used is sufficient.

\subsubsection*{Fluxes in generic $SU(5)$ models}

As was shown in \cite{Krause:2011xj} by generalizing the analysis of \cite{Braun:2011zm}, it is possible to construct a similar set of fluxes for an F-theory model with generic $SU(5)$-singularity. The flux $G_4^{\,\lambda}$ is unaffected by the transition from the restricted to the generic model. On the other hand, the $G_4^X$-flux ceases to exist as a factorisable flux in $H^{2,2}_{\rm vert}(\hat Y_4)$. This is because in the $U(1)$-restricted model it depends on the resolution divisor $S$, which is not present for generic $SU(5)$ models. Physically, the transition between both geometries corresponds to a Higgsing of the extra $U(1)$ and thus affects the explicit form of the associated $U(1)$-fluxes.  
Nonetheless, one can describe an analogous gauge flux $G_4^{\,X}(\cP)$ as the pullback of factorisable classes from the ambient five-fold $\hat X_5$ into which the four-fold $\hat Y_4$ is embedded via the Tate hypersurface constraint. The resulting flux $\hat Y_4$ is then not of factorisable form. In formulae, the fluxes we obtain in this way are
\begin{align}
&&&&&&&&& \tGP    &&= \quad -X \wedge Y \wedge \cP +\left(Z + \aK - \tfrac15 (2,4,6,3)_i E_i \right) \wedge \cP |_{\hat Y_4}, &&&&&&\\
&&&&&&&&& \tG^{\,\lambda} &&= \quad \lambda \left( \,E_2 \wedge E_4 + \tfrac15 (2,-1,1,-2)_i E_i \wedge \aK \right).&&&&&&
\end{align}
In order to define $\tGP$ we assume that the Tate polynomial coefficient $a_{6,5}$ factorises into e.g.~$a_{6,5} \simeq \rho\,  \tau$ and $\cP$ denotes the class of $\rho$. Indeed it was shown in \cite{Braun:2011zm} that, in such a case, the ambient space intersection $\{x=0\} \cap \{y=0\} \cap \{\rho=0\}$ describes a four-cycle in the Calabi-Yau four-fold, denoted by $\sigma_{\rho}$.  The resulting flux is of $(2,2)$-form only on the sublocus in complex structure moduli space for which $a_{6,5} \simeq \rho \tau$. This is because, unlike fluxes in $H^{2,2}_{\rm vert}(\hat Y_4)$,  $\tGP$ generates a superpotential whose critical locus  precisely corresponds to $a_{6,5} \simeq \rho \tau$.

The intersections of $X\cdot Y$ with $\cP$ on the $SU(n)$-resolution manifold are mostly the same as those of the divisor class $S$ with $\cP$ on the $SU(5) \times U(1)$-resolution manifold \cite{Krause:2011xj}. In particular, for intersection with two base divisors one finds that
\beq
 \int_{X_5^{su_n}} X Y \cP \cB_a \cB_b = \int_{\hat{Y}_4^{su_n\times u_1}} S \cP \cB_a \cB_b,
\eeq
where we leave the pullback map implicit on the right hand side.
 The only difference occurs for the two expressions $S^2$ versus $\sigma_{\rho} \cdot \sigma_{\rho}$,
\beq
 \int_{\hat{Y}_4^{su_n\times u_1}} \left(S \cP\right)^2 = - \int_{B_3} {\aK \cP}^2,   \qquad {\rm but} \qquad   \int_{\hat{Y}_4^{su_n}} \sigma_{\rho} \cdot \sigma_{\rho} = - \int_{B_3} \aK \cP^2 + \int_{C_{34}} \cP, 
\eeq
where $C_{34}$ lies in the class $(3\aK - 2 \cW)(4\aK - 3 \cW)$. Then the chirality of the $\mathbf{10}$-state is the same as in the above case, while the chirality of the single $\mathbf{5}$-state is the weighted sum of the chiralities of the two $\mathbf{5}$-states in the above case (see \cite{Krause:2011xj}). The major difference occurs for the $D3$-tadpole contribution, which obtains an additional term scaling linearly with $\cP$,
\beq \label{QgaugeF-non-restricted}
 Q_{\rm gauge, \cP}(G_4^X) = -  \, \int_{B_3} (\aK - \tfrac35 \cW)  \cP^2 + \tfrac12  \int_{C_{34}}  \cP.\\
\eeq
One notes that an expression defined in terms of the square of the four-form flux depends linearly on $\cP$. A linear rescaling of the $\cP$  thus has different effects than a linear rescaling of $G_4^X(\cP)$. It is this property which, upon comparison to the IIB-picture in section \ref{sec:IIBSU5U1} as well the process of recombination in section \ref{sec:BraneRecomb}, fixes the overall scaling of both $G_4^X(\cP)$ and $G_4^{\lambda}$.

\subsection{Comparison of \texorpdfstring{$G_4$}{G4}-Fluxes and Spectral Cover Fluxes} \label{sec:FSCC}

Before moving on to the Type IIB picture let us take some breath and compare the factorisable $G_4$ fluxes of the previous two sections to the gauge fluxes obtained via spectral covers \cite{Friedman:1997yq,DonagiSC}. Motivated by duality with the heterotic string, the spectral cover or Higgs bundle construction encodes the  neighbourhood of a 7-brane with non-abelian gauge group --- here $SU(n)$ \cite{Donagi:2009ra,Hayashi:2008ba,Wijnholt:2012fx}. Among the possible fluxes constructed in this way is a so-called universal gauge flux defined everywhere in complex structure moduli space. Note that the spectral cover approach is exact in models with heterotic dual --- see e.g.~\cite{Curio:1998bva, Andreas:1999ng} for early comparisons of the moduli spaces in such situations ---  but yields only a semi-local description for more general fibrations. 

Specifically, generic $SU(5)$-models admit a Higgs bundle description based on an $SU(5)_\perp$ spectral cover in which the visible gauge group emerges as the commutant of $SU(5)_\perp$ in an underlying $E_8$. The associated universal spectral cover flux depends on a single parameter. 
Indeed the flux $G_4^\lambda$ given in (\ref{universal_flux_1}) is the precise analogue of this universal $SU(5)_\perp$ spectral cover flux. 
This has already been observed in \cite{Marsano:2011hv}. In fact both the chiral spectrum and the 3-brane tadpole match. This is possible because for non-restricted $SU(n)$ models these quantities localise on the $SU(n)$ divisor, which is correctly captured by the semi-local spectral cover. As we have seen, the spectral cover fluxes completely exhaust the set of factorisable fluxes in such models.

F-theory models for $SU(n)$ with  $n = 4,3,2$, on the other hand, correspond to commutant structure groups $SO(10)$, $E_6$ and $E_7$. For these no spectral cover exists, and the corresponding fluxes, e.g.~in heterotic compactifications, are constructed by different means such as del Pezzo fibrations or via the parabolic construction \cite{Friedman:1997yq}. This seems to be the underlying reason why for $SU(n), n < 5$ no simple factorisable fluxes of the form $G_4^\lambda$ have been found.

The local version of the restricted $SU(n) \times U(1)$ models is given by a split spectral cover \cite{Hayashi:2009ge,Marsano:2009gv,Blumenhagen:2009yv} (for values of $n$ where spectral covers exist), generalising heterotic constructions with abelian gauge groups \cite{Andreas:2004ja,Blumenhagen:2005ga,Tatar:2006dc,Blumenhagen:2006ux}. Again to be specific,  the $SU(5) \times U(1)$ models are the global extension of a split spectral cover with structure group $S[U(4) \times U(1)]_\perp$. Indeed, an extra class of universal fluxes arises, which has been matched in \cite{Krause:2011xj} with $G_4^X({\cal F})$. These fluxes are \emph{not} localised entirely on the $SU(5)$ brane as is obvious already from the fact that they induce a chiral spectrum for the $SU(5)$ singlets ${\bf 1}_{5}$. Correspondingly, quantities like this chiral index  and the 3-brane tadpole do receive contributions in the restricted Tate model which are not correctly captured by the split spectral cover.

Just as an example, consider the  split spectral cover fluxes in the form presented in \cite{Blumenhagen:2009yv} with the independent chiralities of $SU(5)$ charged matter  of the form
\beq
 \chi_{\mathbf{10}_1} = \mu \int_{C_{\mathbf{10}_1}} 5\aK - 4 \cW   \qquad  \chi_{\mathbf{5}_{3}} = \mu \int_{C_{\mathbf{5}_{3}}} \aK.
\eeq
It is a simple enough task to find a linear combination of $G_4^{\,X}({\cal F})$ and $G_4^{\,\lambda}$ which reproduces these topological indices. The flux in question corresponds to a combination with integer coefficients
\beq
 G_4^{\,\mu} \, := \, \mu \,  \left( - G_4^{\,X}(\aK) + 4 \, G_4^{\,\lambda=1} \right) \, = \,\mu \Big( 4 \,E_2 E_4 + \big((S-{ Z}-\aK) + (2,0,2,-1)_i E_i \big) \aK \Big).
\eeq
This flux induces precisely the chiralities listed above and, in addition, it induces a chirality of $\int_{C_{\mathbf{1}_{-5}}} -\aK$ for the state $\mathbf{1}_{-5}$.

\section{Gauge Fluxes in Type IIB Orientifolds and their Match with F-theory} \label{sec:GFIIB}

We are now in a position to address our main objective, a quantitative comparison  between $G_4$ fluxes and their Type IIB counterparts. To this end we first classify all consistent, generic gauge fluxes in Type IIB models with $U(n) \times U(1)$ brane configurations. Specialising again to $n=5$ we find an intriguing match with the gauge fluxes described previously on the F-theory side. We also comment on the role of massive $U(1)$s and their associated fluxes.

\subsection{Generic Flux Configurations in Type IIB \texorpdfstring{$U(n) \times U(1)$}{U(n) x U(1)} Models}
\label{sec:generic_flux_config}
In this section, we describe the set of consistent gauge fluxes of the Type IIB orientifold models under consideration which do not break the non-abelian gauge symmetry. 
The latter constraint implies that only gauge fluxes associated with the diagonal $U(1)_a$ symmetry and with the second $U(1)_b$  due to the extra brane-image pair are of relevance. We, furthermore, focus on fluxes which arise as the pullback of two-forms from the ambient space $X_3$ onto the brane divisors as these are the flux components that induce nontrivial chirality on the branes.

Recall that the brane set-up of the $SU(n) \times U(1)_a \times U(1)_b$ model under consideration consists of a stack of $n$ branes on a divisor $W = \{w=0\}$ (along with an image stack on $\tilde W$) and a single brane on a second divisor $V$ (along with its image on $\tilde{V}$). By D7-brane tadpole cancellation the latter is in the class
\begin{align} \label{Vclass}
 &V = 4\,D_{O7} - \left[k W + (k+1)\tilde{W}\right],   && |\,n = 2k+1, \\
 &V = 4\,D_{O7} - k\,\left[W+\tilde{W}\right],         && |\,n = 2k.
\end{align}
Again, it is useful to express the divisor classes in terms of the combinations $ V_{\pm} = V \pm \tilde{V}$. Then $V_{+} = 8\,D_{O7} - n\,W_{+}$, while $V_- = W_-$ in the $n$-odd case and $V_- = 0$, when $n$ is even.

We reiterate that the diagonal $U(1)_a$ is massive even in absence of gauge fluxes \cite{Jockers:2004yj,Plauschinn:2008yd}. This is because the divisors $W$ and $\tilde W$ lie in different homology classes, i.e.~$W_- \neq 0$, as emphasized  in particular in the recent discussion \cite{Grimm:2011tb}. For $n= 2k$, the extra $U(1)$ is massless because $V_- =0$, while for $n = 2k+1$ with $V_- \neq 0$ also the second $U(1)$ is massive by itself.
However, it is easy to see that the mass matrix is of rank one so that there is a massless linear combination of $U(1)$s given by
\bea \label{U1X}
U(1)_X =   \tfrac{1}{2} \left( U(1)_a - n \, U(1)_b   \right),
\eea  
where the overall normalization has been chosen for later convenience.

Next, let us consider a general flux on this brane set-up. 
Given a 7-brane wrapping a holomorphic divisor $D$, the relevant, gauge invariant quantity is not the curvature of the $U(1)$-bundle per se, but its sum with the pullback of the B-field onto the brane.
We choose the symbol $F$ to denote precisely this gauge invariant combination. More precisely,
\bea \label{F+B}
F|_D = \frac{\ell_s^2}{2\pi}  \langle dA \rangle + B^{(+)}|_D     \in H^2(D, \mathbb Z/2)
\eea
in terms of the orientifold even, discrete piece of the $B$-field $B^{+}$ with components $0$ or $\frac{1}{2}$.
Note that it is this combination that enters all topological quantities such as induced brane charges and chiral indices. We will comment on the relevance of the orientifold-odd component $B^{-}$ in section \ref{sec:Massive}.

We denote by $F_a$ the so-defined flux on the $SU(n)$ divisor stack and by $F_b$ the flux on the additional brane. 
The orientifold action $\sigma$ maps the flux $F_a$ on $W$ to the image flux $F_a' = - \sigma^* F_a = - \tilde F_a$ on $\tilde W$, and similarly for $F_b$.
Oftentimes, it is useful to introduce the notation
\bea
F_a^\pm = \tfrac12 \left(F_a \pm   \sigma^* F_a \right).
\eea
In particular, $F^+$ contains the pullback of the discrete B-field $B_+  = b^\alpha\, \omega_\alpha$, where $\omega_\alpha$ span a basis of $H^2_+(X_3, \mathbb Z)$ and   $b^\alpha = 0$ or $\frac{1}{2}$.

In order for the flux configuration to lead to a consistent string vacuum, the induced D5-tadpole  must vanish. This guarantees a non-anomalous spectrum.
Recall that in orientifold models with non-trivial $H^2_-(X_3)$, spanned by a basis $\omega_a$ of orientifold-odd two-forms, the induced D5-tadpole is proportional to
\bea
\Gamma_a = \int_{X_3} \omega_a \wedge \sum_i n_i \left(D_{+,i} \wedge F^-_i + D_{-,i} \wedge F^+_i \right).
\eea
Here $D_{\pm,i}$ labels the combinations $D_i \pm \tilde D_i$ of brane divisors and image divisors, each carrying a stack of $n_i$ 7-branes and corresponding fluxes. 
D5-brane tadpole cancellation requires that all $h^{1,1}_-(X_3)$ components of $\Gamma_a$ vanish. 
Applied to the specific brane configuration under consideration, this amounts to the constraint
\begin{align}
  &0 = n\,W_+\,\left(F_a^- - F_b^-\right) + W_-\,\left(n\,F_a^+ + F_b^+\right),    && |\,n = 2k+1, \\
  &0 = n\,W_+\,\left(F_a^- - F_b^-\right) + n \, W_-\,F_a^+,                       && |\,n = 2k,
\end{align}
where we have used the fact that the pullback of any involution-odd class to $D_{O7}$ is zero.

 It is immediately clear that the following flux choices do not induce any $D5$-tadpole (where we express each flux as a tuple $\left(F_a, F_b\right)$):
\begin{equation} \label{genericF}
 \begin{array}{l l l c l}
              &        & n = 2k+1                       &\qquad& n = 2k \\
  \hline
  F_X    &:=\qquad & \left( \frac{1}{2n} F, - \frac{1}{2} F \right)            && \left(0, \frac{1}{n} F \right)\\
  F_Y    &:=\qquad & \left(B^-, B^-\right)          && \left(B^-, B^-\right) \\
  F_{\lambda} &:=\qquad & \left(\frac{2 \lambda}{n} D_{O7}, 0\right) && \left(\frac{2 \lambda}{n} D_{O7}, 0\right)
 \end{array}
\end{equation}

Here $F$ and $B^-$ are general elements of $H^2_+(X_3)$ and $H^2_-(X_3)$, respectively, whose  pullback to the divisors represents the corresponding fluxes as described above.  It is understood that all quantities are chosen in a manner consistent with the Freed-Witten quantization condition.  Our normalisation has been picked in order to facilitate match with F-theory.
It turns out that $F_Y$ is entirely trivial --- it does not contribute to the  $D3$-tadpole, to any chiral index  and to the D-term of the massless $U(1)_X$. This is clear because it can be absorbed into the orientifold odd component of the $B$-field since the same amount of $B^-$ is switched on along all branes in the set-up. That the orientifold-odd component of the $B$-field does not enter any of the above topological quantities in consistent set-ups satisfying D7- and D5-brane tadpole cancellation has been demonstrated in \cite{Blumenhagen:2008zz}, and we will comment on its contribution to the D-terms in section \ref{sec:Massive}. On the other hand, $F_X$ is the flux associated with the massless combination $U(1)_X$ given by (\ref{U1X}) for  odd $n$ and, respectively, with the massless $U(1)_b$ on the additional brane for even $n$.

We will now show that  the fluxes $F_X$ and $F_\lambda$ completely exhaust the \emph{generically} possible D5-tadpole free flux configurations.
To see this let us make the following general ansatz for the fluxes
\begin{equation}
 \begin{aligned}
  F_a &=\quad k_a\,D_{O7} \, + \, w_a^+\,W_+  \, + \, G_a^+ \quad + \, w_a^-\,W_- \, + \, G_a^- ,\\
  F_b &=\quad k_b\,D_{O7} \, + \, w_b^+\,W_+  \, + \, G_b^+ \quad + \, w_b^-\,W_- \, + \, G_b^- ,\\
 \end{aligned}
\end{equation}
where the $G_i^+$ do not involve $D_{O7}$ and $W_+$ and the $G_i^-$ do not involve $W_-$. Substituting these into the above expression for the D5-tadpole results in three conditions which constrain $G_i^{\pm}$ and relate the $w_i^{\pm}$.

Of course for special properties of the intersection matrix extra solutions may exist; these, however, are not among the \emph{generically} possible flux configurations.
Sticking to the generic case, the most general form of flux can then be written as the sum of two positive flux configurations:
\beq
  F_{gen} = (F_a, F_b) =  F_X  + F_{\lambda},
\eeq
where $\lambda$ depends on the $k_i$s and the $w_i$s and the flux appearing in $F_X$ is the inolution even piece of $F_b$. Here we have used the fact that $W_+ - W_- = 2\,\tilde{W}$ along with the restriction (\ref{eq:brane-image-inter-relation}) that $D_{O7}\,W = \tilde{W}\,W$. Furthermore, we discard a possible contribution $F_Y$ as this is indistinguishable from a trivial flux configuration. \\

\subsection{Specialisation to Type IIB \texorpdfstring{$U(5) (\times U(1))$}{U(5) (x U(1))} Models and Match with F-theory} \label{sec:IIBSU5U1}
Having classified the \emph{generic} D5-tadpole free flux configurations we can compare these with the generically possible $G_4$-fluxes in the F-theory picture.
Our strategy to establish such a correspondence is by comparing the flux-dependent topological quantities on both sides. For definiteness we now specialise to the case of $SU(5)$ models, but the analysis could equally well be carried out for other values of $SU(n)$.

\subsubsection*{Chiralities in \texorpdfstring{$U(5) \times U(1)$}{U(5) x U(1)} Models}

We first turn to the chiral index of charged fields. Type IIB orientifold models with $SU(5) \times U(1)_a \times U(1)_b$ symmetry are discussed extensively in \cite{ Blumenhagen:2008zz}. The matter content is listed in the left column of table \ref{IIBindexb}, where the subscripts denote the $U(1)_a$ and $U(1)_b$ charges. From the general chirality formula for the bifundamental matter $(\ov N_a, N_b)$ between two stacks of branes along $D_a$ and $D_b$ (and similar expressions for the antisymmetric matter in $\mathbf{10}_{(2,0)}$ and symmetric matter in $\mathbf{1}_{(0,2)}$, see e.g.~\cite{ Blumenhagen:2008zz}), 
\bea
I_{ab} = - \int_{X_3} D_a \wedge D_b \wedge (F_a - F_b),
\eea
one deduces the chiral indices for $F_{\lambda}$ and $F_X$ as summarised in table \ref{IIBindexb}. In particular, $F_Y$ does not induce any chirality.

\begin{table}[htb]
 \begin{center}
  \begin{tabular}{ c | c  c| c c c} 
  {State}                &\quad&   Chirality under \,  $F_\lambda$    &\quad&  Chirality under \, $F_X$ \\
  \hline \\[-8pt]
 $ \mathbf{10}_{(2,0)}  $&   &$     \frac{\lambda}{5}  \int_{X_3}    D_{O7} W_+^{\,2}       $&&$    \frac{1}{10}  \int_{X_3}    D_{O7} W_+ F   $       \\
 $ \mathbf{5}_{(1,-1)}   $&   &$    -\tfrac{\lambda}{10}  \int_{X_3}    D_{O7} W_+^{\,2}      $&&  $    \frac{1}{10}    \int_{X_3}    \left(9 D_{O7} W_+ - 6 W_+^{\,2}\right)F     $       \\
 $ \mathbf{5}_{(1,1)}    $&   & $   - \tfrac{\lambda}{10}   \int_{X_3}    D_{O7} W_+^{\,2}      $&&    $    \frac{1}{10}  \int_{X_3}     -\left(10 D_{O7} W_+ - 6 W_+^{\,2}\right)F      $     \\
 $ \mathbf{1}_{(0,2)}    $&   &  $  0 $                                                                                          &&     $     \frac{1}{10}  \int_{X_3}        -5 \left(12 D_{O7}^{\,2} - 17 D_{O7} W_+ + 6 W_+^{\,2}\right) F $
\end{tabular}
 \caption{Chiral index with respect to generic fluxes in Type IIB orientifolds.}
 \label{IIBindexb}
 \end{center}
\end{table}

These expression are to be compared with the chiral index induced by the universal flux $G_4^{\,\lambda}$ and the $U(1)_X $ flux $G_4^{\,X}({\cal F})$ as collected in table \ref{Fchir}. 
Indeed the F-theory $U(1)_X$ charges are reproduced by the Type IIB charges if we set
\bea \label{U1X-def}
U(1)_X = \tfrac{1}{2} \left( U(1)_a - 5 \, U(1)_b \right),
\eea
which is precisely the massless combination of abelian factors, see eq.~(\ref{U1X}).
With the help of the definitions (\ref{Fmattercurves}) of the curves on the F-theory side and the F-theory/Type IIB dictionary  \eqref{piD}, (\ref{pibarK}) and (\ref{integral-match}),
one observes that the chiral indices in  tables \ref{IIBindexb} and \ref{Fchir} match precisely if we  identify
\bea
G_4^\lambda \leftrightarrow  F_\lambda, \qquad G_4^{\,X}({\cal F}) \leftrightarrow F_X =  (\tfrac{1}{10}  F, -\tfrac12 F)  \quad {\rm with} \quad F = \pi^*({\cal F}).
\eea

\subsubsection*{D3-tadpole in $U(5) \times U(1)$ models}

This picture receives additional support from comparison of the induced D3-tadpole charges.
For general fluxes, the Type IIB flux induced tadpole $Q_{\rm gauge, \, IIB}$ appearing in (\ref{D3tad-gen}) is 
\bea
Q_{\rm gauge, \, IIB} &=&  - \frac{1}{4}  \sum_i n_i  \Big( \int_{D_i} F_i^2  + \int_{\tilde D_i} (\sigma^*F_i)^2  \Big). 
\eea
Specialising to $F_X$ and $F_{\lambda}$ as above, one can read off the following $D3$-charges
\beq
 Q_{\rm gauge, IIB} = - \frac12 \int_{X_3} \left[ \left(D_{O7} - \tfrac35  W_+ \right) F^2 + \tfrac15 \lambda \, D_{O7} \, W_+ \, F + \tfrac25  \lambda^2 \, D_{O7}^{\,2} \,  W_+\right] .
\eeq
These agree with the analogous charges (\ref{QgaugeF}) on the F-theory side. \\

\subsubsection*{The D-term in $U(5) \times U(1)$ models}

The identification (\ref{U1X-def}) allows us to match also the flux induced D-terms for the massless $U(1)_X$ symmetry.
In Type IIB, the $U(1)_X$ Fayet-Iliopoulos term is the corresponding linear combination of D-terms for the diagonal $U(1)_a$ and  $U(1)_b$, i.e.
\bea \label{xiXIIB}
\xi_X \simeq \frac{1}{2{\cal V}_X} \Big(  \int_{X_3} D_a \wedge J \wedge {\rm tr} \, F_a - 5   \int_{X_3} D_b \wedge J \wedge {\rm tr} \, F_b \Big). 
\eea
Evaluated for the combinations $F_X$ and $F_\lambda$ this reproduces the F-theory D-terms  (\ref{DXGX}) if we  take into account factor of $5$ from the trace of $F_a$.

\subsubsection*{Quantisation condition in $U(5) \times U(1)$ models}

Finally, we turn to flux quantisation in Type IIB orientifolds.
According to the analysis of  Freed and Witten for a single brane in oriented Type II string theory, the curvature $\langle dA \rangle$ of what is sloppily referred to as the $U(1)$-bundle on the brane is half-integer quantised whenever the brane worldvolume is non-spin and integer otherwise \cite{Freed:1999vc}. This is generalised by demanding that the flux on each single brane satisfy this constraint individually.  Applied to the case at hand, this reasoning leads to the two independent constraints
\bea \label{FWIIB}
&& \left( \tfrac{1}{10} F + \tfrac{2}{5}\,  \lambda \, D_{O7} + B      + \tfrac{1}{2} W \right)\Big |_W \in H^2(W, \mathbb Z), \\
&&\left( - \tfrac{1}{2} F  + B  + \tfrac{1}{2} W \right)\Big |_W \in H^2(W, \mathbb Z), 
\eea
where the first constraint relates to the $U(5)$ brane stack and the second is due to the fluxes along the $U(1)$-brane $V = 4 D_{O7} - 2 \tilde W - 3 W$.
Recall that in our conventions the gauge flux is defined as the gauge invariant combination (\ref{F+B}) including the $B^+$-field, and in addition we must allow for a non-trivial $B^-$-field corresponding to the flux $F_Y$ in (\ref{genericF}). In effect, the full $B$-field appears in the quantisation condition.

\subsubsection*{Generic $U(5)$ models}

Generic $U(5)$-models are related to $U(5) \times U(1)$ models by a  recombination process involving the $U(1)_b$ brane along divisor $V$ and its image along $\tilde V$.
The resulting invariant brane  in the class $V_+$  is of Whitney type \cite{Collinucci:2008pf}.  We will carefully analyse this recombination process in appendix \ref{app:Whitney_brane_charges}. 
A set of gauge fluxes along the Whitney brane are described in the language of  $D9/\bar{D9}$ tachyon condensation, as discussed in \cite{Collinucci:2008pf,Collinucci:2010gz, Braun:2011zm}       for models with at most $SU(2)/Sp(1)$ gauge groups. For such set-ups the Type IIB fluxes were identified with the analogue of the non-factorisable flux $G_4^X({\cal P})$.

In appendix \ref{app:Whitney_brane_charges} we generalise this to all $U(n)$ models. As an important novelty that becomes relevant for $U(n)$ models with $n >2$, we show that the Whitney brane, despite being invariant, `knows' about the $V_-$-contribution: While the $D7$-charge only depends on the involution-invariant part of the $D7$-charges of the brane/image-brane pair, the $D5$- and $D3$-charges also depend on the anti-invariant part. For example, the 
D3-brane tadpole-contribution of the above flux set-up in the $U(5)$-case is
\bea
Q_{\rm gauge, IIB} =  - \frac{1}{2} \int_{X_3}  P^2  ( D_{O7} - \frac{3}{5} W_+) +  \frac{P}{2} \left( 4 D_{O7} - 3 W_+ \right) \left(3 D_{O7} - 2 W_+ \right).
\eea

This matches the 3-brane charge of $G_4^X({\cal P})$ in generic F-theory SU(5) models, eq.~(\ref{QgaugeF-non-restricted}), if we   identify ${P} = - \pi^*{\cal P}$.

\subsection{Massive \texorpdfstring{$U(1)$}{U(1)}s and their Fluxes in Type IIB and F-theory }\label{sec:Massive}

To conclude this analysis we would like to address the question of geometrically massive $U(1)$s in the present context and compare our findings with the discussion in \cite{Grimm:2011dj,Grimm:2011tb}.

In addition to the abelian gauge symmetry   (\ref{U1X-def}) with massless gauge potential the Type IIB orientifold enjoys also a massive $U(1)$-symmetry given by the orthogonal combination
\bea
U(1)_{X'} = \tfrac{1}{2} \left( 5 \, U(1)_a + \, U(1)_b \right).
\eea
This is because both $U(1)_a$ and $U(1)_b$ individually receive a mass by what has been dubbed in \cite{Grimm:2011tb} the geometric St\"uckelberg mechanism. This St\"uckelberg mechanism operates even in absence of gauge flux due to the gauging of the shift symmetry of the axionic fields $c^a$ obtained by expanding $C_2 = c^a \omega_a$ with $\omega_a$ a basis of $H^{1,1}_-(X_3)$.
Massive $U(1)$-symmetries are known to give rise to global perturbative symmetries respected by the perturbative Yukawa couplings and broken only non-perturbatively by D-brane instantons.\footnote{Recent investigations of M5/D3-instantons in this context include~\cite{Blumenhagen:2010ja,Cvetic:2010rq,Donagi:2010pd,Grimm:2011dj,Marsano:2011nn,Cvetic:2011gp,Bianchi:2011qh}.} The F-theory uplift of such a massive $U(1)$ was discussed in \cite{Grimm:2011tb} in terms of  a non-harmonic, more precisely non-closed, two-form $\tw_{X'}$. This allowed for a detailed match of the St\"uckelberg mechanism in the framework of gauged supergravity.
Based on this field theoretic agreement, it was suggested that the uplift of the Type IIB gauge fluxes associated with such massive $U(1)$-symmetries involves non-harmonic forms with the important caveat that such non-harmonic fluxes have to satisfy the F-theory uplift  of the Type IIB D5-tadpole cancellation condition. This condition relates various components of the fluxes, in particular the non-harmonic ones, to each other. Note that while harmonic $G_4$-fluxes automatically satisfy this constraint, it could not be decided at a general level whether the converse is also true, i.e.~whether every gauge flux associated with a massive $U(1)$ admits an alternative description in terms of harmonic forms which is equivalent to the proposed description in terms of non-harmonic forms plus the D5-tadpole condition.

Let us compare this picture to the results of the present work: One of our main findings has been to identify the generic D5-tadpole-free $U(1)_a$ flux $F_\lambda$ with the \emph{harmonic} universal flux $G_4^\lambda$. In view of the above discussion this is not in contradiction with the proposal of \cite{Grimm:2011tb}; rather, it demonstrates that --- at least for the fluxes discussed here ---  the F-theory D5-tadpole condition does remove the non-harmonic parts of the fluxes such that an effective description in terms of harmonic fluxes \emph{is} possible. It would be important to further investigate if this phenomenon applies to all F-theory fluxes associated with a geometrically massive $U(1)$ in the Type IIB picture.

Another interesting question concerns the D-term of the massive $U(1)$.
In Type IIB orientifolds, where the mass of the geometrically massive $U(1)$  is $g_s$-suppressed with respect to the Kaluza-Klein scale,
the associated D-terms are kept in the effective field theory analysis. The field-dependent Fayet-Iliopoulos term of such a massive $U(1)$ receives a contribution from the gauge flux and also from the dynamical B-field moduli $b^a$. These arise by expanding $B= b^a \, \omega_a$ and combine with the above mentioned $c^a$ into $h^{1,1}_-(X_3)$ chiral multiplets $G^a = c^a - \tau b^a$ with $\tau$ the axio-dilaton. Note that these moduli are present whenever a $U(1)$-symmetry becomes geometrically massive because the geometric St\"uckelberg mechanisms hinges on $h^{1,1}_- \neq 0$. 
In the case at hand, the Type IIB D-term for $U(1)_{X'}$ is therefore the linear combination of the terms\footnote{In  $\xi_{X}$, eq.~(\ref{xiXIIB}), the $b^a$-dependent terms have have cancelled because of the specific homological relation between $D_a$ and $D_b$ which renders the combination $U(1)_X$ massless.}
\bea
\xi_{X'} \simeq \frac{1}{2{\cal V}_X} \Big( 5  \int_{X_3} D_a \wedge J \wedge (F_a + B^-)  +    \int_{X_3} D_b \wedge J \wedge (F_b +B^-) \Big).
\eea
 
In the unhiggsed phase of vanishing vacuum expectation values of all charged open string fields the D-term supersymmetry conditions read
\bea
\xi_X = 0, \qquad \qquad \xi_{X'}=0.
\eea 

On the other hand, the F-theory D-term supersymmetry condition involves only the vanishing of $\xi_X$ given by (\ref{xiXF}). So are we missing the extra D-term $\xi_{X'}=0$?
Indeed, the gauged supergravity analysis of \cite{Grimm:2011tb} suggests the existence of a non-closed two-form $\tw_{X'}$ describing the massive $U(1)_{X'}$ in F-theory. Dimensional reduction of the M-theory action including this non-harmonic form precisely reproduces the D-term $\xi_{X'}$. To understand why the constraint $\xi_{X'}=0$ never appears in F-theory we must take into account that in Type IIB theory it is always possible to achieve $\xi_{X'}=0$ by a suitable choice of B-field moduli $b^a$,  see \cite{Grimm:2011dj} for a more detailed discussion. Due to their appearance in the D-term these moduli acquire a mass of the order of the $U(1)_{X'}$ mass. In F-theory, where this mass is no longer suppressed with respect to the Kaluza-Klein scale, 
both the gauge boson and the $B$-field moduli have been integrated out. Therefore, the usual F-theory description which ignores the D-term $\xi_{X'}$ corresponds to the Type IIB configuration with the $B$-field moduli canceling the flux contribution to the D-term. If, on the other hand, we insist on including the $B$-field moduli into the effective action we also need to take into account the massive $U(1)_{X'}$ gauge boson. This leads to the dimensional reduction of \cite{Grimm:2011tb} including the non-harmonic two-form $\tw_{X'}$.

\section{Brane Recombination} \label{sec:BraneRecomb}

In this section we shed some more light on the brane recombination process that interpolates between the $U(1)$-restricted F-theory model with $a_6=0$ on the one hand and the generic $SU(n)$ model on the other.  
We have seen in section \ref{sec:geometry_topology} that the $U(1)$-restricted and non-restricted models correspond to IIB brane set-ups with a brane/image-brane pair in the former and a Whitney-type brane in the latter case. The intermediate process of recombination is expected to be describable as a smooth deformation in the complex structure moduli space in F-theory, as described in this specific context in \cite{Grimm:2010ez,Braun:2011zm}.  

Brane recombination is known to lead to a jump in the flux quanta \cite{Gaiotto:2005rp}. In fact, since the overall $D3$-charge is to be invariant under such a smooth deformation, the change in the flux should compensate precisely for the change in the geometrically induced $D3$-charge. This has been exploited e.g.~in \cite{Collinucci:2008pf} in similar contexts. 
From the analysis around eq.~(\ref{tadchange}) we know that the change in the geometric D3-brane charge is given by 
\beq
\Delta Q_{D3}^g = \tfrac18 \chi\left(C_{34}\right).
\eeq
The four-form fluxes, on the other hand, can be expressed as
\beq
 \bal
  G_4^{su_n } = \tGP + \tG^{\lambda}, \qquad  G_4^{su_n \times u_1} = G_4^X(\cF) + G_4^{\lambda^{\prime}}
 \eal
\eeq
and their $D3$-tadpole contribution is given in (\ref{QgaugeF}) and (\ref{QgaugeF-non-restricted}), respectively. Both expressions are integrals over the base manifold, which is the same for both resolution manifolds. It is, therefore, possible to simply subtract one of the expressions from the other. Let us define
\beq \label{flux_change}
 A = \cP - \cF,  \qquad  \sigma = \lambda - \lambda^{\prime}.
\eeq
Since the resulting expression for $\Delta Q_{\rm gauge}$ should only depend on the differences of the fluxes, not on the specific choice of $\cP$, $\cF$ or $\lambda$, $\lambda^{\prime}$, this fixes
\beq \label{flux_change2}
 A = \tfrac12 \left[a_{6,5} \right] =  - \frac{1}{2} c_1(C_{34}), \qquad \sigma = \tfrac12.
\eeq
Substituting these values back into the expression for $\Delta Q_{\rm gauge}$ confirms that this contribution does indeed cancel $\Delta Q_{D3}^g$.\\

The values for the change in the fluxes fits in nicely with the quantisation conditions. Since modulo two, we have $c_2^{su_n } = \tfrac{5}{2} \tG^{\lambda}$, while $c_2^{su_n \times u_1} = \tfrac{5}{2} G_4^X(W)$ (see the discussion in section \ref{sec:FSU5U1}), the following configurations are \emph{sufficient} to meet the Freed-Witten quantisation condition:
\beq
 \bal
  SU(n) \times U(1): \qquad & \frac{1}{5} \, \cF + \tfrac12 W \in \bbZ, \qquad \frac{\lambda^{\prime}}{5} \in \bbZ, \\
  SU(n): \qquad & \frac{1}{5} \cP  \in \bbZ, \qquad\qquad\quad \frac{\lambda}{5} + \tfrac12 \in \bbZ.
 \eal
\eeq
We observe that the change in the fluxes as given by (\ref{flux_change}) interpolates between these two configurations.\\

In field theory  the recombination process is interpreted as a D-flat vacuum expectation value of vector-like open string recombination modes $\mathbf{1}_{{n}} + \mathbf{1}_{{-n}}$ localised on the recombination curve $C_{34}$. Before analyzing the condition for such vector-like pairs to exist, let us note that the flux $\tilde G_4^X({\cal P})$ in the restricted and $ G_4^X({\cal F})$ in the non-restricted models differ in that $G_4^X({\cal F})$ does not induce any superpotential in the effective action. This is because, being described by an element of $H^{2,2}_{\rm vert.}(\hat Y_4)$, it is always of $(2,2)$ type. By contrast, $\tilde G_4^X(\cal P)$ does induce a superpotential for the complex structure moduli of the four-fold. In Type IIB language, the superpotential fixes some of the brane deformation moduli describing the geometry of the Whitney brane.

This is consistent with the following picture:
Before recombination there exist pairs of vector-like recombination modes $\Phi_i, \tilde \Phi_i$ localized on $C_{34}$ which acquire a VEV in a D-flat manner if $a_6$ is switched on.\footnote{In addition the spectrum generically contains a chiral excess of either $\Phi$ or $\tilde \Phi$-fields.} 
The recombination modes participate in Yukawa or higher-order flux-dependent F-term couplings of the schematic form
\bea
\sum \Phi_i \tilde \Phi_j  W_k(\zeta) ,
\eea
where $\zeta$ represent the open string moduli of the 7-brane. As the recombination moduli condense, the brane moduli $\zeta$ are constrained by a flux-dependent F-term. In particular if $W_k(\zeta)$ is quadratic, they acquire a mass after recombination.

The superpotential is such that after recombination $a_{6,5}$ factories into $a_{6,5} = \rho \tau$ \cite{Braun:2011zm}. As the two-form $\cal P$ appearing in the flux $\tilde G_4^X({\cal P})$ is in the class $\rho$ it is constrained to lie in the domain
\bea \label{Pconstr}
0 \leq {\cal P} \leq  a_{6,5}.
\eea
This constraint admits a neat interpretation from the recombination picture: Together with the relations (\ref{flux_change}) and (\ref{flux_change2}) between the fluxes before and after recombination, eq.~(\ref{Pconstr})  ensures that vector-like pairs of recombination modes transforming as ${\bf 1_n} + \mathbf{1}_{{-n}}$ exist in the non-recombined phase $a_{6,5}=0$. Put differently, fluxes outside the domain (\ref{Pconstr}) would correspond to fluxes $G_4^X(\cal F)$ in the non-recombined phase for which the spectrum of recombination modes on $C_{34}$ is purely chiral. In this case no recombination in the above sense is possible and the flux $G_4^X({\cal F})$ acts as an obstruction. (See \cite{Donagi:2011jy} for a recent discussion of processes with chiral 'recombination' modes in the language of brane gluings).

This can be seen as follows: Since the massless modes in question are localized along the curve $C_{34}$ the discussion can be phrased in terms of the line bundle  $L$ on $C_{34}$ obtained by pulling back the gauge flux ${\cal F}$ from the base $B$ to $C_{34}$. In other words we define ${L} $ as the line bundle on $C_{34}$ with  first Chern class or degree
\bea
d({L} ) = {\cal F}|_{C_{34}}.
\eea
The massless modes ${\bf 1_n}$ and $\mathbf{1}_{{-n}}$ on $C_{34}$ are given by the cohomology groups
\bea
H^i(C_{34}, L  \otimes \sqrt{K_{C_{34}}} )
\eea
with $i=0$ and $i=1$, respectively.
The chiral index which follows by Riemann-Roch, 
\bea
\chi(C_{34}, L \otimes \sqrt{K_{C_{34}}} ) =  h^0(C_{34}, L \otimes \sqrt{K_{C_{34}}} )  - h^1(C,  L \otimes \sqrt{K_{C_{34}}}  )  =   d( L),
\eea
 is in agreement  with the result $\chi({\bf 1_n}) = \int_{C_{34}} {\cal F }$ as given in table \ref{Fchir} for the special case $n=5$.
 
 Now, a necessary condition for both $h^0(C_{34}, L \otimes \sqrt{K_{C_{34}}} )$ and  $h^1(C,  L \otimes \sqrt{K_{C_{34}}}  ) $ to be non-vanishing --- corresponding to the existence of vector-like pairs of recombination modes ---  is that 
 \bea
 -\frac{1}{2} d(K_{C_{34}})  \leq d(L) \leq \frac{1}{2} d(K_{C_{34}}), 
 \eea
where  the degree $d(K_{C_{34}})$ is just
\bea
d(K_{C_{34}}) =  - c_1(C_{34})|_{C_{34}} = a_{6,5} |_{C_{34}}.
\eea
 
First, by Kodaira's vanishing theorem,  line bundles of negative degree on a curve cannot have any sections, hence for $h^0(C_{34}, L \otimes \sqrt{K_{C_{34}}} ) $ to be non-zero we need $0 \leq d(L \otimes \sqrt{K_{C_{34}}})$. By Serre duality, $h^1 (L \otimes \sqrt{K_{C_{34}}} ) = h^0 (L^\vee \otimes \sqrt{K_{C_{34}}} ) $, from which the second inequality follows by the same argument.

 Therefore, in order for brane recombination to be possible we need that
 \bea
 -\frac{1}{2} a_{6,5}|_{C_{34}} \leq {\cal F} |_{C_{34}} \leq \frac{1}{2} a_{6,5} |_{C_{34}}.
  \eea
 In view of our findings that ${\cal F} = {\cal P} - \frac{1}{2} a_{6,5}$, see (\ref{flux_change}) and (\ref{flux_change2}), this is just the restriction of the constraint (\ref{Pconstr}) onto $C_{34}$, as promised.

\section{Summary and Open Questions} \label{sec:Concl}

In this article we have established a detailed dictionary between $G_4$ gauge fluxes in F-theory compactifications and Type IIB orientifolds.
We have focused on F-theory Tate models
 with $SU(n)$ and $SU(n) \times U(1)$ gauge groups for $n \leq 5$. The technology for resolving the associated Calabi-Yau four-folds generalises our methods in \cite{Krause:2011xj}
 and provides all necessary topological data of the F-theory geometry in a manner independent of the base space.
 
 Key to our analysis of the Type IIB analogue of these F-theory compactifications has been to avoid the conifold singularity \cite{Donagi:2009ra} typically encountered in the Sen limit of an F-theory model.  In order for the $SU(n)$ F-theory model to be smoothly connected to a Type IIB orientifold 
certain topological restrictions on the base space $B$ of the F-theory four-fold must be imposed. These restrictions amount to absence of an $E_6$ point in $SU(5)$ models and its analogue for other values of $n$. 
This demonstrates that, as expected, the class of F-theory compactifications is much larger than the class of Type IIB orientifolds.

Gauge fluxes in F-theory are given by $G_4$-fluxes and are thus encoded in $H^{2,2}(\hat Y_4)$. The simplest such four-forms factorise into two two-forms and lie in the primary vertical subspace $H^{2,2}_{\rm vert}(\hat Y_4)$.
For the Tate models under consideration we have found two types of factorisable gauge fluxes  in  $H^{2,2}_{\rm vert}(\hat Y_4)$. First,  in $SU(n) \times U(1)$ Tate models \cite{Grimm:2010ez} the flux associated with the $U(1)$-factor is of the form $G_4^X = {\cal F} \wedge \tw_X$ \cite{Grimm:2010ez,Braun:2011zm,Krause:2011xj,Grimm:2011fx}. Our analysis identifies the corresponding Type IIB fluxes as a linear combination of flux on the $U(n)$ and the extra $U(1)$-brane/image-brane  stack. Second, for $SU(n)$ with $n=5$ one finds one extra type of gauge flux in $H^{2,2}_{\rm vert}(\hat Y_4)$ called $G_4^\lambda$. This flux has been identified in \cite{Marsano:2011hv} as the type of universal flux provided via spectral covers in the Higgs bundle approach. Our detailed match with Type IIB shows that this flux is associated with the diagonal $U(1) \subset U(5)$, which is massive by the geometric St\"uckelberg mechanism.   This result is rather surprising given the very different group theory underlying both types of constructions. An important ingredient to establish the correspondence with Type IIB is the D5-tadpole cancellation condition, which is automatic for harmonic $G_4$-fluxes but must be imposed by hand in Type IIB. 

By deforming the $SU(n) \times U(1)$ model into a generic $SU(n)$ model, the factorisable flux $G_4^X$ turns into flux given by an element of $H^4(\hat Y_4)$ which cannot be written as the sum of products of two-forms \cite{Braun:2011zm,Krause:2011xj}. It induces a superpotential that fixes some of the brane moduli. We have identified the related constraint on the flux quantum numbers as the condition for existence of vector-like pairs of recombination moduli. Furthermore, a careful analysis of the Whitney brane resulting from brane recombination in Type IIB establishes a quantitative match of the associated fluxes.

Apart from this type of non-factorisable fluxes, our analysis has focused on fluxes in $H^{2,2}_{\rm vert}(\hat Y_4)$. An important question for future work is to get a better handle on the remaining elements in $H^4(\hat Y_4)$ and the associated fluxes. 
As one of the most pressing open questions we point out that 
for $SU(n)$ models with $n <5$, no universal fluxes of type $G_4^\lambda$ exist as elements in $H^{2,2}_{\rm vert}(\hat Y_4)$. 
From the perspective of the spectral cover construction, this does not come as a complete shock. To understand this recall that for $SU(5)$,  $G_4^\lambda \in H^{2,2}_{\rm vert}(\hat Y_4)$ is the global extension of the spectral cover $SU(5)_\perp$-flux, where $SU(5)_\perp$ describes the orthogonal complement of $SU(5)$ within $E_8$. Gauge groups $SU(n), n= 4,3,2$  would be associated with orthogonal structure groups $G= SO(10), E_6, E_7$, for which no spectral covers exist. Indeed, in heterotic models the corresponding flux is constructed by different methods such as del Pezzo fibrations or via extensions \cite{Friedman:1997yq}. 
On the other hand, from a Type IIB perspective it is not obvious what distinguishes $SU(5)$ from $SU(n)$ with $n<5$.
Our result that the $SU(5)$ spectral cover fluxes represent merely the diagonal $U(1) \subset U(5)$ fluxes  in Type IIB
suggests that an analogue of the universal fluxes should exist also for $n<5$.
One possibility is that the missing fluxes in F-theory are not described by elements in $H^{2,2}_{\rm vert}(\hat Y_4)$, but rather by non-factorisable fluxes. The corresponding four-forms must, however, exist as integral $(2,2)$ forms for all values of complex structure moduli (unlike the ones obtained by recombination from $G_4^X$ as sketched above) because the Type IIB fluxes do not induce a superpotential; this property should not change in F-theory. 

A better understanding of non-factorisable four-forms is also required for a  complete evaluation of the quantisation condition, which necessitates a basis of $H^{4}(\hat Y_4, \mathbb Z) \cap H^{2,2}(\hat Y_4)$. Once such a basis is available it will be possible to compare the F/M-theory quantisation condition in detail with the Freed-Witten condition in Type IIB. We hope to return to these questions in the future.

% At least for F-theory models with a smooth Type IIB limit, our preliminary analysis suggests that the flux quantisation is not shifted by $\frac{1}{2}$
%  because $c_2(\hat Y_4)$ is even, in agreement with the findings of \cite{Collinucci:2010gz} for models with up to $Sp(1)/SU(2)$  singularities. If substantiated further, this result would imply that the Type IIB limit of such models corresponds to configurations with non-trivial $B$-field such as to balance possible half-integer shifts in the Freed-Witten condition in case the brane divisors are non-spin. 
% On the other hand, what is currently surprising is that the Type IIB Freed-Witten condition treats the flux on each brane independently, while $G_4$ fluxes in F-theory provide a combined description of all fluxes. The presence of extra four-forms could single out the flux integral over individual branes and render the quantisation prescriptions in both pictures equivalent. 

\section*{Acknowledgements}

We thank Thomas Grimm, Arthur Hebecker, Max Kerstan, Eran Palti and Roberto Valan\-dro for discussions. S.K. acknowledges financial support from the Klaus-Tschira-Stiftung. This work was furthermore supported by the Transregio TR33 "The Dark Universe".

\newpage

\appendix
\section{Detailed Geometry of F-theory Resolutions for \texorpdfstring{$SU(n) \left( \times U(1) \right)$}{SU(n) (x U(1))} Models} \label{app:DetGeom}

In this appendix  we provide the details of the resolution of singular Calabi-Yau four-folds $\hat Y_4$ describing $SU(n)$ and $SU(n) \times U(1)$ models in F-theory. Special emphasis will be put on the computation of intersection numbers involving the resolution divisors as well as topological quantities such as $c_2(\hat Y_4)$ the Euler characteristic $\chi(\hat Y_4)$. 

\subsection{Resolution Structure}
\label{app:res_struc}  
\subsubsection*{$SU(n)$-Models}

Consider an elliptically fibered CalabiYau four-fold $Y_4$ given as a Weierstrass model in Tate form as in (\ref{tate_poly}),
\begin{equation}
\label{PT1}
 P_T = \{y^2 + a_1 x y z + a_3 y z^3 = x^3 + a_2 x^2 z^2 + a_4 x z^4 + a_6 z^6\}.
\end{equation}
The Tate algorithm \cite{Bershadsky:1996nh} identifies the vanishing orders of the sections $a_i$ of $\aK^i$ along a divisor $\cW: w=0$ in $B$ for a model with $SU(n)$ singularity along $\cW$ as follows:
\beq \label{app_vi}
 \begin{array}{c | c c c c c}
       & \quad a_1 \quad & \quad a_2 \quad & \quad a_3 \quad & \quad a_4 \quad & \quad a_6 \quad \\
  \hline
  \quad v_i^n \quad &  0  &  1  &  k  & n-k &  n
 \end{array}
\eeq
Here $v_i$ stands for the vanishing order of $a_i = w^{v_i} a_{i, v_i}$, and $k$ is defined via $n = 2\,k$ or $n = 2\,k + 1$ if $n$ is even or odd respectively. Explicitly, we list the third and fourth vanishing orders for $n = 2,...,5$ in table \ref{v3v4}.

  \begin{table}[htbp]
 \begin{center}
  \begin{tabular}{c | c c}
 \quad $n$ \quad &  $\quad v_3^n \quad  $&$ v_4^n $ \\
  \hline
 $ 2$ &$   1    $ &$   1$    \\
$  3 $&$   1     $&   $2 $   \\
  $4 $& $  2  $   &  $ 2$    \\
  $5$ &$   2  $   &$   3   $  \\
 \end{tabular}
 \caption{Tate coefficients for $SU(n)$.}
 \label{v3v4}
 \end{center}
\end{table}

In \cite{Krause:2011xj} the resolution of such an $SU(n)$ singularity was worked out for  $n=5$. 
Here we generalise the procedure, which was inspired by the Tate algorithm \cite{Bershadsky:1996nh}, and explicitly apply it to $n=2,3,4,5$.
The resolution manifold $\hat Y_4$ follows by a sequence of $n-1$ blow-ups of the fibre over the singular divisor $\cW: w=0$.
This blow-up induces $n-1$ blow-up divisors $d_k$, where we let $k$ run from $2$ to $n$. Denoting by $d_0$ the proper transform of $w=0$, the scaling relations induced by the blow-up can be shown to take the values given in table \ref{scalings}. See \cite{Krause:2011xj} for details of how to determine these scalings.
  \begin{table}[htbp]
\begin{center}
 \begin{tabular}{c | c c c c}
  $k\,$   &   $\,x$   &   $y$   &   $d_0$   &   $d_k$  \\
  \hline
  $2$     &   $1$     &   $1$   &   $1$     &   $-1$   \\
  $3$     &   $1$     &   $2$   &   $1$     &   $-1$   \\
  $4$     &   $2$     &   $2$   &   $1$     &   $-1$   \\
  $5$     &   $2$     &   $3$   &   $1$     &   $-1$   \\
 \end{tabular} 
 \caption{Scaling relations for $SU(n)$ resolution divisors.}  \label{scalings}
\end{center} 
\end{table}
One observes that the charges of $x$ and $y$ under the induced scaling relations are precisely given by the vanishing orders $v_3^k$ and $v_4^k$ respectively. This property will be connected in the forth-coming section on $SU(n)\times U(1)$-models to the fact that for such models, the additional singularity resides on the curve $a_{3,v_3} = a_{4,v_4} = 0$. \\

The blow-up divisors each are $\mathbb P^1$-fibrations over the divisor $w=0$ in the base. Their intersection matrix reproduces the Cartan matrix of $SU(n)$.
The order of the blow-up process ($n = 2,3,4,5$), however, does not always co-incide with the intersection structure of these divisors. For example, in the $SU(5)$-case the divisors intersect in the order $d_0, d_2, d_4, d_5, d_3, d_0$ --- that is, $d_0$ intersects $d_2$ and $d_3$ but not $d_4$ or $d_5$, etc.. One can then define coordinates $e_i$ so as to represent the standard intersection structure 
\bea
\int_{\hat Y_4} E_i\,E_j\,{\cal B}_a\,{\cal B}_b = C_{ij} \int_B {\cW} {\cal B}_a\,{\cal B}_b,
\eea
where $C_{ij}$ takes the usual form of the $SU(n)$-Cartan matrix and $B_a$ are arbitrary divisors on the base. The identification between the $d_n$ and the $e_i$ for the various singularities is listed in table \ref{index_conversion} (in all cases one defines $e_0 := d_0$).

\begin{table}[htbp]
\begin{center}
 \begin{tabular}{c | c c c c}
         &  $d_2$  & $ d_3$  &  $d_4 $ &  $d_5 $ \\
  \hline
  $SU(2)$  &  $e_1$  &       &       &       \\
  $SU(3)$  &  $e_1$  & $ e_2$  &       &       \\
  $SU(4)$  &  $e_1$  & $ e_3$  &  $e_2$  &       \\
  $SU(5)$  &  $e_1$  &  $e_4$  & $ e_2$  & $ e_3 $ 
 \end{tabular} 
 \caption{Labeling of resolution divisors.}
 \label{index_conversion}
\end{center} 
\end{table}

Let us further denote by $N$ the index of the divisor of the last blow-up. In the $d_k$-basis, this is simply $N = n = 2,3,4,5$, while, in the $e_i$-basis, one finds from the above table that $N = 1,2,2,3$ for the various singularities considered here. We will observe later on that the divisor $e_N$ plays a special role in $SU(n) \times U(1)$-models.\\

\subsubsection*{$SU(n) \times U(1)$-Models}

As described already in \ref{sec:topological_invariants}, for the $U(1)$-restricted $SU(n) \times U(1)$ models with $a_6 \equiv 0$ an $SU(2)$ singularity appears in the fiber above the curve
\beq \label{C34}
 C_{34} = \{a_{3,v_3} = 0\} \cap \{a_{4,v_4} = 0\}.
\eeq
This requires
an additional resolution divisor $\{s=0\}$ with scaling 
$(x, y, s) \sim (\lambda x, \lambda y, \lambda^{-1} s)$.\\

\subsection{Intersection Structure}
\label{app:intersec_struc}

\subsubsection*{$SU(n)$-models}

With these basic properties at hand, let us consider the intersection structure on the Calabi-Yau four-fold. The resolution blow-ups can be expressed in terms of toric geometry; in particular, the subspace spanned by the coordinates $\{x, y, z, e_0\} \cup \{e_i\}$ is a toric variety. This allows us to analyze part of the intersection structure by triangulating the polygon spanned by these variables and analyzing the resulting Stanley-Reisner ideal. This ideal encodes the set of variables which are not allowed to mutually vanish. Put differently, it encodes part of the patch structure of the ambient manifold, encoding e.g.\ that there is no patch on which $x$, $y$, and $z$ may vanish. In order to deduce the intersection structure on the Calabi-Yau four-fold one has to furthermore take into account the proper transform of the Tate polynomial. \\

\begin{figure} 
 \centering
 \def\svgwidth{0.5 \textwidth}
 \executeiffilenewer{polytope.svg}{polytope.pdf}%
 {inkscape -z -D --file=polytope.svg %
  --export-pdf=polytope.pdf --export-latex}%
   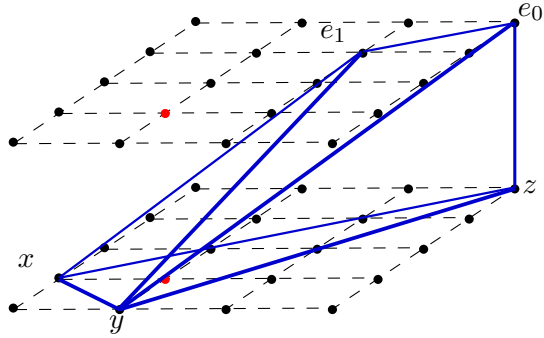%

 \caption{Toric Lattice Polytope corresponding to the sub-manifold of the $SU(2)$-resolution manifold spanned by the coordinate sub-set $\{x, y, z, e_0, e_1\}$ (the lower red dot denotes the origin of the lattice)}
\label{fig:polygon}
\end{figure}

As an example let us consider the $SU(2)$-case. There are two scaling relations,
\beq
(x,y,z,e_0,e_1) \sim (\lambda^2 x, \lambda^3 y, \lambda z,e_0,e_1) \sim (\sigma x, \sigma y, z, \sigma e_0, \sigma^{-1} e_1),
\eeq
and the polygon is subsequently given by
\beq
x = (-1,0,\underline{0}), \quad y = (0,-1,\underline{0}), \quad z = (2,3,\underline{0}), \quad e_0 = (2,3,\underline{v}), \quad e_1 = (1,2,\underline{v}).
\eeq
We can draw a 3-dimensional, schematic representation of the polygon, by mapping $\underline{0} \rightarrow 0$ and $\underline{v} \rightarrow 1$, see Figure~\ref{fig:polygon}.  It is clear that all sides of the polygon except for one are triangular, and the remaining side allows for two triangulations. For each of the two triangulations we can then analyse the resultant cone structure, which encodes the patch structure of the corresponding toric variety. It is a simple matter to deduce that the following combinations of variables are not allowed to vanish simultaneously in the ambient space
\beq
 \left\{x y z,\, x y e_0,\,y z e_1\right\} \otimes \left\{ \begin{aligned} &x e_0 \\ &z e_1 \end{aligned} \right\},
\eeq
where the choice between $x e_0$ and $z e_1$ is a choice between the two possible triangulations.
Furthermore, the proper transform of the Tate polynomial becomes
\beq
 y^2 + a_1 x y z + a_{3,1} y z^3 e_0 = x^3 e_1 + a_{2,1} x^2 z^2 e_0 e_1 + a_{4,1} x z^4 e_0 + a_6 z^6 e_0^2.
\eeq
From this it is clear that, on the Calabi-Yau four-fold, both $x=e_0=0$ and $z = e_1 = 0$ imply $y=0$. However, both $x y e_0$ and $y z e_1$ are not allowed to vanish simultaneously. Then on the Calabi-Yau four-fold both $x e_0$ and $z e_1$ become elements of the generator set of the SR-ideal, and, since these two generate the terms $x y e_0$ and $y z e_1$, we can remove the latter two from the generator set. We thus arrive at the following set of SR-ideal generators, which is a subset of the total generator set: $\{xyz, z e_1, x e_0\}.$\\

\begin{table}[htbp]
 \begin{center}
  \begin{tabular}{c | l | c}
            &  non-intersection divisors                                                         &  $N_{double}$  \\
   \hline
   $SU(2)$  &  $x y z$, $z e_{i | i \geq 1}$, $x e_0$                                                                  &  1             \\
   $SU(3)$  &  $x y z$, $z e_{i | i \geq 1}$, $x e_0$, $y e_0$, $y e_1$                                                &  3             \\
   $SU(4)$  &  $x y z$, $z e_{i | i \geq 1}$, $x e_0$, $y e_0$, $x e_1$, $y e_1$, $x e_3$, $e_0 e_2$                   &  6             \\
   $SU(5)$  &  $x y z$, $z e_{i | i \geq 1}$, $x e_0$, $y e_0$, $x e_1$, $y e_1$, $y e_2$, $x e_4$, $e_0 e_2$, $T$     &  9             \\
            &  $T \in \left\{(e_0 e_3, e_1 e_3),\, (e_0 e_3, e_2 e_4),\, (e_1 e_4, e_2 e_4)\right\}$ &
  \end{tabular}
 \caption{Part of the Stanley-Reisner ideal on the resolution four-folds.}
 \label{sr_ideal_sun}
 \end{center}
\end{table}
In table \ref{sr_ideal_sun} we list the generator set of combinations of variables whose zero loci do not intersect on the four-fold. Note that for $n=5$ the polygon described above allows for 18 different triangulations, which reduce to three different Stanley-Reisner ideals on the four-fold.
In the last column we count the number of elements which involve double intersections of two exceptional divisors, $E_i \, E_j$. For example, the divisor $\{x=0\}$ lies in the divisor class $2\,(Z + \aK) - v_3^i\,E_i$, while the divisor $\{e_0 = 0\}$ lies in the class $\cW - \sum_i E_i$. The vanishing of their intersection can thus be used to deduce \beq \sum_j v_3^i \, E_i \, E_j = 2\,\aK \sum_i E_i + \cW\,v_3^i\,E_i - 2\,(Z + \aK)\,\cW. \eeq We are interested in the number of linearly independant expressions for these double intersections in order to determine how many elements we can re-express in terms of base divisor intersections. By the above, $x e_0$ is one such relation, whilst, e.g.~$z e_1 \rightarrow Z\,E_1 = 0$ does not involve any such double intersection.\\

In a next step, we would like to re-express two-forms of the form $E_i \, E_j$ as linear combinations of $E_i \aK$, $E_i \, \cW$, $(Z+\aK) \, \cW$ and possibly some remaining $E_i \, E_j$-terms. The number of $E_i \, E_j$-terms for the various $SU(n)$-resolutions are $\binom{n}{2}$, i.e.~$1$, $3$, $6$, and $10$ for $n = 2,3,4,5$ respectively. From table \ref{sr_ideal_sun} we thus see that for $n<5$ these types of terms can be re-expressed entirely without any remaining $E_i \, E_j$-terms, while one is left with one such term in the $SU(5)$-case. Which combination of $i$ and $j$ one uses for this remaining term is a matter of choice. For later use, we choose a term that is orthogonal to the intersection of two base divisors because its corresponding entry in the Cartan matrix is zero. Specifically we choose $(i,j) = (2,4)$.

The result of this analysis can be compactly written as
\beq \label{eiej_relations_app}
 E_i\,E_j = C_{ij}\,(Z+\aK)\,\cW + w_m\,E_m\,\cW + k_m\,E_m\,\aK + b\,E_2\,E_4,
\eeq
where $b=0$ for $n < 5$. In tables \ref{doubleSU2} -- \ref{doubleSU5} we list the coefficients of the expression for the various $SU(n)$-cases. Note that in all cases the coefficients of the $(Z+\aK)\,\cW$-term are the Cartan matrix entries $C_{ij}$. This becomes important in finding candidates for gauge fluxes. \\

\begin{table}
\begin{center}
 \begin{tabular}{c || c | c | c}
  $(i,j)$  &  $C_{ij}$   &  $w_m$  & $k_m$    \\
  \hline
  $(1,1)$  &  $-2$  &  $1$    &  $2$
 \end{tabular}
 \caption{Double-intersection coefficients for $SU(2)$-model.}
 \label{doubleSU2}
\end{center}
\end{table}

\begin{table}
\begin{center}
 \begin{tabular}{c || c | c | c }
  $(i,j)$  &  $C_{ij}$     &  {$w_m$}   &  $k_m$     \\
  \hline
  $(1,1)$  &  $-2$    &  $(2,0)$   &  $(-1,2)$  \\
  $(1,2)$  &  $ 1$    &  $(-1,0)$  &  $(2,-1)$  \\
  $(2,2)$  &  $-2$    &  $(1,1)$   &  $(-1,2)$                  
 \end{tabular}
\caption{Double-intersection coefficients for $SU(3)$-model.}
\label{doubleSU3}
\end{center}
\end{table}

\begin{table}
\begin{center}
 \begin{tabular}{c || c | c c c | c c c}
  $(i,j)$  &  $C_{ij}$     &  $w_m$       &  $k_m$    \\
  \hline
  $(1,1)$  &  $-2$    &  $(2,0,0)$   &  $(-1,2,2)$  \\
  $(1,2)$  &  $ 1$    &  $(-1,0,0)$  &  $(1,-1,-1)$  \\
  $(1,3)$  &  $ 0$    &  $(0,0,0)$   &  $(1,0,0)$  \\
  $(2,2)$  &  $-2$    &  $(1,1,1)$   &  $(0,2,0)$  \\
  $(2,3)$  &  $ 1$    &  $(0,0,-1)$  &  $(-1,-1,1)$  \\
  $(3,3)$  &  $-2$    &  $(0,0,2)$   &  $(1,2,0)$  
 \end{tabular}
 \caption{Double-intersection coefficients for $SU(4)$-model.}
\label{doubleSU4}
\end{center}
\end{table}

\begin{table}
\begin{center}
 \begin{tabular}{c || c | c c c c | c c c c | c}
  $(i,j)$  &  $C_{ij}$   &  $w_m$         &  $k_m$          &  $b$  \\
  \hline
  $(1,1)$  &  $-2$  &  $(2,0,0,0)$   &  $(-1,2,2,2)$   &  $ 0$  \\
  $(1,2)$  &  $ 1$  &  $(-1,0,0,0)$  &  $(1,-1,-1,-1)$ &  $ 0$  \\
  $(1,3)$  &  $ 0$  &  $(0,0,0,0)$   &  $(0,0,0,0)$    &  $ 0$  \\
  $(1,4)$  &  $ 0$  &  $(0,0,0,0)$   &  $(1,0,0,0)$    &  $ 0$  \\
  $(2,2)$  &  $-2$  &  $(2,3,0,0)$   &  $(-2,-1,2,2)$  &  $-1$  \\
  $(2,3)$  &  $ 1$  &  $(-1,-2,0,0)$ &  $(1,2,-1,-1)$  &  $ 0$  \\
  $(2,4)$  &  $ 0$  &  $(0,0,0,0)$   &  $(0,0,0,0)$    &  $ 1$  \\
  $(3,3)$  &  $-2$  &  $(1,2,1,1)$   &  $(0,-1,2,0)$   &  $ 1$  \\
  $(3,4)$  &  $ 1$  &  $(0,0,0,-1)$  &  $(-1,-1,-1,1)$ &  $-1$  \\
  $(4,4)$  &  $-2$  &  $(0,0,0,2)$   &  $(1,2,2,0)$    &  $ 0$  
 \end{tabular}
 \caption{Double-intersection coefficients for $SU(5)$-model (for triangulation $T_1$).}
\label{doubleSU5}
\end{center}
\end{table}

\subsubsection*{$SU(n) \times U(1)$-models}
The resolution of the additional singularity along $C_{34}$, (\ref{C34}), also results in new generators of the Stanley-Reisner ideal. In addition to those elements listed in table \ref{sr_ideal_sun} one finds
\beq
 \{x y,\, z s,\, s e_0\} \cup \{s e_k \,|\, i \neq N \}.
\eeq
This implies that the relation (\ref{eiej_relations_app}) and in particular the coefficients listed in tables \ref{doubleSU2} -- \ref{doubleSU5}  carry over to the $U(1)$-restricted case. Furthermore, one finds that the divisor $\{s=0\}$ only intersects the Cartan divisor of the last blow-up,
\beq
S\,E_N = S\,\cW \qquad S\,E_i = 0 \qquad \textmd{for}  \quad i\neq N.
\eeq
Finally, the SR-ideal element $x y$ implies the relation
\beq \label{sr_xy}
 6\,(Z + \aK)^2 - 5\,\aK\,S + S^2 - (3\,v_3^i + 2\,v_4^i)\,E_i \aK + n\,S\,\cW + v_3^i\,v_4^j\,E_i\,E_j = 0.
\eeq
\

\subsection{Topological Invariants}
\label{app:topological_invariants}
In order to analyse the D3-tadpole as well as the flux quantisation condition, we are interested in the second and fourth Chern class of the resolution manifold.
It is a simple enough matter to deduce the former from that of the original manifold by adjunction. The relevant changes in the divisor classes during the resolution process can be read off from the induced scaling relations $(x, y, e_0, e_1) \sim (\lambda^{v_3} x, \lambda^{v_4} y, \lambda e_0, \lambda^{-1} e_1)$. It is given in terms of the powers of $\lambda$. Similarly the change in the class of the Tate polynomial simply depends on the sum of these terms.
Then the Chern class of the resolution manifold at each blow-up step is given by:
\begin{equation}
 \begin{aligned}
  \left[c \left(Y_4^{\rm aft} \right)\right] = \left[c\left(Y_4^{\rm bef} \right)\right] & \cdot \left(1 + E_i \right) \cdot \frac{1 + X^{\rm bef} - v_3^{\,i}\,E_i}{1 + X^{\rm bef}} \cdot \frac{1 + Y^{\rm bef} - v_4^{\,i}\,E_i}{1 + Y^{\rm bef}} \\
  &\cdot \frac{1 + E_{0}^{\rm bef} - E_i}{1 + E_0^{\rm bef}} \cdot \frac{1 + T^{\rm bef}}{1 + T^{\rm bef} - \left(v_3^{\,i} + v_4^{\,i}\right)\,E_i},
 \end{aligned}
\end{equation}
where we have used $T$ to denote the divisor class of the Tate divisor $\{P_T = 0\}$, and left the pullback map implicit for clarity of the expression. From this it is easy enough to derive the change in the second Chern class. We can then use the relations derived above to eliminate as many $E_i E_j$-terms as possible from these expressions. The resulting expressions for the change in $c_2$ are summarised in table \ref{delta_c2_sun}.\\

 \begin{table}[htb]
 \begin{center}
  \begin{tabular}{c | r r r}
  \quad \text{Gauge Group} \qquad  & \quad $\Delta c_2$  & &\\
  \hline
  $Sp(1)$/$SU(2)$ &  \quad $\cW\,E_1  $              &$ - 7\,\aK \, E_1 $ &\\
  $SU(3)$      &  \quad $(0,2)_i \, \cW\,E_i     $&$- (4,10)_i \, \aK \, E_i  $&\\
  $SU(4)$      &$  \quad (1,4,1)_i \, \cW\,E_i   $&$- (6,14,8)_i \, \aK \, E_i  $&\\
  $SU(5)$      &$  \quad (0,2,6,2)_i \, \cW\,E_i $&$- (4,11,17,10)_i \, \aK \, E_i $ &$+ E_2\,E_4$
 \end{tabular}
 \end{center}
 \caption{Change of second Chern class in  $SU(n)$-models compared to smooth Weierstrass models.}
 \label{delta_c2_sun}
\end{table}

\begin{table}[htb]
 \begin{center}
  \begin{tabular}{c | r r r r}
  \, $SU(n)$ \,  & & $\Delta c_2$ & &\\
  \hline
 $ SU(2)\tsp$  &$  \, U_{2} $&$ + 2\,\cW\,E_1              $&$- 10\,\aK \, E_1               $&               \\
$  SU(3)  $    &$  \, U_{3} $&$+ (1,4)_i \, \cW\,E_i      $&$- (6,14)_i \, \aK \, E_i       $&               \\
$  SU(4)$      &$  \, U_{4} $&$+ (3,8,3)_i \, \cW\,E_i    $&$- (9,20,9)_i \, \aK \, E_i     $&               \\
 $ SU(5)$      &$  \, U_{5} $&$+ (2,6,12,3)_i \, \cW\,E_i $&$- (6,16,24,14)_i \, \aK \, E_i $&$ + 2\,E_2\,E_4$
 \end{tabular}
 \end{center}
 \caption{Change of second Chern class in  $SU(n)\times U(1)$-models compared to smooth Weierstrass models. Here $U_{n} = (6\,\aK - n\,\cW)\,(Z + \aK - S)$.}
 \label{delta_c2_sunu1}
\end{table}
 
The situation for  $SU(n) \times U(1)$-models is similar. Since the resolution of the $SU(n)$ singularity is essentially unaffected by the restriction $a_6=0$, all we need to do is to evaluate the
change from the $SU(n)$-resolution to the $SU(n)\times U(1)$-resolution manifolds. This is given by
\beq
 \left[c \left(Y_4^{su_n\times u_1} \right)\right] = \left[c \left(Y_4^{su_n} \right)\right] \cdot \left(1 + S \right) \cdot \frac{1 + X^{su_n} - S}{1 + X^{su_n}} \cdot \frac{1 + Y^{su_n} - S}{1 + Y^{su_n}} \cdot \frac{1 + T^{su_n}}{1 + T^{su_n} - S},
\eeq

Adding this to the change from the original, non-singular manifold to the $SU(n)$-manifold and eliminating as many $E_i\,E_j$-terms as possible again, we find the  change in the second Chern class from the non-singular manifold to the $SU(n)\times U(1)$-resolution manifolds as summarized in table \ref{delta_c2_sunu1}.\\

As is described in section \ref{sec:topological_invariants}, the change in the Euler characteristic can be computed from the change in the second Chern class by making use of the relation
\beq
 \Delta c_4 = 3 \left( \Delta c_2 \right)^2.
\eeq
This holds, provided $\Delta c_2$ is orthogonal to the second Chern class of the original, singular manifold. From tables \ref{delta_c2_sun} and \ref{delta_c2_sunu1} and general considerations of the intersections structure, it is clear that this is the case. In table \ref{chi_FT}, we list the resulting expressions for the change of the Euler number of $SU(n)$-resolution manifolds compared to the original, non-resolved manifold.

\begin{table}[htb]
 \begin{center}
  \begin{tabular}{c | c}
 \quad \text{Gauge Group} \qquad  & \quad $\Delta \chi$ \\
  \hline
 $ Sp(1)$ / $SU(2)  $&  $\quad -6\,\int_{\cal W}\,(49\,\aK^2 - 14\,\aK\,\cW + \cW^2) $ \\
 $SU(3)                     $&$ \quad -24\,\int_{\cal W}\,(19\,\aK^2 - 8\,\aK\,\cW + \cW^2) $ \\
 $SU(4)                     $&$  \quad -12\,\int_{\cal W}\,(50\,\aK^2 - 28\,\aK\,\cW + 5\,\cW^2)$  \\
 $ SU(5)                     $ &$  \quad -15\,\int_{\cal W}\,(50\,\aK^2 - 35\,\aK\,\cW + 8\,\cW^2)$ 
 \end{tabular}
 \end{center}
 \caption{Change of Euler characteristic due to resolution of  $SU(n)$ singularities.}
 \label{chi_FT}
\end{table}

The change in the Euler characteristic from the $SU(n)$-resolution manifold to the $SU(n)\times U(1)$-resolution manifold, on the other hand, becomes
\beq
  \Delta^{su_n \times u_1}_{su_n} \chi(Y_4) = -3\,\int_{B_3} \, (3\,\aK - v_3^n\,\cW)(4\,\aK - v_4^n\,\cW)(6\,\aK - n\,\cW).
\eeq
At this point the importance of the coefficients $v_3$, $v_4$ in the induced scaling relations of the $SU(n)$-resolution becomes clear, as they can now be used to reexpress the above as
\beq \label{chi_change}
 \Delta^{su_n \times u_1}_{su_n} \chi(Y_4) = 3\,\chi(C_{34}).
\eeq
\ \\

\subsection{Matching D3-brane Tadpoles in F-theory and Type IIB}
\label{app_tad}
In this appendix, we list the expressions for the geometric D3-tadpole of Type IIB $U(n)$-configurations and compare them to the related F-theory expressions.

\subsubsection*{$SU(n)$}
From the analysis in section \ref{sec:TopInv}, one finds the D3-tadpole to be of the form
\bea \label{eq:d3_tadpole_non_restricted}
  24\,Q_{D3}^g     = \frac{1}{2} \Big(  \sum_{i=0}^3 a_i  D_{O7}^{\,3-i} W_+^{\,i} - \tfrac34 nW_+\left(W_+^{\,2} - W_-^{\,2}\right)  + 12 c_2(B_3) D_{O7} \Big)
\eea
with
\bea \label{ai}
 a_i = \left(360,\, -147n,\, 21n^2,\, -(n^3-n)\right)_i.
\eea

This expression  $24\,Q_{D3}^g$ is to be compared with $\chi(\hat Y_4)$ of the resolved F-theory four-fold.
The Euler characteristic  of the resolved F-theory geometries for the various $SU(n)$-models, as listed in table \ref{chi_FT}, can be written as
\bea \label{chiY4bi}
 \chi(\hat Y_4) = \sum_{i=0}^3   b_i   \, \aK^{\,3-i} \, \cW^{\,i} + 12 c_2(B_3) \aK.
\eea
The coefficients $b_i$ for the various cases of $SU(n), n= 2,3,4,5$ are collected in table \ref{vi}.
Indeed, (\ref{eq:d3_tadpole_non_restricted}) and  (\ref{chiY4bi}) agree thanks to  (\ref{eq:no-E6-point-relation}).
 
  \begin{table}[htb]
 \begin{center}
  \begin{tabular}{c | c c c c }
                          & \quad $b_0$ \quad & \quad $b_1$ \quad & \quad $b_2$ \quad & \quad $b_3$    \\
  \hline
  $ \quad Sp(1) / SU(2)  $&$  360 $&$ \ -147n $&$  \, 21n^2      \qquad   $&$     -(n^3-n)$  \\
  $ \quad SU(3) $ &$  360 $& $\ -152n $& $ \, 21n^2 + n  $&$     -(n^3-n)$  \\
  $ \quad SU(4)  $&$  360 $& $\ -150n $& $ \, 21n^2       \qquad   $&$     -(n^3-n)$  \\         
  $ \quad SU(5)  $&$  360 $&$ \ -150n $&$  \, 21n^2      \qquad    $&$     -(n^3-n) $ 
 \end{tabular}
 \end{center}
 \caption{Coefficients $b_i$ appearing (\ref{chiY4bi}).}
 \label{vi}
\end{table}

\subsubsection*{$SU(n) \times U(1)$}
The details of the Type IIB  $SU(n) \times U(1)$ differ for even and odd values of $n$ as described in section \ref{sec:sen_limit}.

\emph{Even $n$}\\[4pt]
The geometric tadpole contribution for the Sen limit of $SU(n) \times U(1)$ with even $n$ is 
\bea
 24\,Q_{D3}^g & =& n\,\left( \chi(W) + \chi(\tilde{W})\right) + 2\chi_0(4D_{O7} - \tfrac{n}{2}W_+) + 4\,\chi(D_{O7}) \\
                          &= & a_i D_{O7}^{\,3-i} W_+^{\,i} - \tfrac34 nW_+\left(W_+^{\,2} - W_-^{\,2}\right) + 12 c_2(B_3) D_{O7}, 
 \eea
where  \begin{center}$a_i = \left(144,\, -48n,\, 6n^2,\, -(\tfrac14 n^3-n)\right)_i$.\end{center}
This can be compared with the contribution for non-restricted $SU(n)$ models, for which we have already derived a corresponding formula on the F-theory side (see (\ref{chi_change})).
The difference to the tadpole contribution for non-restricted models, (\ref{eq:d3_tadpole_non_restricted}), is
\beq
 \Delta \,Q_{D3}^g = \left(216,\, -99n,\, 15n^2,\, -\tfrac34 n^3\right)_i D_{O7}^{\,3-i} W_+^{\,i},
\eeq
while the difference in the F-theory case is given by
\beq
 \Delta \,Q_{D3}^g = 3 \chi(C_{34}) = \left(216,\, -99n,\, 15n^2,\, -\tfrac34 n^3\right)_i \aK^{\,3-i} \cW^{\,i}.
\eeq
This matches perfectly upon using the usual identifications $\pi^*(\aK) =  D_{O7}$ and $\pi^*(\cW) = W_+$. Since the IIB and the F-theory description of the $SU(n)$ models match, this also shows agreement for their $SU(n) \times U(1)$ cousins.\\

\emph{Odd $n$}\\[4pt]
For  odd  values of $n$ the geometric tadpole is
\bea
 24\,Q_{D3}^g &=& n\,\left( \chi(W) + \chi(\tilde{W})\right) + 2\chi_0(4D_{O7} - n/2\,W_+) + 4\,\chi(D_{O7}) \\
                          &=& a_i D_{O7}^{\,3-i} W_+^{\,i} + 12 c_2(B_3) D_{O7} + 6D_{O7} W_-^{\,2}
 \eea
 with  \begin{center}$a_i = \left(144,\, -48n,\, 6n^2,\, -\tfrac14 (n^3-n)\right)_i$.\end{center}

We first note that the last term vanishes because the restriction of an involution-odd class to the orientifold plane is zero. Then the remainder can be compared to the contribution for non-restricted models.
The difference to the tadpole contribution for non-restricted models, (\ref{eq:d3_tadpole_non_restricted}), 
\beq
 \Delta \,Q_{D3}^g = \left(216,\, -99n,\, 15n^2,\, -\tfrac34 (n^3-n)\right)_i D_{O7}^{\,3-i} W_+^{\,i} - \tfrac34 nW_+\left(W_+^{\,2} - W_-^{\,2}\right),
\eeq
agrees with the difference in the F-theory case,
\beq
 \Delta \,Q_{D3}^g = 3 \chi(C_{34}) = \left(216,\, -99n + 9,\, 15n^2 - \tfrac32 n -\tfrac92,\, -\tfrac34 (n^3-n)\right)_i \aK^{\,3-i} \cW^{\,i}.
\eeq
upon the usual identifications. 

\newpage
\section{Gauge Fluxes in F-theory and Type IIB}
\subsection{Universal F-theory Fluxes}
Following the analysis in section \ref{subsec_G41}, we list a basis of flux candidates from double intersections of exceptional divisors in table \ref{tab:eiej_basis}.
\begin{table}[htb]
 \begin{center}
  \begin{tabular}{c | l | c}
   Gauge Group  &  Basis of Lin. Comb. of $E_i\,E_j$-terms  $\perp B_a\,B_b$  & count \\
   \hline
   $SU(2)$\tsp  &  none  & $0$ \\ [2pt]
   $SU(3)$      &  $E_{11} - E_{22}$,\, $E_{11} + 2\,E_{12}$ & $2$\\[2pt]
   $SU(4)$      &  $E_{13}$,\, $E_{11} - E_{22}$,\, $E_{11} - E_{33}$,\, $E_{12} - E_{23}$,\, $E_{11} + 2\,E_{12}$ & $5$\\[2pt]
   \multirow{2}{*}{$SU(5)$}  &  $E_{13}$,\, $E_{14}$,\, $E_{24}$,\, $E_{11} - E_{22}$,\, $E_{11} - E_{33}$,\, $E_{11} - E_{44}$,  & \multirow{2}{*}{$9$}\\
                           &  $E_{12} - E_{23}$,\, $E_{12} - E_{34}$,\, $E_{11} + 2\,E_{12}$ &
  \end{tabular}
 \end{center}
 \caption{Basis of linear combinations of double intersections of exceptional divisors which are orthogonal to the intersection of two base divisors.}
 \label{tab:eiej_basis}
\end{table}
These can be re-expressed in terms of Cartan fluxes for all cases except $n=5$, in which case one is left with one term of the form $E_{ij}$. This can be used to define a linear combination of fluxes which leaves the $SU(n)$-gauge group intact,
\beq
G_4^{\,\lambda} = E_2\,E_4 + (\frac{2}{5},-\frac{1}{5},\frac{1}{5},-\frac{2}{5})_i \, E_i \, \aK .
\eeq
\ \\

In the $U(1)$-restricted models, the additional divisor introduces a new class of potential fluxes which may be used in combination with Cartan fluxes to form a class of fluxes leaving the $SU(n)$-gauge group intact. These combinations are listed for $n = 2,..,5$ in table \ref{tab:Cartan_orthogonal}.

\begin{table}[htb]
 \begin{center}
  \begin{tabular}{c | l c l }
   gauge group & \multicolumn{3}{c}{linear combination $\perp$ to all Cartan fluxes}  \\
   \hline
   non-sing. &   $(1)$ & $\cdot$ & $ (S - Z - \aK)$   \\
   $SU(2)$   &   $(1;\, \frac{1}{2})$ & $\cdot$ & $ (S - Z - \aK;\, E_1)$   \\
   $SU(3)$   &   $(1;\, \frac{1}{3},\, \frac{2}{3})$ & $\cdot$ & $ (S - Z - \aK;\, E_1,\, E_2)$   \\
   $SU(4)$   &   $(1;\, \frac{1}{2},\, 1,\, \frac{1}{2})$ & $\cdot$ & $ (S - Z - \aK;\, E_1,\, E_2,\, E_3)$   \\
   $SU(5)$   &   $(1;\, \frac{2}{5},\, \frac{4}{5},\, \frac{6}{5},\, \frac{3}{5})$ & $\cdot$ & $ (S - Z - \aK;\, E_1,\, E_2,\, E_3,\, E_4)$
  \end{tabular}
 \caption{Linear combination of vertical two-forms orthogonal to all Cartan fluxes.}
 \label{tab:Cartan_orthogonal}
 \end{center}
\end{table}

\subsection{Whitney-branes in Type IIB and their Gauge Fluxes}
\label{app:Whitney_brane_charges}
In this section we derive the general expression for the D3-tadpole of a Whitney-brane. The analysis closely follows \cite{Collinucci:2010gz,Braun:2011zm}, but additionally takes into account involution-odd contributions. In this framework, a $D7$-brane arises from tachyon condensation of a $D9$/anti-$D9$-pair. Viewing the Whitney brane as the result of a recombination process in which a brane combines with its image brane, it can then be derived from two $D9$/$\bar{D9}$-pairs, which form the brane and its image. It turns out that the $D7$-charge of such a set-up only depends on the involution-invariant $D7$-charge of the brane/image-brane pair, but the $D5$- and $D3$-charge also depend on the anti-invariant part. In other words, the Whitney brane resulting from a recombination process remembers the anti-invariant contributions of the brane/image-brane pair. We will demonstrate below that this logic is indeed necessary to arrive at the match between the F-theory- and Type IIB-fluxes as proposed in the main body of this paper.\\

Let us start with a single, `ordinary' brane resulting from tachyon condensation of the following $D9/\bar{D9}$-set-up:
\begin{equation}
\label{eq:D9-brane_set-up}
 \begin{array}{c c c}
 D9_1 & \qquad & \overline{D9}_2 \\
 \mathcal{O}(D + E) & \qquad & \mathcal{O}(E)\\
 \end{array}
\end{equation}
The charges of the resulting $D7$-brane can be computed from
\beq
 \Gamma_{D9} = \left(e^{D + E} - e^{E}\right) \cdot \left( 1 + \frac{c_2}{24} \right)
\eeq
and, defining $F := E + \tfrac12 D$, are readily evaluated to be
\beq
 \left( Q_{D9}, Q_{D7}, Q_{D5}, Q_{D3},  \right) =  \left( 0,\, D,\, DF,\, \tfrac{1}{24} (D^3 + D c_2) + \tfrac12 D F^2 \right)
\eeq
Similarly, one can construct an image-brane by appropriately inverting the expression (\ref{eq:D9-brane_set-up}), resulting  in the following charge vector:
\beq
\left( 0,\, \tD,\, -\tD \tF,\, \tfrac{1}{24} (\tD^3 + \tD c_2) + \tfrac12 \tD \tF^2 \right)
\eeq
where $\tD = \sigma^{\ast} D$ denotes the image of $D$ under the orientifold involution.

Upon brane recombination, these two 7-branes will combine to form a singular brane, whose singularity resembles that of a Whitney umbrella. It is thus usually called a Whitney brane. Let us make the following notational conventions:
\beq
\bal
 D_{\pm} &= D \pm \sigma^{\ast} D, \qquad \left[ D_1 \cdots D_n \right]_{\pm} &= \left(D_1 \cdots D_n\right) \pm \sigma^{\ast}\left(D_1 \cdots D_n\right)
\eal
\eeq
Then the charge vector of the Whitney brane, which is simply the sum of the charge vectors of the brane and its image, becomes
\beq \label{chargevector}
\left( 0,\, D_+,\, [DF]_-,\, \tfrac{1}{24} \left([D^3]_+ + [D]_+ c_2\right) + \tfrac12 [D F^2]_+ \right).
\eeq
For a general $U(n)$-configuration, the Whitney brane will have $D7$-charge $8 D_{O7} - n (W + \tilde{W})$. From the configuration we started with, (\ref{eq:D9-brane_set-up}), we know that $D$ must be integer quantised. Then, if $n$ is odd, it is not possible for $D_-$ to be trivial unless $W_-$ is trivial or $W$ is spin.\\

In the case of a Whitney-brane without any fluxes, the only $D3$-tadpole contribution should be of geometric type and depend on the Euler characteristic of the brane. Since the brane is singular, the usual formula for the Euler characteristic requires an adjustment, which was derived in \cite{Collinucci:2008pf} to be
\beq
 \chi_0(D_+) = \int_{X_3} D_+^{\,3} + c_2 \cdot D_+ + 3\, D_+ D_{O7} (D_{O7} - D_+),
\eeq 
Then we can split the $D3$-charge into the geometric and the flux contributions as
\beq
\label{eq:D3-splitting}
 Q_{D3}^g = \tfrac{1}{24} \chi_0(D_+) + Q_{D3}^{rem},
\eeq
where the remainder term should vanish for a Whitney-brane without fluxes.
This is achieved by the following two combinations
\beq
 F^0 + \tilde F^0 = \pm \left( D_{O7} - \tfrac12 D_+\right),   \qquad    F^0 - \tilde F^0 = \pm \tfrac12 D_-,
\eeq
where the overall sign must be the same in both expressions. Note that this automatically meets the $D5$-tadpole condition.\\

In order to calculate the flux contribution to the $D3$-charge of a Whitney brane, one should then use (\ref{eq:D3-splitting}) and subtract the contribution of the $F^0$-flux. This leads to the following expression for the flux contribution to the D3-tadpole
\beq
\label{eq:D3-flux}
 Q_{D3}^f = \tfrac12 [DF^2]_+ - \tfrac18 \left\{ D_+ \left( D_+ - 2 D_{O7} \right) - \left( D_- \right)^2 \right\} (F + \tilde F).
\eeq
A similar formula was derived in \cite{Braun:2011zm}; here we add the contribution of involution-odd fluxes.
Note that there are again two flux configurations, for which this term becomes zero:
\beq
 (F + \tilde F,\, F - \tilde F) = (0,\, 0),   \qquad   (F + \tilde F,\, F - \tilde F) = \left(D_+ - 2 D_{O7},\, -D_-\right).
\eeq
In other words, the flux contribution to the D3-tadpole vanishes precisely when the positive flux part lies in the trivial class or in that of $\left[a_{6,5}\right]$. This agrees with \cite{Collinucci:2010gz, Braun:2011zm}, where it is noted that fluxes are only allowed on non-generic Whitney branes, in particular those
 for which $a_{6,5}$ factorises non-trivially.

\subsubsection*{Specialisation to $U(5)$-Models}
Let us apply the above analysis to the $SU(5)$-set-up described in section \ref{sec:generic_flux_config}. Here the Whitney brane is such that $D_+ = 8 D_{O7} - 5 W_+$ and $D_- = W_-$, where $W, \tilde{W}$ denote the divisor class of the GUT stack divisor and its image, respectively. 
We consider a flux configuration $\frac{1}{10} {P}$ on $W$, where $ {P} \in H^2_+(X_3)$. D5-tadpole cancellation then requires that the induced D5-charge of the Whitney brane corresponds to setting $F =  - \frac{P}{2}$ in the charge vector (\ref{chargevector}).
Summing up the contribution of the $SU(5)$ and the Whitney brane to the D3-brane tadpole eventually yields 
\bea
 - \frac{1}{2} \int_{X_3}  P^2  ( D_{O7} - \frac{3}{5} W_+) +  \frac{P}{2} \left( 4 D_{O7} - 3 W_+ \right) \left(3 D_{O7} - 2 W_+ \right).
\eea
To arrive at this result we have used the relation $2 D_{O7} W_+ = W_+^{\,2} - W_-^{\,2}$.

\newpage
\section{Intersections of Exceptional Divisors on \texorpdfstring{$SU(5)$}{SU(5)}-Resolution Manifold} \label{app:IntED}
In the following we list the triple and quadruple intersections of the exceptional divisors of the $SU(5)$/$SU(5)\times U(1)$-resolution manifolds for the standard triangulation used in this and preceeding paper, \cite{Krause:2011xj}. These can be derived from the Stanley-Reisner ideal elements in table \ref{sr_ideal_sun} and are useful in the calculation of various flux-induced properties such as the $D3$-tadpole for fluxes involving the wedge product of two dual forms of exceptional divisors.\\

We express the triple intersections as $E_{ijk} = (a, b)$, which we use as short-hand notation for 
\beq
\int_{\tY}\,E_i\,E_j\,E_k\,D = \int_{\cal W}\,(a \aK + b \cW)\,D.
\eeq
The non-vanishing combinations are:
\begin{equation}
\label{eq:triple_Eis}
 \begin{array}{lcrrr c lcrrr c lcrrr}
  E_{111}  &=&  (& 4,&-4)&  \qquad  & E_{112}  &=& (&-3,& 2)& \qquad  & E_{114}  &=& (&-2,& 0),   \\
  E_{122}  &=&  (& 2,&-1)&  \qquad  & E_{124}  &=& (& 1,& 0)& \qquad  & E_{222}  &=& (& 3,&-4),   \\
  E_{223}  &=&  (&-4,& 3)&  \qquad  & E_{224}  &=& (&-1,& 0)& \qquad  & E_{233}  &=& (& 3,&-2),   \\
  E_{234}  &=&  (& 1,& 0)&  \qquad  & E_{244}  &=& (&-1,& 0)& \qquad  & E_{333}  &=& (&-4,& 1),   \\
  E_{334}  &=&  (& 1,&-1)&  \qquad  & E_{344}  &=& (&-2,& 2)& \qquad  & E_{444}  &=& (& 2,&-4).
 \end{array}
\end{equation}

Similarly, we express the quadruple intersections as $E_{ijkl} = (a, b, c)$, which we use as short-hand notation for
\beq
\int_{\tY} E_i E_j E_k E_l = \int_{\cal W}\,a\,\aK^2 + b\,\aK \cW + c\,\cW^2.
\eeq
The non-vanishing combinations are:
\begin{equation}
\label{eq:quadruple_Eis}
 \begin{array}{lcrrrr c lcrrrr c lcrrrr}
  E_{1111}  &=&  (&-14,&  20,&  -8), &  \quad  & E_{1112}  &=& (& 9,& -12,&  4),  &  \quad  & E_{1114}  &=& (& 4,& -4,&  0),     \\
  E_{1122}  &=&  (& -6,&   7,&  -2), &  \quad  & E_{1124}  &=& (&-3,&   2,&  0),  &  \quad  & E_{1144}  &=& (&-2,&  0,&  0),     \\
  E_{1222}  &=&  (&  4,&  -4,&   1), &  \quad  & E_{1224}  &=& (& 2,&  -1,&  0),  &  \quad  & E_{1244}  &=& (& 1,&  0,&  0),     \\
  E_{2222}  &=&  (&-16,&  29,& -14), &  \quad  & E_{2223}  &=& (&13,& -22,&  9),  &  \quad  & E_{2224}  &=& (&-1,&  0,&  0),     \\
  E_{2233}  &=&  (&-12,&  17,&  -6), &  \quad  & E_{2234}  &=& (&-1,&   1,&  0),  &  \quad  & E_{2244}  &=& (& 0,& -1,&  0),     \\
  E_{2333}  &=&  (&  9,& -12,&   4), &  \quad  & E_{2334}  &=& (& 3,&  -2,&  0),  &  \quad  & E_{2344}  &=& (&-2,&  2,&  0),     \\
  E_{2444}  &=&  (&  1,&  -2,&   0), &  \quad  & E_{3333}  &=& (&-8,&   9,& -4),  &  \quad  & E_{3334}  &=& (&-1,& -1,&  1),     \\
  E_{3344}  &=&  (& -2,&   4,&  -2), &  \quad  & E_{3444}  &=& (& 4,&  -8,&  4),  &  \quad  & E_{4444}  &=& (&-6,& 12,& -8).
 \end{array}
\end{equation}\\

\newpage
\section{Quantisation Condition} \label{app:Quantisation}
Here we consider the quantisation condition for $U(1)$-restricted models, focussing for definiteness on $SU(5) \times U(1)$. Firstly, we note the following intersection properties for various powers of the resolution divisor $S$ necessary to resolve the singularities over the curve $C_{34}$:
\beq
 \int_{\hat{Y}_4} S (S+\aK) {\cal B}_a {\cal B}_b = 0, \qquad \int_{\hat{Y}_4} S^2 (S+\aK) {\cal B}_a = -\int_{C_{34}} {\cal B}_a, \qquad \int_{\hat{Y}_4} S^2 (S+\aK)^2 = \chi\left(C_{34}\right).
\eeq

Then from $c_2(\hat Y_4) = 5 \tw_X \wedge \cW \text{ mod } 2$ we find that the intersection  $I_{AB} := \int_{\hat{Y}_4} A \, B \, c_2(\hat Y_4)$ with two divisor classes $A$ and $B$  take the following form mod 2:
\beq
 \bal
  I_{Z \cB_a} = I_{Z E_i} = I_{Z S} = I_{\cB_a \cB_b} = I_{\cB_a E_i} = I_{\cB S} = I_{E_i S} = 0 \text{ mod } 2, \\
  I_{E_i E_j} = \left\{\begin{split} 0 \text{ mod } 2, \\ \int_{\cW} \aK \cW \text{ mod } 2, \end{split}\right\} \qquad I_{SS} = \int_{\cW} \aK \cW \text{ mod } 2.
 \eal
\eeq
Here $\cB_a$ is a base divisor class, $E_i$ are exceptional classes, and $Z$ is the section $z=0$. We observe that the only non-vanishing integeral which one can obtain in this way takes the form $\int_{\cW} \aK \cW$. 
The constraint $2 \int_B \aK^2 \cW = \int_B \aK \cW^2$, which is necessary in order for an F-theory $SU(n)$ model to possess a smooth Type IIB limit, then implies
that the integral of $c_2({\hat Y_4})$ with any of the above divisor classes is even. Similarly, no further restrictions arise from integrating $c_2(\hat Y_4)$ against the matter surfaces.  To determine if really no shift in the quantisation condition occurs one needs a basis of $H^4(\hat Y_4, \mathbb Z)$, which might include extra four-forms in addition to the ones tested in this appendix. This is beyond the scope of this article.

\newpage

\bibliography{rev-paperfluxesnew}  
\bibliographystyle{utphys}

\end{document}

%% file: double-cover.pdf_tex
%% Creator: Inkscape inkscape 0.48.2, www.inkscape.org
%% Accompanies image file 'double-cover.pdf' (pdf, eps, ps)
\begingroup%
  \makeatletter%
  \providecommand\color[2][]{%
    \errmessage{(Inkscape) Color is used for the text in Inkscape, but the package 'color.sty' is not loaded}%
    \renewcommand\color[2][]{}%
  }%
  \providecommand\transparent[1]{%
    \errmessage{(Inkscape) Transparency is used (non-zero) for the text in Inkscape, but the package 'transparent.sty' is not loaded}%
    \renewcommand\transparent[1]{}%
  }%
  \providecommand\rotatebox[2]{#2}%
  \ifx\svgwidth\undefined%
    \setlength{\unitlength}{640.265625bp}%
    \ifx\svgscale\undefined%
      \relax%
    \else%
      \setlength{\unitlength}{\unitlength * \real{\svgscale}}%
    \fi%
  \else%
    \setlength{\unitlength}{\svgwidth}%
  \fi%
  \global\let\svgwidth\undefined%
  \global\let\svgscale\undefined%
  \makeatother%
  \begin{picture}(1,0.65097982)%
    \put(0,0){\includegraphics[width=\unitlength]{double-cover.pdf}}%
    \put(0.44113039,0.59350367){\makebox(0,0)[lb]{\smash{$W$}}}%
    \put(0.29119262,0.2961271){\makebox(0,0)[lb]{\smash{$\tilde W$}}}%
    \put(0.00631086,0.13619347){\makebox(0,0)[lb]{\smash{$B$:}}}%
    \put(-0.00118603,0.6322376){\makebox(0,0)[lb]{\smash{$X_3$:}}}%
    \put(0.31493277,0.1411914){\makebox(0,0)[lb]{\smash{$\cW$}}}%
    \put(0.64104742,0.15118725){\makebox(0,0)[lb]{\smash{$\aK$}}}%
    \put(0.60981038,0.55227079){\makebox(0,0)[lb]{\smash{$D_{O7}$}}}%
    \put(0.46112209,0.22365717){\makebox(0,0)[lb]{\smash{$\pi$}}}%
  \end{picture}%
\endgroup%

%% file: non_restricted.pdf_tex
%% Creator: Inkscape inkscape 0.48.2, www.inkscape.org
%% Accompanies image file 'non_restricted.pdf' (pdf, eps, ps)
\begingroup%
  \makeatletter%
  \providecommand\color[2][]{%
    \errmessage{(Inkscape) Color is used for the text in Inkscape, but the package 'color.sty' is not loaded}%
    \renewcommand\color[2][]{}%
  }%
  \providecommand\transparent[1]{%
    \errmessage{(Inkscape) Transparency is used (non-zero) for the text in Inkscape, but the package 'transparent.sty' is not loaded}%
    \renewcommand\transparent[1]{}%
  }%
  \providecommand\rotatebox[2]{#2}%
  \ifx\svgwidth\undefined%
    \setlength{\unitlength}{751.36015625bp}%
    \ifx\svgscale\undefined%
      \relax%
    \else%
      \setlength{\unitlength}{\unitlength * \real{\svgscale}}%
    \fi%
  \else%
    \setlength{\unitlength}{\svgwidth}%
  \fi%
  \global\let\svgwidth\undefined%
  \global\let\svgscale\undefined%
  \makeatother%
  \begin{picture}(1,0.6639106)%
    \put(0,0){\includegraphics[width=\unitlength]{non_restricted.pdf}}%
    \put(0.46262767,0.59922575){\makebox(0,0)[lb]{\smash{\scriptsize $D_{V + \tilde{V}}$}}}%
    \put(0.81292573,0.51085268){\makebox(0,0)[lb]{\smash{\scriptsize $W$}}}%
    \put(0.82357308,0.23828034){\makebox(0,0)[lb]{\smash{\scriptsize $\tilde{W}$}}}%
    \put(0.82783203,0.42247962){\makebox(0,0)[lb]{\smash{\scriptsize $D_{O7}$}}}%
  \end{picture}%
\endgroup%

%% file: restricted.pdf_tex
%% Creator: Inkscape inkscape 0.48.2, www.inkscape.org
%% Accompanies image file 'restricted.pdf' (pdf, eps, ps)
\begingroup%
  \makeatletter%
  \providecommand\color[2][]{%
    \errmessage{(Inkscape) Color is used for the text in Inkscape, but the package 'color.sty' is not loaded}%
    \renewcommand\color[2][]{}%
  }%
  \providecommand\transparent[1]{%
    \errmessage{(Inkscape) Transparency is used (non-zero) for the text in Inkscape, but the package 'transparent.sty' is not loaded}%
    \renewcommand\transparent[1]{}%
  }%
  \providecommand\rotatebox[2]{#2}%
  \ifx\svgwidth\undefined%
    \setlength{\unitlength}{751.36015625bp}%
    \ifx\svgscale\undefined%
      \relax%
    \else%
      \setlength{\unitlength}{\unitlength * \real{\svgscale}}%
    \fi%
  \else%
    \setlength{\unitlength}{\svgwidth}%
  \fi%
  \global\let\svgwidth\undefined%
  \global\let\svgscale\undefined%
  \makeatother%
  \begin{picture}(1,0.55966627)%
    \put(0,0){\includegraphics[width=\unitlength]{restricted.pdf}}%
    \put(0.32314729,0.54590997){\makebox(0,0)[lb]{\smash{\scriptsize $V$}}}%
    \put(0.46262767,0.5437805){\makebox(0,0)[lb]{\smash{\scriptsize $\tilde{V}$}}}%
    \put(0.81292573,0.45540743){\makebox(0,0)[lb]{\smash{\scriptsize $W$}}}%
    \put(0.82357308,0.18283509){\makebox(0,0)[lb]{\smash{\scriptsize $\tilde{W}$}}}%
    \put(0.82783203,0.36703437){\makebox(0,0)[lb]{\smash{\scriptsize $D_{O7}$}}}%
  \end{picture}%
\endgroup%

%% file: polytope.pdf_tex
%% Creator: Inkscape inkscape 0.48.2, www.inkscape.org
%% Accompanies image file 'polytope.pdf' (pdf, eps, ps)
\begingroup%
  \makeatletter%
  \providecommand\color[2][]{%
    \errmessage{(Inkscape) Color is used for the text in Inkscape, but the package 'color.sty' is not loaded}%
    \renewcommand\color[2][]{}%
  }%
  \providecommand\transparent[1]{%
    \errmessage{(Inkscape) Transparency is used (non-zero) for the text in Inkscape, but the package 'transparent.sty' is not loaded}%
    \renewcommand\transparent[1]{}%
  }%
  \providecommand\rotatebox[2]{#2}%
  \ifx\svgwidth\undefined%
    \setlength{\unitlength}{302.46015625bp}%
    \ifx\svgscale\undefined%
      \relax%
    \else%
      \setlength{\unitlength}{\unitlength * \real{\svgscale}}%
    \fi%
  \else%
    \setlength{\unitlength}{\svgwidth}%
  \fi%
  \global\let\svgwidth\undefined%
  \global\let\svgscale\undefined%
  \makeatother%
  \begin{picture}(1,0.59195813)%
    \put(0,0){\includegraphics[width=\unitlength]{polytope.pdf}}%
    \put(0.88606712,0.55778525){\makebox(0,0)[lb]{\smash{$e_0$}}}%
    \put(0.89400205,0.24567802){\makebox(0,0)[lb]{\smash{$z$}}}%
    \put(0.54222018,0.5181106){\makebox(0,0)[lb]{\smash{$e_1$}}}%
    \put(0.17456845,0.01027511){\makebox(0,0)[lb]{\smash{$y$}}}%
    \put(0.01586986,0.11607417){\makebox(0,0)[lb]{\smash{$x$}}}%
  \end{picture}%
\endgroup%

%% file: paperfluxesv15.bbl
\providecommand{\href}[2]{#2}\begingroup\raggedright\begin{thebibliography}{10}

\bibitem{Vafa:1996xn}
C.~Vafa, ``{Evidence for F-Theory},'' {\em Nucl. Phys.} {\bf B469} (1996)
  403--418,
\href{http://www.arXiv.org/abs/hep-th/9602022}{{\tt hep-th/9602022}}.
%%CITATION = HEP-TH/9602022;%%.

\bibitem{Morrison:1996na}
D.~R. Morrison and C.~Vafa, ``{Compactifications of F theory on Calabi-Yau
  threefolds. 1},'' {\em Nucl.Phys.} {\bf B473} (1996) 74--92,
\href{http://www.arXiv.org/abs/hep-th/9602114}{{\tt hep-th/9602114}}.
%%CITATION = HEP-TH/9602114;%%.

\bibitem{Morrison:1996pp}
D.~R. Morrison and C.~Vafa, ``{Compactifications of F theory on Calabi-Yau
  threefolds. 2.},'' {\em Nucl.Phys.} {\bf B476} (1996) 437--469,
\href{http://www.arXiv.org/abs/hep-th/9603161}{{\tt hep-th/9603161}}.
%%CITATION = HEP-TH/9603161;%%.

\bibitem{Donagi:2008ca}
R.~Donagi and M.~Wijnholt, ``{Model Building with F-Theory},''
\href{http://www.arXiv.org/abs/0802.2969}{{\tt 0802.2969}}.
%%CITATION = 0802.2969;%%.

\bibitem{Beasley:2008dc}
C.~Beasley, J.~J. Heckman, and C.~Vafa, ``{GUTs and Exceptional Branes in
  F-theory - I},'' {\em JHEP} {\bf 01} (2009) 058,
\href{http://www.arXiv.org/abs/0802.3391}{{\tt 0802.3391}}.
%%CITATION = 0802.3391;%%.

\bibitem{Beasley:2008kw}
C.~Beasley, J.~J. Heckman, and C.~Vafa, ``{GUTs and Exceptional Branes in
  F-theory - II: Experimental Predictions},'' {\em JHEP} {\bf 01} (2009) 059,
\href{http://www.arXiv.org/abs/0806.0102}{{\tt 0806.0102}}.
%%CITATION = 0806.0102;%%.

\bibitem{Donagi:2008kj}
R.~Donagi and M.~Wijnholt, ``{Breaking GUT Groups in F-Theory},''
\href{http://www.arXiv.org/abs/0808.2223}{{\tt 0808.2223}}.
%%CITATION = ARXIV:0808.2223;%%.

\bibitem{Hayashi:2008ba}
H.~Hayashi, R.~Tatar, Y.~Toda, T.~Watari, and M.~Yamazaki, ``{New Aspects of
  Heterotic--F Theory Duality},'' {\em Nucl. Phys.} {\bf B806} (2009) 224--299,
\href{http://www.arXiv.org/abs/0805.1057}{{\tt 0805.1057}}.
%%CITATION = 0805.1057;%%.

\bibitem{Bershadsky:1996nh}
M.~Bershadsky {\em et al.}, ``{Geometric singularities and enhanced gauge
  symmetries},'' {\em Nucl. Phys.} {\bf B481} (1996) 215--252,
\href{http://www.arXiv.org/abs/hep-th/9605200}{{\tt hep-th/9605200}}.
%%CITATION = HEP-TH/9605200;%%.

\bibitem{Katz:1996xe}
S.~H. Katz and C.~Vafa, ``{Matter from geometry},'' {\em Nucl.Phys.} {\bf B497}
  (1997) 146--154, \href{http://www.arXiv.org/abs/hep-th/9606086}{{\tt
  hep-th/9606086}}.

\bibitem{Hayashi:2009ge}
H.~Hayashi, T.~Kawano, R.~Tatar, and T.~Watari, ``{Codimension-3 Singularities
  and Yukawa Couplings in F-theory},'' {\em Nucl.Phys.} {\bf B823} (2009)
  47--115,
\href{http://www.arXiv.org/abs/0901.4941}{{\tt 0901.4941}}.
%%CITATION = ARXIV:0901.4941;%%.

\bibitem{Denef:2008wq}
F.~Denef, ``{Les Houches Lectures on Constructing String Vacua},''
\href{http://www.arXiv.org/abs/0803.1194}{{\tt 0803.1194}}.
%%CITATION = 0803.1194;%%.

\bibitem{Weigand:2010wm}
T.~Weigand, ``{Lectures on F-theory compactifications and model building},''
  {\em Class. Quant. Grav.} {\bf 27} (2010) 214004,
\href{http://www.arXiv.org/abs/1009.3497}{{\tt 1009.3497}}.
%%CITATION = 1009.3497;%%.

\bibitem{Blumenhagen:2009yv}
R.~Blumenhagen, T.~W. Grimm, B.~Jurke, and T.~Weigand, ``{Global F-theory
  GUTs},'' {\em Nucl. Phys.} {\bf B829} (2010) 325--369,
\href{http://www.arXiv.org/abs/0908.1784}{{\tt 0908.1784}}.
%%CITATION = 0908.1784;%%.

\bibitem{Grimm:2009yu}
T.~W. Grimm, S.~Krause, and T.~Weigand, ``{F-Theory GUT Vacua on Compact
  Calabi-Yau Fourfolds},'' {\em JHEP} {\bf 07} (2010) 037,
\href{http://www.arXiv.org/abs/0912.3524}{{\tt 0912.3524}}.
%%CITATION = 0912.3524;%%.

\bibitem{Chen:2010ts}
C.-M. Chen, J.~Knapp, M.~Kreuzer, and C.~Mayrhofer, ``{Global SO(10) F-theory
  GUTs},'' {\em JHEP} {\bf 1010} (2010) 057,
  \href{http://www.arXiv.org/abs/1005.5735}{{\tt 1005.5735}}.

\bibitem{Knapp:2011wk}
J.~Knapp, M.~Kreuzer, C.~Mayrhofer, and N.-O. Walliser, ``{Toric Construction
  of Global F-Theory GUTs},'' {\em JHEP} {\bf 03} (2011) 138,
\href{http://www.arXiv.org/abs/1101.4908}{{\tt 1101.4908}}.
%%CITATION = 1101.4908;%%.

\bibitem{mayrhofer:diss}
C.~Mayrhofer, {\em Compactifications of Type IIB String Theory and F-Theory
  Models by Means of Toric Geometry}.
\newblock PhD thesis, Vienna University of Technology, 11, 2010.

\bibitem{Marsano:2009ym}
J.~Marsano, N.~Saulina, and S.~Schafer-Nameki, ``{F-theory Compactifications
  for Supersymmetric GUTs},'' {\em JHEP} {\bf 08} (2009) 030,
\href{http://www.arXiv.org/abs/0904.3932}{{\tt 0904.3932}}.
%%CITATION = 0904.3932;%%.

\bibitem{Esole:2011sm}
M.~Esole and S.-T. Yau, ``{Small resolutions of SU(5)-models in F-theory},''
  \href{http://www.arXiv.org/abs/1107.0733}{{\tt 1107.0733}}.

\bibitem{Marsano:2011hv}
J.~Marsano and S.~Schafer-Nameki, ``{Yukawas, G-flux, and Spectral Covers from
  Resolved Calabi- Yau's},'' {\em JHEP} {\bf 11} (2011) 098,
\href{http://www.arXiv.org/abs/1108.1794}{{\tt 1108.1794}}.
%%CITATION = 1108.1794;%%.

\bibitem{Krause:2011xj}
S.~Krause, C.~Mayrhofer, and T.~Weigand, ``{$G_4$ flux, chiral matter and
  singularity resolution in F- theory compactifications},'' {\em Nucl. Phys.}
  {\bf B858} (2012) 1--47,
\href{http://www.arXiv.org/abs/1109.3454}{{\tt 1109.3454}}.
%%CITATION = 1109.3454;%%.

\bibitem{Grimm:2011fx}
T.~W. Grimm and H.~Hayashi, ``{F-theory fluxes, Chirality and Chern-Simons
  theories},'' \href{http://www.arXiv.org/abs/1111.1232}{{\tt 1111.1232}}.
53 pages, 5 figures.
%%CITATION = ARXIV:1111.1232;%%.

\bibitem{Kumar:2009ac}
V.~Kumar, D.~R. Morrison, and W.~Taylor, ``{Mapping 6D N = 1 supergravities to
  F-theory},'' {\em JHEP} {\bf 1002} (2010) 099,
\href{http://www.arXiv.org/abs/0911.3393}{{\tt 0911.3393}}.
%%CITATION = ARXIV:0911.3393;%%.

\bibitem{Grassi:2011hq}
A.~Grassi and D.~R. Morrison, ``{Anomalies and the Euler characteristic of
  elliptic Calabi-Yau threefolds},''
  \href{http://www.arXiv.org/abs/1109.0042}{{\tt 1109.0042}}.
63 pages, 15 tables. v2: minor changes.
%%CITATION = ARXIV:1109.0042;%%.

\bibitem{Braun:2011ux}
V.~Braun, ``{Toric Elliptic Fibrations and F-Theory Compactifications},''
\href{http://www.arXiv.org/abs/1110.4883}{{\tt 1110.4883}}.
%%CITATION = ARXIV:1110.4883;%%.

\bibitem{Park:2011ji}
D.~S. Park, ``{Anomaly Equations and Intersection Theory},'' {\em JHEP} {\bf
  01} (2012) 093,
\href{http://www.arXiv.org/abs/1111.2351}{{\tt 1111.2351}}.
%%CITATION = 1111.2351;%%.

\bibitem{Bonetti:2011mw}
F.~Bonetti and T.~W. Grimm, ``{Six-dimensional (1,0) effective action of
  F-theory via M- theory on Calabi-Yau threefolds},''
\href{http://www.arXiv.org/abs/1112.1082}{{\tt 1112.1082}}.
%%CITATION = 1112.1082;%%.

\bibitem{Morrison:2012td}
D.~R. Morrison and W.~Taylor, ``{Classifying bases for 6D F-theory models},''
\href{http://www.arXiv.org/abs/1201.1943}{{\tt 1201.1943}}.
%%CITATION = ARXIV:1201.1943;%%.

\bibitem{Braun:2011zm}
A.~P. Braun, A.~Collinucci, and R.~Valandro, ``{G-flux in F-theory and
  algebraic cycles},'' {\em Nucl. Phys.} {\bf B856} (2012) 129--179,
\href{http://www.arXiv.org/abs/1107.5337}{{\tt 1107.5337}}.
%%CITATION = 1107.5337;%%.

\bibitem{Grimm:2011tb}
T.~W. Grimm, M.~Kerstan, E.~Palti, and T.~Weigand, ``{Massive Abelian Gauge
  Symmetries and Fluxes in F-theory},'' {\em JHEP} {\bf 12} (2011) 004,
\href{http://www.arXiv.org/abs/1107.3842}{{\tt 1107.3842}}.
%%CITATION = 1107.3842;%%.

\bibitem{Donagi:2009ra}
R.~Donagi and M.~Wijnholt, ``{Higgs Bundles and UV Completion in F-Theory},''
\href{http://www.arXiv.org/abs/0904.1218}{{\tt 0904.1218}}.
%%CITATION = 0904.1218;%%.

\bibitem{Marsano:2009gv}
J.~Marsano, N.~Saulina, and S.~Schafer-Nameki, ``{Monodromies, Fluxes, and
  Compact Three-Generation F-theory GUTs},'' {\em JHEP} {\bf 0908} (2009) 046,
  \href{http://www.arXiv.org/abs/0906.4672}{{\tt 0906.4672}}.

\bibitem{Marsano:2011nn}
J.~Marsano, N.~Saulina, and S.~Sch{\"a}fer-Nameki, ``{On G-flux, M5 instantons,
  and U(1)s in F-theory},'' \href{http://www.arXiv.org/abs/1107.1718}{{\tt
  1107.1718}}.

\bibitem{Wijnholt:2012fx}
M.~Wijnholt, ``{Higgs Bundles and String Phenomenology},''
\href{http://www.arXiv.org/abs/1201.2520}{{\tt 1201.2520}}.
%%CITATION = ARXIV:1201.2520;%%.

\bibitem{Lust:2005bd}
D.~Lust, P.~Mayr, S.~Reffert, and S.~Stieberger, ``{F-theory flux,
  destabilization of orientifolds and soft terms on D7-branes},'' {\em
  Nucl.Phys.} {\bf B732} (2006) 243--290,
\href{http://www.arXiv.org/abs/hep-th/0501139}{{\tt hep-th/0501139}}.
%%CITATION = HEP-TH/0501139;%%.

\bibitem{Alim:2009bx}
M.~Alim, M.~Hecht, H.~Jockers, P.~Mayr, A.~Mertens, {\em et al.}, ``{Hints for
  Off-Shell Mirror Symmetry in type II/F-theory Compactifications},'' {\em
  Nucl.Phys.} {\bf B841} (2010) 303--338,
\href{http://www.arXiv.org/abs/0909.1842}{{\tt 0909.1842}}.
%%CITATION = ARXIV:0909.1842;%%.

\bibitem{Grimm:2009ef}
T.~W. Grimm, T.-W. Ha, A.~Klemm, and D.~Klevers, ``{Computing Brane and Flux
  Superpotentials in F-theory Compactifications},'' {\em JHEP} {\bf 1004}
  (2010) 015,
\href{http://www.arXiv.org/abs/0909.2025}{{\tt 0909.2025}}.
%%CITATION = ARXIV:0909.2025;%%.

\bibitem{Grimm:2012rg}
T.~W. Grimm, D.~Klevers, and M.~Poretschkin, ``{Fluxes and Warping for Gauge
  Couplings in F-theory},''
\href{http://www.arXiv.org/abs/1202.0285}{{\tt 1202.0285}}.
%%CITATION = ARXIV:1202.0285;%%.

\bibitem{Grimm:2010ez}
T.~W. Grimm and T.~Weigand, ``{On Abelian Gauge Symmetries and Proton Decay in
  Global F- theory GUTs},'' {\em Phys. Rev.} {\bf D82} (2010) 086009,
\href{http://www.arXiv.org/abs/1006.0226}{{\tt 1006.0226}}.
%%CITATION = 1006.0226;%%.

\bibitem{Collinucci:2008pf}
A.~Collinucci, F.~Denef, and M.~Esole, ``{D-brane Deconstructions in IIB
  Orientifolds},'' {\em JHEP} {\bf 02} (2009) 005,
\href{http://www.arXiv.org/abs/0805.1573}{{\tt 0805.1573}}.
%%CITATION = 0805.1573;%%.

\bibitem{Greene:1993vm}
B.~R. Greene, D.~R. Morrison, and M.~Plesser, ``{Mirror manifolds in higher
  dimension},'' {\em Commun.Math.Phys.} {\bf 173} (1995) 559--598,
  \href{http://www.arXiv.org/abs/hep-th/9402119}{{\tt hep-th/9402119}}.

\bibitem{Mayr:1996sh}
P.~Mayr, ``{Mirror symmetry, N=1 superpotentials and tensionless strings on
  Calabi-Yau four folds},'' {\em Nucl.Phys.} {\bf B494} (1997) 489--545,
\href{http://www.arXiv.org/abs/hep-th/9610162}{{\tt hep-th/9610162}}.
%%CITATION = HEP-TH/9610162;%%.

\bibitem{Collinucci:2010gz}
A.~Collinucci and R.~Savelli, ``{On Flux Quantization in F-Theory},''
\href{http://www.arXiv.org/abs/1011.6388}{{\tt 1011.6388}}.
%%CITATION = 1011.6388;%%.

\bibitem{Katz:2011qp}
S.~Katz, D.~R. Morrison, S.~Schafer-Nameki, and J.~Sully, ``{Tate's algorithm
  and F-theory},'' {\em JHEP} {\bf 1108} (2011) 094,
\href{http://www.arXiv.org/abs/1106.3854}{{\tt 1106.3854}}.
%%CITATION = ARXIV:1106.3854;%%.

\bibitem{Morrison:2011mb}
D.~R. Morrison and W.~Taylor, ``{Matter and singularities},'' {\em JHEP} {\bf
  01} (2012) 022,
\href{http://www.arXiv.org/abs/1106.3563}{{\tt 1106.3563}}.
%%CITATION = 1106.3563;%%.

\bibitem{Sen:1997gv}
A.~Sen, ``{Orientifold limit of F theory vacua},'' {\em Phys.Rev.} {\bf D55}
  (1997) 7345--7349,
\href{http://www.arXiv.org/abs/hep-th/9702165}{{\tt hep-th/9702165}}.
%%CITATION = HEP-TH/9702165;%%.

\bibitem{Collinucci:2008zs}
A.~Collinucci, ``{New F-theory lifts},'' {\em JHEP} {\bf 0908} (2009) 076,
\href{http://www.arXiv.org/abs/0812.0175}{{\tt 0812.0175}}.
%%CITATION = ARXIV:0812.0175;%%.

\bibitem{Collinucci:2009uh}
A.~Collinucci, ``{New F-theory lifts. II. Permutation orientifolds and enhanced
  singularities},'' {\em JHEP} {\bf 1004} (2010) 076,
\href{http://www.arXiv.org/abs/0906.0003}{{\tt 0906.0003}}.
%%CITATION = ARXIV:0906.0003;%%.

\bibitem{Blumenhagen:2009up}
R.~Blumenhagen, T.~W. Grimm, B.~Jurke, and T.~Weigand, ``{F-theory uplifts and
  GUTs},'' {\em JHEP} {\bf 0909} (2009) 053,
\href{http://www.arXiv.org/abs/0906.0013}{{\tt 0906.0013}}.
%%CITATION = ARXIV:0906.0013;%%.

\bibitem{Witten:1996md}
E.~Witten, ``{On flux quantization in M theory and the effective action},''
  {\em J.Geom.Phys.} {\bf 22} (1997) 1--13,
  \href{http://www.arXiv.org/abs/hep-th/9609122}{{\tt hep-th/9609122}}.

\bibitem{Grimm:2010ks}
T.~W. Grimm, ``{The N=1 effective action of F-theory compactifications},'' {\em
  Nucl.Phys.} {\bf B845} (2011) 48--92,
  \href{http://www.arXiv.org/abs/1008.4133}{{\tt 1008.4133}}.

\bibitem{Friedman:1997yq}
R.~Friedman, J.~Morgan, and E.~Witten, ``{Vector bundles and F theory},'' {\em
  Commun.Math.Phys.} {\bf 187} (1997) 679--743,
\href{http://www.arXiv.org/abs/hep-th/9701162}{{\tt hep-th/9701162}}.
%%CITATION = HEP-TH/9701162;%%.

\bibitem{DonagiSC}
R.~Donagi, ``{Principal bundles on elliptic fibrations},'' {\em Asian J. Math}
  {\bf 1} (1997) 214--223,
\href{http://www.arXiv.org/abs/alg-geom/9702002}{{\tt alg-geom/9702002}}.
%%CITATION =ALG-GEOM/9702002 ;%%.

\bibitem{Curio:1998bva}
G.~Curio and R.~Y. Donagi, ``{Moduli in N=1 heterotic / F theory duality},''
  {\em Nucl.Phys.} {\bf B518} (1998) 603--631,
\href{http://www.arXiv.org/abs/hep-th/9801057}{{\tt hep-th/9801057}}.
%%CITATION = HEP-TH/9801057;%%.

\bibitem{Andreas:1999ng}
B.~Andreas and G.~Curio, ``{On discrete twist and four flux in N=1 heterotic /
  F theory compactifications},'' {\em Adv.Theor.Math.Phys.} {\bf 3} (1999)
  1325--1413,
\href{http://www.arXiv.org/abs/hep-th/9908193}{{\tt hep-th/9908193}}.
%%CITATION = HEP-TH/9908193;%%.

\bibitem{Andreas:2004ja}
B.~Andreas and D.~Hernandez~Ruiperez, ``{U(n) vector bundles on Calabi-Yau
  threefolds for string theory compactifications},'' {\em Adv.Theor.Math.Phys.}
  {\bf 9} (2005) 253--284,
\href{http://www.arXiv.org/abs/hep-th/0410170}{{\tt hep-th/0410170}}.
%%CITATION = HEP-TH/0410170;%%.

\bibitem{Blumenhagen:2005ga}
R.~Blumenhagen, G.~Honecker, and T.~Weigand, ``{Loop-corrected
  compactifications of the heterotic string with line bundles},'' {\em JHEP}
  {\bf 0506} (2005) 020,
\href{http://www.arXiv.org/abs/hep-th/0504232}{{\tt hep-th/0504232}}.
%%CITATION = HEP-TH/0504232;%%.

\bibitem{Tatar:2006dc}
R.~Tatar and T.~Watari, ``{Proton decay, Yukawa couplings and underlying gauge
  symmetry in string theory},'' {\em Nucl.Phys.} {\bf B747} (2006) 212--265,
\href{http://www.arXiv.org/abs/hep-th/0602238}{{\tt hep-th/0602238}}.
%%CITATION = HEP-TH/0602238;%%.

\bibitem{Blumenhagen:2006ux}
R.~Blumenhagen, S.~Moster, and T.~Weigand, ``{Heterotic GUT and standard model
  vacua from simply connected Calabi-Yau manifolds},'' {\em Nucl.Phys.} {\bf
  B751} (2006) 186--221, \href{http://www.arXiv.org/abs/hep-th/0603015}{{\tt
  hep-th/0603015}}.

\bibitem{Jockers:2004yj}
H.~Jockers and J.~Louis, ``{The Effective action of D7-branes in N = 1
  Calabi-Yau orientifolds},'' {\em Nucl.Phys.} {\bf B705} (2005) 167--211,
\href{http://www.arXiv.org/abs/hep-th/0409098}{{\tt hep-th/0409098}}.
%%CITATION = HEP-TH/0409098;%%.

\bibitem{Plauschinn:2008yd}
E.~Plauschinn, ``{The Generalized Green-Schwarz Mechanism for Type IIB
  Orientifolds with D3- and D7-Branes},'' {\em JHEP} {\bf 0905} (2009) 062,
\href{http://www.arXiv.org/abs/0811.2804}{{\tt 0811.2804}}.
%%CITATION = ARXIV:0811.2804;%%.

\bibitem{Blumenhagen:2008zz}
R.~Blumenhagen, V.~Braun, T.~W. Grimm, and T.~Weigand, ``{GUTs in Type IIB
  Orientifold Compactifications},'' {\em Nucl. Phys.} {\bf B815} (2009) 1--94,
\href{http://www.arXiv.org/abs/0811.2936}{{\tt 0811.2936}}.
%%CITATION = 0811.2936;%%.

\bibitem{Freed:1999vc}
D.~S. Freed and E.~Witten, ``{Anomalies in string theory with D-branes},''
  \href{http://www.arXiv.org/abs/hep-th/9907189}{{\tt hep-th/9907189}}.

\bibitem{Grimm:2011dj}
T.~W. Grimm, M.~Kerstan, E.~Palti, and T.~Weigand, ``{On Fluxed Instantons and
  Moduli Stabilisation in IIB Orientifolds and F-theory},'' {\em Phys.Rev.}
  {\bf D84} (2011) 066001,
\href{http://www.arXiv.org/abs/1105.3193}{{\tt 1105.3193}}.
%%CITATION = ARXIV:1105.3193;%%.

\bibitem{Blumenhagen:2010ja}
R.~Blumenhagen, A.~Collinucci, and B.~Jurke, ``{On Instanton Effects in
  F-theory},'' {\em JHEP} {\bf 1008} (2010) 079,
  \href{http://www.arXiv.org/abs/1002.1894}{{\tt 1002.1894}}.

\bibitem{Cvetic:2010rq}
M.~Cveti{\v c}, I.~Garcia-Etxebarria, and J.~Halverson, ``{Global F-theory
  Models: Instantons and Gauge Dynamics},'' {\em JHEP} {\bf 1101} (2011) 073,
  \href{http://www.arXiv.org/abs/1003.5337}{{\tt 1003.5337}}.

\bibitem{Donagi:2010pd}
R.~Donagi and M.~Wijnholt, ``{MSW Instantons},''
  \href{http://www.arXiv.org/abs/1005.5391}{{\tt 1005.5391}}.

\bibitem{Cvetic:2011gp}
M.~Cveti{\v c}, J.~Halverson, and I.~Garcia-Etxebarria, ``{Three Looks at
  Instantons in F-theory -- New Insights from Anomaly Inflow, String Junctions
  and Heterotic Duality},'' \href{http://www.arXiv.org/abs/1107.2388}{{\tt
  1107.2388}}.

\bibitem{Bianchi:2011qh}
M.~Bianchi, A.~Collinucci, and L.~Martucci, ``{Magnetized E3-brane instantons
  in F-theory},'' \href{http://www.arXiv.org/abs/1107.3732}{{\tt 1107.3732}}.

\bibitem{Gaiotto:2005rp}
D.~Gaiotto, M.~Guica, L.~Huang, A.~Simons, A.~Strominger, {\em et al.},
  ``{D4-D0 branes on the quintic},'' {\em JHEP} {\bf 0603} (2006) 019,
\href{http://www.arXiv.org/abs/hep-th/0509168}{{\tt hep-th/0509168}}.
%%CITATION = HEP-TH/0509168;%%.

\bibitem{Donagi:2011jy}
R.~Donagi and M.~Wijnholt, ``{Gluing Branes, I},''
\href{http://www.arXiv.org/abs/1104.2610}{{\tt 1104.2610}}.
%%CITATION = ARXIV:1104.2610;%%.

\end{thebibliography}\endgroup
